\documentclass[12pt, leqno]{article}

\usepackage[left=3cm,right=3cm,top=2cm,bottom=2cm]{geometry}

\usepackage{iftex}

\usepackage{indentfirst}

\usepackage[normalem]{ulem}

\usepackage{setspace}

\usepackage[english]{babel}

\usepackage{kotex}

\usepackage{hyphenat}

\usepackage[defaultlines=2,all]{nowidow}

\usepackage{color}
\usepackage{xcolor}

\usepackage{csquotes}

\usepackage{titlecaps}

\usepackage[leqno]{amsmath}

\usepackage{fouriernc}

\PassOptionsToPackage{hyphens}{url}

\usepackage{xr-hyper}

\usepackage[hidelinks]{hyperref}

\makeatletter

\newcommand*{\addFileDependency}[1]{
\typeout{(#1)}
\@addtofilelist{#1}
\IfFileExists{#1}{}{\typeout{No file #1.}}
}\makeatother

\usepackage{titlesec}
\usepackage{capt-of} 

\renewcommand{\thesection}{\Roman{section}}
\titleformat{\section}
  {\centering\normalfont\large\scshape}
  {\thesection.}
  {1em}
  {} 
\titlespacing{\subsection}
  {0pt}                
  {6ex} 
  {1ex plus .25ex minus 0ex} 
  [1em]                

\renewcommand{\thesubsection}{\thesection.\Alph{subsection}}
\titleformat{\subsection}
  {\normalfont\itshape}
  {\textit{\thesubsection}.}
  {1em}
  {}
\titlespacing{\subsection}
  {0pt}                
  {3.5ex plus 1ex minus .2ex} 
  {1ex plus .25ex minus 0ex} 
  [1em]                

\renewcommand{\thesubsubsection}{\arabic{subsubsection}}
\titleformat{\subsubsection}[runin]
  {\normalfont\itshape}
  {\thesubsubsection.}
  {1em}
  {\titlecap}
  [.~~] 
\titlespacing{\subsubsection}
  {2em} 
  {3ex plus 1ex minus .2ex} 
  {0pt} 

\renewcommand{\theparagraph}{\roman{paragraph}}
\titleformat{\paragraph}[runin]
  {\normalfont\itshape} 
  {\theparagraph.} 
  {1em}
  {}
  [.~~] 

\titlespacing{\paragraph}
  {2em} 
  {0pt} 
  {0pt} 

\usepackage{caption}
\usepackage{subcaption}  

\renewcommand{\thetable}{\Roman{table}}
\renewcommand{\thefigure}{\Roman{figure}}

\DeclareCaptionLabelFormat{custom}{\textsc{#1}~#2}
\makeatletter
\newcommand{\begina}{%
  \renewcommand{\thetable}{A.\arabic{table}}%
  \renewcommand{\theHtable}{A.\arabic{table}}%
  \setcounter{table}{0}%
  \renewcommand{\thefigure}{A.\arabic{figure}}%
  \renewcommand{\theHfigure}{A.\arabic{figure}}%
  \setcounter{figure}{0}%
}
\newcommand{\beginb}{%
  \renewcommand{\thetable}{B.\arabic{table}}%
  \renewcommand{\theHtable}{B.\arabic{table}}%
  \setcounter{table}{0}%
  \renewcommand{\thefigure}{B.\arabic{figure}}%
  \renewcommand{\theHfigure}{B.\arabic{figure}}%
  \setcounter{figure}{0}%
}
\newcommand{\beginc}{%
  \renewcommand{\thetable}{C.\arabic{table}}%
  \renewcommand{\theHtable}{C.\arabic{table}}%
  \setcounter{table}{0}%
  \renewcommand{\thefigure}{C.\arabic{figure}}%
  \renewcommand{\theHfigure}{C.\arabic{figure}}%
  \setcounter{figure}{0}%
}
\newcommand{\begind}{%
  \renewcommand{\thetable}{D.\arabic{table}}%
  \renewcommand{\theHtable}{D.\arabic{table}}%
  \setcounter{table}{0}%
  \renewcommand{\thefigure}{D.\arabic{figure}}%
  \renewcommand{\theHfigure}{D.\arabic{figure}}%
  \setcounter{figure}{0}%
}
\newcommand{\begine}{%
  \renewcommand{\thetable}{E.\arabic{table}}%
  \renewcommand{\theHtable}{E.\arabic{table}}%
  \setcounter{table}{0}%
  \renewcommand{\thefigure}{E.\arabic{figure}}%
  \renewcommand{\theHfigure}{E.\arabic{figure}}%
  \setcounter{figure}{0}%
}
\newcommand{\beginf}{%
  \renewcommand{\thetable}{F.\arabic{table}}%
  \renewcommand{\theHtable}{F.\arabic{table}}%
  \setcounter{table}{0}%
  \renewcommand{\thefigure}{F.\arabic{figure}}%
  \renewcommand{\theHfigure}{F.\arabic{figure}}%
  \setcounter{figure}{0}%
}
\newcommand{\beging}{%
  \renewcommand{\thetable}{G.\arabic{table}}%
  \renewcommand{\theHtable}{G.\arabic{table}}%
  \setcounter{table}{0}%
  \renewcommand{\thefigure}{G.\arabic{figure}}%
  \renewcommand{\theHfigure}{G.\arabic{figure}}%
  \setcounter{figure}{0}%
}

\makeatother
\PassOptionsToPackage{norule, ragged, hang}{footmisc}
\RequirePackage{fnpct}
\RequirePackage{footmisc}

\makeatletter
\renewcommand\@makefntext[1]{
    \@thefnmark.~#1}
\makeatother

\usepackage{subfiles}
\usepackage{catchfile}
\usepackage{graphicx} 
\usepackage{calc} 
\usepackage{placeins}
\usepackage{rotating}
\usepackage{float}
\usepackage{adjustbox}
\usepackage{wrapfig}
\usepackage{lscape}
\usepackage{pdflscape}
\usepackage{changepage}

\usepackage{etoolbox}
\makeatletter
\patchcmd\longtable{\par}{\if@noskipsec\mbox{}\fi\par}{}{}
\makeatother

\newcommand{\blandscape}{\begin{landscape}}
\newcommand{\elandscape}{\end{landscape}}

\renewcommand{\arraystretch}{1.3}

\usepackage{longtable,booktabs,array}
\usepackage{colortbl}
\usepackage{tabu}
\usepackage{tabularx}
\usepackage{multicol}
\usepackage{makecell}
\usepackage{multirow}
\usepackage{threeparttable}
\usepackage{threeparttablex}
\usepackage{hhline}

\IfFileExists{footnotehyper.sty}{\usepackage{footnotehyper}}{\usepackage{footnote}}
\makesavenoteenv{longtable}
\usepackage{graphicx}
\makeatletter
\def\maxwidth{\ifdim\Gin@nat@width>\linewidth\linewidth\else\Gin@nat@width\fi}
\def\maxheight{\ifdim\Gin@nat@height>\textheight\textheight\else\Gin@nat@height\fi}
\makeatother

\setkeys{Gin}{width=\maxwidth,height=\maxheight,keepaspectratio}

\makeatletter
\def\fps@figure{htbp}
\makeatother
\usepackage{natbib}
\makeatletter
\newcommand{\@arxivsupplabel}[2]{\expandafter\gdef\csname r@#1\endcsname{{#2}{0}{#2}{}{}}} 

\@arxivsupplabel{sec:furthernixonshock}{S.1.1}
\@arxivsupplabel{sec:suppappendixlegalacts}{S.1.2.1}
\@arxivsupplabel{sec:suppappendixjapan}{S.1.2.2}
\@arxivsupplabel{sec:supphistoryselection}{S.1.2.3}
\@arxivsupplabel{sec:suppappendixplans}{S.1.2.4}
\@arxivsupplabel{sec:suppappendixtradepolicy}{S.1.2.5}
\@arxivsupplabel{sec:supplementtaxratecalc}{S.1.3}
\@arxivsupplabel{sec:supplementalliberalization}{S.1.4}
\@arxivsupplabel{sec:supplementharmonization}{S.2.1}
\@arxivsupplabel{tab:supptablerollingoutput}{Table S.1}
\@arxivsupplabel{fig:suppappendixmicrotfp}{Figure S.1}
\@arxivsupplabel{fig:suppcontinuousdev}{Figure S.2}
\@arxivsupplabel{tab:supptablerollingdevelopment}{Table S.2}
\@arxivsupplabel{sec:suppunidorobustness}{S.4.1}
\@arxivsupplabel{fig:suppddaltrca}{Figure S.3}
\@arxivsupplabel{tab:supptablerollingcapital2}{Table S.3}
\@arxivsupplabel{tab:supprollingforwardoutput}{Table S.4}
\@arxivsupplabel{tab:supprcaforwardlink}{Table S.5}

\makeatother

\usepackage{catchfile}

\makeatletter
\newcommand{\readFileContent}[2]{%
  \begingroup
  \catcode`\^^M=12
  \gdef\@temp{#2}%
  \expandafter\endgroup
  \expandafter\CatchFileDef\csname #1\endcsname{\@temp}{}
}
\makeatother

\readFileContent{resultstfpmax}{results_tfpcrossection_maxtfp.tex}
\readFileContent{resultstfpmin}{results_tfpcrossection_mintfp.tex}

\readFileContent{resultssemiship}{results_semi_ship.tex}
\readFileContent{resultsolsship}{results_ols_ship.tex}

\readFileContent{resultslaborprodmax}{results_semi_max_yn.tex}
\readFileContent{resultslaborprodmin}{results_semi_min_yn.tex}

\readFileContent{resultsrcahcimean}{results_hci_rca_mean.tex}
\readFileContent{resultsrcanonhcimean}{results_nonhci_rca_mean.tex}

\readFileContent{resultslogrca}{results_log_rca.tex}

\readFileContent{resultsexportshare}{results_export_share.tex}
\readFileContent{resultsprobrca}{results_prob_rca.tex}

\readFileContent{resultssemiprice}{results_semi_prices.tex}
\readFileContent{resultsolsprice}{results_ols_prices.tex}

\readFileContent{resultsminprice}{results_min_price.tex}
\readFileContent{resultsmaxprice}{results_max_price.tex}

\readFileContent{resultssemilabor}{results_semi_labor.tex}
\readFileContent{resultsolslabor}{results_ols_labor.tex}
\readFileContent{resultssemilaborfour}{results_semi_labor_4digit.tex}

\readFileContent{resultsprobrcamin}{results_tradeprob_minrca.tex}
\readFileContent{resultsprobrcamax}{results_tradeprob_maxrca.tex}

\readFileContent{resultsmeanprobrca}{results_tradeprob_meanrcahci.tex}

\readFileContent{resultssemiinvest}{results_semi_invest.tex}
\readFileContent{resultssemicosts}{results_ols_invest.tex}

\readFileContent{resultssemicosts}{results_semi_costs.tex}
\readFileContent{resultsolscosts}{results_ols_costs.tex}

\readFileContent{resultforwardoutnonhci}{results_forwardoutput_nonhci.tex}
\readFileContent{resultforwardoutall}{results_forwardoutput_all.tex}

\readFileContent{resultforwardpricesnonhci}{results_forwardprices_nonhci.tex}
\readFileContent{resultforwardpricesall}{results_forwardprices_all.tex}

\title{\Large MANUFACTURING REVOLUTIONS: \linebreak INDUSTRIAL POLICY AND INDUSTRIALIZATION IN SOUTH KOREA\thanks{I am grateful to the editor, Nathan Nunn, Lawrence Katz, and the anonymous referees. I thank Daron Acemoglu, Robert Allen, Sam Bazzi, Sascha O. Becker, Timo Boppart, Eric Chaney, David Cole, Arin Dube, Alice Evans, Réka Juhász, Mounir Karadja, Max Kesy, Oliver Kim, Changkeun Lee, Weijia Li, Ernest Liu, Andreas Madestam, Javier Mejia, Matti Mitrunen, Arieda Muço, Aldo Musacchio, Suresh Naidu, Dwight Perkins, Erik Prawitz, Pablo Querubín, Dani Rodrik, Martin Rotemberg, Todd Tucker, Eric Verhoogen, Robert Wade, and Lisa Xu. I thank audiences at American University, ANU, Azim Premji University, Berkeley, Central European University, College de France, Columbia, Geneva Graduate Institute, EBRD, Econometric Society European Meetings, Harvard, IMT--Lucca, INSEAD, Kellogg, KCL, Korea Development Institute, LSE ID, MIT, NBER SI, NEUDC, University of Nottingham, NYU--Abu Dhabi, University of Oxford, OzClio, Peking University, QMUL, Seoul National University, University of Sussex, UMass--Amherst, University of Melbourne, UNSW, University of Technology Sydney, and University of Wollongong for valuable comments. I am especially indebted to my advisors Melissa Dell, Torsten Persson, James Robinson, and David Strömberg for their support. I benefited from the excellent assistance of Ida Brzezinska, Lottie Field, Seung Yeon Han, Shehryar Hasan, BoSuk Hong, Chan Kim, Véronica Pérez, Esha Vase, Hannah Wei, Cheongyeon Won, Hye Jin Won, Stephen Xu, and Kristen Yang. I thank the staff of the Bank of Korea and the Korea Development Institute.}}
\author{NATHAN LANE\thanks{University of Oxford, CESifo, and SoDa Labs. Correspondence: \url{nathaniel.lane@economics.ox.ac.uk}. Department of Economics, Manor Road Building, 10 Manor Pl, Oxford OX1 3UQ, United Kingdom.}}
\date{May 2025}

\begin{document}
\maketitle

\begin{abstract}
I study the impact of industrial policies on industrial development by considering an important episode during the East Asian miracle: South Korea's heavy and chemical industry (HCI) drive, 1973--1979. Based on newly assembled data, I use the introduction and termination of industrial policies to study their impacts during and after the intervention period. (1) I reveal that heavy-chemical industrial policies promoted the expansion and dynamic comparative advantage of directly targeted industries. (2) Using variation in exposure to policies through the input-output network, I demonstrate that the policy indirectly benefited downstream users of targeted intermediates. (3) The benefits of HCI persisted even after the policy ended, as some results were slower to appear. The findings suggest that the temporary drive shifted Korean manufacturing into more advanced markets and supported durable change. This study helps clarify the lessons drawn from the East Asian growth miracle. \textit{JEL} Codes: L5, O14, O25, N6. 
\linebreak
\end{abstract}

\section{Introduction}\label{sec:introduction}

Miracles by nature are mysterious. The forces behind the East Asian growth miracle are no exception. Industrial policy (IP) has defined Asia's postwar transformation \citep{Rodrik1995}. Historically, economists focusing on development saw these policies as essential for industrial development \citep{Rosenstein-Rodan1943,Nurkse1953,Hirschman1958}, and some have argued that they were instrumental to East Asia's ascent \citep{Wade1990,Amsden1992}. However, many economists have been skeptical of their use \citep{Baldwin1969,Krueger1990}, and others have argued that they played a counterproductive role in Asia \citep{Pack2000}. South Korea exemplifies both Asia's rapid transformation and controversies around industrial policy. At the beginning of the 1960s, South Korea was a politically unstable industrial laggard; however, by the 1980s, it had undergone the kind of manufacturing transformation that had taken Western economies over a century to achieve \citep{Nelson1999}. What role did industrial policy play in this transformation? As conversations about industrial policy have returned \citep{Juhasz2022}, empirical evidence surrounding its efficacy is scant, especially for the East Asian miracle \citep{Lane2019}.

I use the context of the heavy industrial drive to employ a dynamic differences-in-differences (DD) estimation strategy. I evaluate the impact of South Korean industrial policy by comparing changes in outcomes between targeted (treated) and non-targeted (untreated) manufacturing industries each year before and after the policy's launch. My baseline DD results are based on traditional two-way fixed effect (TWFE) estimators. I then build on these results in two ways: First, I show that the core results are robust to using a double-robust DD estimator \citep{Callaway2020,SantAnna2020a} that combines outcome regression and propensity score models to adjust the counterfactual. Second, I employ a cross-country, triple differences (DDD) estimation strategy, comparing Korean manufacturing sectors to foreign placebo manufacturing sectors. 

The main DD estimates capture the impacts of heavy industry policies, which emphasized directed credit and, to a lesser extent, trade policy. Industry pre-trends inform Korea's counterfactual sectoral structure. Absent these interventions, industries would have evolved according to an earlier pattern of comparative advantage. Thus, I refer to Korea's comparative advantage without intervention as its static comparative advantage. Differences after 1973 reveal the effect of industrial policy in promoting dynamic comparative advantage, where the overarching policy was associated with the ascent of new industries and new patterns of specialization.\footnote{These definitions build on \citet{Redding1999}, who defines dynamic comparative advantage more generally as a time-varying version of classic static comparative advantage.}

To estimate these effects, I construct a new data set on industrial outcomes spanning Korea's miracle period (1967--1986). I harmonize material from digitized industrial surveys and historical machine-readable statistics into consistent panel data and then combine this industry-level data with digitized input-output (IO) accounts. This process results in panel data that covers a key episode of industrial development.

I highlight three empirical results. First, I find that the policy package resulted in significant positive impacts across industrial development outcomes in targeted industries. Relative to pre-intervention levels, targeted heavy-chemical industries expanded their output by more than 100\% over non-treated manufacturing sectors. Furthermore, labor productivity was more than 15\% higher. This divergence was not driven by a decline in non-treated industries. Moreover, since industrial development is multidimensional, I consider it across outcomes and find impacts on employment growth, export performance, and output prices. HCI not only appears to have durable, longer-run effects on treated industries, but I also find evidence of persistent impacts on plant-level, total-factor productivity (TFP) in the post-1979 period. 

I emphasize the role of investment policy and find evidence that supports dynamic learning-by-doing. Reduced-form estimates show that HCI sectors are correlated with stronger learning-by-doing forces, and the results are consistent with industry-wide, cross-plant learning spillovers. Importantly, I do not find that the HCI policy crowded out investment in non-treated industries.

Second, HCI coincided with a shift in the longer-term dynamic comparative advantage of the targeted export industry. Post-1979 outcomes, such as the share of activity in manufacturing sectors (employment or output), remained significantly higher than in non-treated sectors. Additionally, treated industries were 10 percentage points more likely to achieve comparative advantage in global markets after 1973. Indeed, the revealed comparative advantage (RCA) of HCI products increased 13\% more than other manufacturing exports over the same period, and I observe similar patterns using gravity-based methods \citep{Costinot2012a}. However, these patterns only emerged over time. Consistent with infant industry theory (e.g., \citep{Bardhan1971}), shorter-term evaluations may fail to capture the full, dynamic impacts of a given policy.

Third, heavy and chemical industry drive policies correspond to the development of downstream industries. I find that downstream sectors with strong links to targeted industries expanded during the policy period. During the drive, comparative advantage emerged among downstream exporters and fully materialized after the end of the policy period (1979). However, given that policies targeted more upstream industries, the backward linkage effects of the policy appear limited. Hence, I find evidence that the policy may have supported network spillovers. These results are consistent with quantitative research on optimal policy approaches within networks, such as \citet{Liu2019}, which uses IO data from this study. Accounting for linkages reduces the precision of the main effects yet preserves the core pattern of industrial development estimates. 

This study makes three contributions. First, I build on emerging research that uses contemporary econometrics to study the impact of industrial policies, including cross-country explorations of trade policy by \citet{Nunn2010}, seminal case studies by \citet{Aghion2015} and \citet{Criscuolo2012}, and sector-specific studies by \citet{Blonigen2016}. It also complements the structural literature in industrial organization, which analyzes sector-specific industrial policies \citep{Kalouptsidi2018,Barwick2019}, including earlier calibration-based evaluations \citep{Baldwin1988,Irwin1991,Head1994}. Similarly, relevant research in development economics by \citet{Rotemberg2019} and \citet{Martin2017} has explored industrial policy in the context of India's small and medium-sized enterprise policy. 

I also contribute to the empirical study of industrial policy via natural experiments. This paper is the first study to deploy modern empirical techniques to evaluate the heavy and chemical industry drive episode. I join \citet{Juhasz2018} and related work by \citet{Inwood2013} and \citet{Harris2015}, who use historical experiments to estimate the impacts of output market protection on manufacturing development. I consider the efficacy of infant industry policy in (a) a contemporary setting and (b) with outward-oriented (e.g., export-facing) policies. My findings align with studies that use temporary historical episodes to explore the process of dynamic comparative advantage \citep{Hanlon2020,Mitrunen2019,Pons-Benaiges2017}. For example, \citet{Jaworski2018} and \citet{Giorcelli2019} explore the impact of temporary government policies on industrial development. More broadly, I complement historical research highlighting the potential of transitory policy to promote the longer-run development of nascent industries. I do so by examining a purposeful, targeted intervention. By considering targeted policies, I connect with studies evaluating place-based policies, notably \citet{Criscuolo2012} and \citet{Becker2010}, who use exogenous spatial variation to study the impact of targeted support on distressed regions.

Finally, I add to debates on the role of industrial policy in development, especially those surrounding the East Asian miracle. On the one hand, rich qualitative research has emphasized the role of industrial strategies in newly industrializing economies \citep{Johnson1982,Wade1990,Amsden1992,Chang1993}. On the other hand, economists have generally expressed skepticism of such interventions \citep{Pack2006, Noland2003}, especially their role in East Asia's ascent \citep{Krueger1995,Pack2000}. This study is the first modern empirical attempt to revisit debates on the East Asian episode, which is summarized in Section \ref{sec:theorysection}). By employing contemporary econometrics, I build on early correlational studies \citep{Weinstein1995,Beason1996} and more recent quantitative research \citep{Liu2019}.

To summarize, this study uses variations from the heavy and chemical industry drive to study the impacts of industrial policy in South Korea. It attempts to provide a disciplined, data-driven account of the episode. My analysis is organized in the following way. In Section \ref{sec:historysection}, I discuss the institutional setting of the heavy industry drive and detail the policies. In Section \ref{sec:theorysection}, I describe the general theoretical case for industrial policies. Section \ref{sec:datasection} then provides an overview of the data. In Section \ref{sec:directsection}, I present estimates of the direct impact of the heavy industry push on targeted industries, and in Section \ref{sec:policysection}, I turn to policy mechanisms. Finally, I consider the estimates of HCI's spillovers into external sectors through input-output linkages in Section \ref{sec:networksection}. I conclude in Section \ref{sec:conclusionsection} with a discussion of my findings.

\section{Institutional Context}\label{sec:historysection}

I first consider the institutional and historical context of the heavy and chemical industry drive. This section describes the policy's launch, sectoral choice, and variation over time. Finally, I synthesize my use of these features in the empirical research design.

\subsection{The Nixon Shock and Launch}\label{sec:hcicauses}

Political crises in South Korea catalyzed its 1973 industrial drive, which was fundamentally security-driven \citep{Kim2011c,Moon2011}. Among the factors behind Korea's crisis were (a) North Korea's increasing militarization and offensive actions \citep{Kim1997,Moon2009} and, critically, (b) a shift in U.S. foreign policy toward Asia. In 1969, President Richard Nixon declared that the United States would no longer provide direct military support to its allies in the Asia-Pacific region, creating the risk of U.S. troop withdrawal from the Korean Peninsula \citep{Nixon2010,Kim1970,Kwak2003}. Unfortunately, this U.S. pivot coincided with North Korea's growing military antagonism. Like its South Vietnamese allies, South Korea believed it would need to defend itself against an impending Communist-backed invasion. However, South Korea had no domestic arms industry, and the North rivaled the South militarily, having pursued a military industrialization campaign through the 1960s \citep{Hamm1999}---South Korea had not kept up. Without U.S. troops, South Korean armaments would not be able to absorb a North Korean blitz \citep{Cushman1979,Eberstadt1999}.\footnote{Appendix \ref{sec:troophistorysection} describes the so-called Nixon shock and the subsequent political crisis. Online Appendix \ref{sec:furthernixonshock} describes Korea's military status.} Military exigencies drove not only the timing of the heavy industry push, but also shaped its sectoral scope. I turn to this next.

\subsection{Sectoral Choice}\label{sec:historysectorvariation}

\subsubsection{Sectoral rationale and selection}\label{treatment-rationale} The heavy and chemical industry drive targeted six strategic sectors: steel, nonferrous metals, shipbuilding, machinery, electronics, and petrochemicals \citep{Stern1995,Castley1997}. Throughout this study, I define \textit{treated} or \textit{targeted} (I use the terms interchangeably throughout) industries as those listed in major policy acts---specifically, the enforcement decrees and national sectoral acts that undergirded the drive. Section \ref{sec:datasection} specifies how I coded policy treatment separately from legislation, and I provide legislative details in the Online History Appendix \ref{sec:suppappendixlegalacts}.

Why were these sectors chosen, and what might deliberations over their selection tell us about expectations for their success? Two rationales dominated the choice of heavy-chemical sectors, and both have been documented by scholars and policymakers.

First, heavy industrial intermediates were seen as key for the military and industrial modernization \citep{Lee1991,Woo-Cumings1998,Kim2011c}. In the early 1970s, unlike the North, direct military production was largely beyond the South's capabilities. Early failures in arms manufacturing were specifically mired by inputs of ``inadequate'' quality \citep[375]{Horikane2005b}. One former government official, Kim Chung-yum, reported it was ``apparent that the development of the heavy and chemical industries to the level of advanced countries was required to develop the defense industry'' \citep[409]{Kim2011h}. Hence, industrial intermediates were a means to promote military industrialization and future hardware production. For planners, the steel and nonferrous metals sectors supplied crucial upstream materials for basic defense components, electronic components for electronic weaponry, and machinery for precision military production (\emph{ibid}). Former officials from the government of South Korean President Park Chung-hee echoed these rationales \citep{Kim2015}. Thus, unlike downstream weaponry, upstream inputs were within Korea's capabilities and less controversial to lenders.

Second, relative to advanced military hardware, South Korea saw a potential advantage in targeting upstream production. Where Korea lacked the prerequisites to manufacture arms at scale, upstream intermediates were more practical---they were technologically within reach and featured economies of scale that could be supported through export markets \citep{Kim2011h}. To consider feasibility, the regime studied its contemporaries, including those in Western Europe and Japan \citep{Perkins2013}, though the latter was less a metaphor than a blueprint. Japan's experience gave South Korea a guide to the sectors in which they had potential \citep{Kong2000,Moon2011,Kim2011f}, and components of Korea's drive were borrowed from the New Long-Range Economic Plan of Japan (1958--1968). I discuss the overlap between Japanese and Korean policies in Online Appendix \ref{sec:suppappendixjapan}.

\subsubsection{Selection skepticism by foreign lenders}\label{selection-skepticism-by-lenders.} Yet, \emph{ex-ante}, the potential of Korea's heavy-chemical industries was not obvious, and international investors expressed doubts, famously rejecting the financing of erstwhile heavy industrial projects \citep{Amsden1992,Redding1999}. The International Monetary Fund (IMF), the U.S. Agency for International Development (USAID), and multiple European nations refused to provide loans for less ambitious, proto-HCI schemes \citep{Rhyu2011,Woo1991}. In 1969, both the U.S. Export-Import (EXIM) Bank and the World Bank blocked an early integrated steel mill, with the World Bank concluding that Korea ``had \textit{no comparative advantage} [emphasis my own] in the production of steel'' \citep[324]{Kim2011h,Rhyu2011}. The skepticism of lenders toward Korea's proto-HCI projects continued through the early 1970s, and such practicalities constrained early forays into heavy industrial projects---that is, until South Korea's political turn in late 1972\footnote{See Online History Appendix \ref{sec:supphistoryselection} for details on these constraints.}, when President Park's autocratic self-coup and breakthroughs due to international capital finally enabled a heavy industrial push, which I detail next.

\subsection{Policy Instruments and Variation: Before, During, and After HCI}\label{sec:policyvariation}

The drive's January 1973 announcement broke with Korea's earlier horizontal export-first industrial policy regime \citep{FrankJr1975,Krueger1979,Westphal1982,Westphal1990}, which famously was not sectoral but rather aimed at export activity \textit{writ large} \citep[28]{Hong1977}.\footnote{Before 1973, Korea implemented no less than 38 different incentives to promote exports \citep[18]{Lim1981}. See Online Appendix \ref{sec:suppappendixplans} for details on Korean economic planning.} Before the HCI drive, export incentives were essentially ``administered uniformly across all industries'' (\citep[217--218]{Westphal1982}\citep[44]{Westphal1990}). Exporters were exempted from so many restrictions that scholars have argued that the export drive essentially ``allowed exporters to operate under a virtual free trade regime'' (\citep[91]{Nam1980}; \citep{Lim1981}). In other words, the heavy-chemical drive represented a pivot to a fundamentally sector-specific strategy.

Yet, what was the industrial policy bundle? I consider two classes of policies in detail---(i) investment policy and (ii) trade policy---and their variation across the period.

\subsubsection{The bias of lending and investment incentives}
\label{sec:lendinghistory} Directed lending was a central lever of the heavy industrial drive \citep{Woo1991,Lee1991}.\footnote{Woo summarizes that Korean policy sought to ``hemorrhage as much capital as possible into the heavy industrialization program'' (\citeyear[159]{Woo1991}).} Half of all domestic credit consisted of subsidized ``policy loans,'' which were allocated by financial institutions---both less traditional non-banking institutions (development banks)  and more traditional commercial banks \citep{Koo1984,Lee1996}. Broadly defined, policy-targeted loans were designed to advance government objectives and were automatically re-discounted by the central bank at a preferential rate.\footnote{Historically, Korean policy loans have served rural and infrastructural development objectives and were a prominent lever of heavy industry targeting.} For example, over this period, policy loans had longer repayment schedules, and average interest rates were five percentage points lower than benchmark loans \citep{Cho1995}. 

Figure \ref{fig:lendingpolicyplot} illustrates the shift from (pre-1973) sector-agnostic policies to (post-1973) sector-specific investment policies. Panels B and C of Figure \ref{fig:lendingpolicyplot} track the rise in new credit to the heavy-chemical sector after 1973 and the decline in direct state lending after 1979. Specifically, Panels B and C plot the change in loans issued by the Korea Development Bank (KDB), the source of around 90\% of attractive policy loans lent by non-banking financial institutions \citep[42]{Cho1995}. Panel B presents the real value of all new KDB loans by industry, and Panel C presents these values for machinery and intermediates, a major focus of the industrial drive policy. The thin lines correspond to two-digit industries, and the thick lines are averages for targeted (red) and non-targeted (gray) industries. Parallel lines denote the average lending for each period.

The sectoral bias of lending by state institutions is also seen in more traditional commercial deposit banks, which also allocated a significant share of policy loans \citep{TheWorldBank1993,Cho1995}. Appendix Figure \ref{fig:appendixloans} shows similar growth for total credit and intermediate equipment loans by commercial banks. Indeed, across lending institutions, aggregate data in both figures (Figure \ref{fig:lendingpolicyplot} and Appendix Figure \ref{fig:appendixloans}) plot a trend-break in sectoral-specific lending after 1979, which marks the beginning of policy liberalization.

Similarly, Panel A of Figure \ref{fig:lendingpolicyplot} traces the sectoral bias of tax policy over the period, using the estimated effective marginal tax rate. These estimates account for a myriad of period-specific investment incentives, notably tax breaks, investment tax credits, and special depreciation rates \citep{Kwack1985, Stern1995, Lee1996}; see Online Appendix \ref{sec:supplementtaxratecalc} for details. Panel A presents the divergence in rates after 1973, when tax laws were reformed to concentrate investment in heavy industry \citep{Kwack1984,Kim1990}.\footnote{Packages included the ``Special Tax Treatment for Key Industries'' (Tax Exemption and Reduction Control Law), which gave strategic industries the choice of a five-year tax holiday, an 8\% tax credit toward machinery investment, or a 100\% special depreciation allowance \citep[395]{Lee1996}.} Like directed credit above, tax policies converged after the start of the liberalization period in 1979.

\subsubsection{The bias of trade policy}\label{sec:historyoftradepolicy} The heavy and chemical industry drive also altered biases related to trade policy. Pre-1973, government policies broadly exempted exporters from import restrictions \citep{Nam1980,Westphal1990}. Indeed, measures of nominal protection were lower for heavy industry during this period (see Online Appendix \ref{sec:suppappendixtradepolicy} for pre-1973 trade policy). After 1973, exemptions were aimed at heavy industry \citep{Woo1991,Cho1995}, and HCI producers were exempted from up to 100\% of import duties. \citet{Park1977} estimates that ``key industries,'' on average, enjoyed 80\% tariff exemptions \citeyearpar[212]{Park1977}. Although the post-1973 trade policy was refocused toward heavy industry, the nominal protection of output markets did not appear to rise substantially, especially relative to other policies (see Section \ref{sec:tradepolicysection}). 

\subsubsection{Post-1979 liberalization}\label{sec:liberalizationhistory} President Park Chung-hee's assassination in 1979 prompted the withdrawal of his signature policy. With the fall of Park's regime, South Korea dismantled heavy-chemical industrial incentives and pursued structural economic reforms. I provide details of the post-1979 liberalization in Online Appendix \ref{sec:supplementalliberalization}. For example, the state-controlled banking sector was liberalized, with notable reforms in 1981 and 1983. Special rates on policy loans were eliminated, and they took a different form over the post-1979 period (see Appendix \ref{sec:appendixcommercialbanks}). While the role of government policy loans shrank \citep{Cho1986,Nam1992}, fiscal reforms closed the gap in effective marginal corporate tax rates between strategic and non-strategic industries \citep{Kwack1992}. Meanwhile, the post-Park autocracy only accelerated Korea's trend toward trade policy liberalization.

\subsection{Summary: Features for Empirical Study}\label{sec:historyempirical}

The policy context above informs the research design of this study, which I summarize in four points: 

First, the episode introduced sectoral variation over time, as the heavy and chemical industry drive shifted national policy toward a discrete set of nascent industries. This shift began and ended because of external political events: the Nixon Doctrine and Park's assassination, respectively. The liberalization of HCI is also useful, as theoretical justifications often entail temporary policy.

Second, policy variation was purposeful. Notably, I consider an actual policy and not a random variation mimicking industrial policy. Given the complications of estimating the impact of industrial policies, researchers have used important natural experiments that mimic policy variations \citep{Juhasz2018,Hanlon2020,Mitrunen2019}. Nevertheless, the case for industrial strategy hinges on the policy being intentional \citep{Juhasz2022}, and it may be difficult to glean insights from a random, accidental policy variation \citep{Rodrik2004}.

Third, although targeted, Korea did not believe heavy-chemical industries would develop without intervention, and financiers doubted the viability of the Korean heavy industry sector. Foreign lenders rejected financing for early prototype projects on the grounds of comparative advantage. Korean planners countered that investment could cultivate comparative advantage in targeted sectors.

Fourth, the political context of the heavy and chemical industry drive reduces the role of political confounders. This setting, including the existential threat facing South Korea, meant industrial policies were binding and coherent. Clientelism and political demands often divert resources to industries with a comparative disadvantage \citep{Rodrik2005,Lin2011}, and policy estimates may reflect political failures rather than the potential of a given policy. Korea's heavy and chemical industry drive was driven by top-down changes in national economic and defense strategy---the sectoral bias was not driven by lobbying or heavy industrial constituents.

\section{Conceptual Cases for Industrial Policy}\label{sec:theorysection}

Mainstream neoclassical justifications often rely on the existence of externalities \citep{Corden1997,Juhasz2023}. In this section, I discuss two externalities relevant to the South Korean policy episode: 1) dynamic economies of scale and 2) linkage effects. I consider each in the context of earlier empirical work on East Asia.

\subsubsection{Dynamic economies of scale}\label{sec:lbdtheory} First, dynamic externalities have long guided justifications for infant industry policy \citep{Bardhan1971,Succar1987,Young1991}. Intra-industry learning-by-doing externalities embody this class of justifications, whereby firms accumulate production experience over time and, in turn, this experience benefits other firms within the same industry. Hence, individual firms may not internalize the benefits of learning, producing, or under-investing in a socially beneficial activity. Interventions may also be justified even without across-firm spillovers, such as when firm-level learning occurs alongside other imperfections \citep{Lucas1984,Corden1997}. For instance, a firm may have strong learning-by-doing forces, yet if it faces capital constraints, it may be unable to survive turbulent nascent periods. 

Such dynamic economies of scale are the means by which industrial policy can, in theory, cultivate a dynamic comparative trade advantage \citep{Redding1999}. Theoretically, if learning-by-doing conditions are suitable (i.e., within-industry learning spillovers or firm-level learning combined with imperfections), a successful infant industrial policy in a new sector can promote the evolution of comparative advantage on the international market.

Correlational studies of East Asian industrial policy have suggested that interventions may not correspond to industrial development or externalities. For Korea, \citet{Lee1996} identifies a negative relationship between postwar interventions and industry-level outcomes, specifically, protection and manufacturing productivity (see also \citep{Dollar1990}). \citet{Beason1996} find that Japanese industrial policy is not positively correlated with industry development. Similarly, \citet{Yoo1990} argues that HCI may have harmed South Korea's export development performance relative to its contemporaries.

\subsubsection{Linkage effects}\label{sec:linkagetheory} Second, pecuniary externalities through linkages have been another justification for implementing a particular industrial policy \citep{Krueger1982, Grossman1989,Krugman1993}, where policies targeting one sector benefit external industries through input-output (IO) connections. Development economists have long considered how industrial interventions impart benefits beyond the direct targets of the policies through IO linkages \citep{Scitovsky1954,Rasmussen1956,Hirschman1958}. They argue that intuitive targeting is likely justified where the social benefits conferred to others are considerable. These benefits are transmitted in two directions. The first is through \emph{backward linkages} to upstream industries selling inputs to targeted sectors. For example, if an industrial policy increases the size of targeted industries, it increases the demand for upstream producers. Second, industrial policy can confer benefits through \emph{forward linkages} to downstream industries purchasing inputs from targeted sectors. For example, if a given policy increases the productivity of a treated industry, it may lower prices, which benefits firms using those inputs.

As with dynamic externalities, tests of industrial policy justifications with linkage spillovers have attempted to explore the relationship between targeting---or often, policy levers---and the existence of linkage spillovers. Incisive studies of East Asia, in particular, have rejected industrial policy on the grounds that it has not corresponded to these externalities. \citet{Noland2004} argues that Korean policy did not target sectors with high linkage spillovers. Using measures of IO linkages, \citet{Pack2000} finds that industries targeted by South Korea and Japan had low linkages with non-targeted industries and questions whether the policy targeted externalities. Taken together, \citet{Noland2003} and \citet{Pack2006} argue that industrial development and targeting seem uncorrelated with growth in key historical episodes. A recent applied theoretical study by \citet{Liu2019} reveals that common features of IO tables may correspond to optimal targeting, using evidence for South Korea and China.

\section{Data}\label{sec:datasection} 

I use newly assembled industry-level data on industrial development during South Korea's miracle period, 1967--1986. Industry-level panels are constructed using digitized data from the Economic Planning Board's (EPB) Mining and Manufacturing Surveys and Census (MMS). MMS data are suitable for studying the heavy and chemical industry drive, which was fundamentally a sectoral policy. The survey is high quality and reports consistent manufacturing census outcomes over the study period. The MMS census data are published nearly every five years, with annual intercensal surveys. Manufacturing outcomes are published at the five-digit industry level and aggregated from establishment (or plant) level surveys.\footnote{I supplement digitized MMS statistics with early machine-readable MMS data (1977--1986).} In addition to industry-level data, I also use post-1979 plant-level microdata from the MMS. Price data are digitized from historical and contemporary Bank of Korea producer price index publications and yearbooks.

\paragraph{Long and short industry panels}\label{sec:longshortpanel} This study uses two harmonized industry panels. Appendix Table \ref{tab:appendixtabledescriptive} presents pre-1973 statistics (mean and standard deviation, non-normalized values) for key industrial variables. Part B of Appendix Table \ref{tab:appendixtabledescriptive} reports values from the ``short'' granular five-digit industry panel, harmonized from 1970 to 1986. Part A reports values from the ``long'' more aggregated four-digit panel, harmonized from 1967 to 1986. The terminal date of the study is 1986, the year before Korea's consequential democratic transition. 

Creating these consistent industry panels from MMS data is not trivial and requires harmonization across multiple code revisions. Between 1967 and 1986, the EPB updated Korea's industrial codes (KSIC) four times, with a major revision in 1970. Thus, harmonizing MMS data alone requires multiple crosswalk schemas and their digitization. I describe this process in the Online Data Appendix \ref{sec:supplementharmonization} and the concordance within the MMS and across other data series.

The harmonization process introduces a trade-off between the two panels above. The short panel (1970--1986) contains more industry observations (five-digit level) but covers a more limited timeline. The shorter panel requires less harmonization and thus is closer to the original MMS publication statistics. In contrast, the long panel (1967--1986) contains fewer industries (four-digit level) but covers a longer timeline. Thus, the longer panel requires more harmonization but encompasses the critical pre-1973 (``pre-treatment'') period. Although the long panel adds three years to the pre-treatment period, (i) four-digit observations and (ii) the harmonization process significantly reduce the number of industry observations relative to the short, disaggregated panel.

\paragraph{Defining treatment}\label{sec:definingtreatdata} I define treated or targeted industries as those appearing in major industry legislation. Section \ref{sec:historysection} described the industry scope of the HCI drive, which was built from six major sectoral acts. For sectors such as shipbuilding, aggregate sectors from the acts and census industries are closely aligned. However, care is required for more complex industries, such as chemicals, and the relevant legislation. Consequently, I hand-match the industries in the legislation to the harmonized data, both long and short panels. This process entails matching industry labels in legislation to industries in the five-digit KSIC industry codes. See Online Appendix \ref{sec:suppappendixlegalacts} for legislation and matching.

\paragraph{Linkages}\label{sec:linkagedata} Inter-industry linkage data are constructed from the Bank of Korea's 1970 ``basic'' input-output tables, which I digitized. These are the most disaggregated tables for the period, covering approximately 320 sectors. I used these tables to create the measures of exposure to industrial policy through linkages, which I detail in Section \ref{sec:appendixlinkcalculations}. However, the Bank of Korea data and MMS surveys use different coding schemes. Thus, combining IO accounts with industry data requires further harmonization (see Online Data Appendix \ref{sec:supplementharmonization}).

\paragraph{Trade flows and trade policy}\label{sec:tradedata} I also use international trade flow and trade policy data. The ``long'' four-digit industry panels are hand-matched to the Standard International Trade Classification (SITC, Rev.1) four-digit-level trade data. The trade flow data come principally from the UN Comtrade database. For trade policy, the product-level measures of quantitative restrictions (QRs), e.g., tariffs, are digitized from \citet{Luedde-Neurath1988} and connected to modern nomenclatures. These data are available for 1968, 1974, 1976, 1978, 1980, and 1982, representing the most disaggregated, readily available data for the period \citep{Westphal1990}. These statistics contain measures of core non-tariff barriers, notably QRs. Most empirical studies of Korean trade policy use highly aggregated data. In terms of QRs, \citet{Luedde-Neurath1988} codes the severity of restrictions from least to most severe (0 to 3).

I use trade policy data to calculate separate measures for output and input market protection exposure. Output protection for industry $i$ is simply the average tariff (or quantitative restriction) score for that sector: $\textrm{output-tariff}_i$. Heavy industry policy also used exemptions from import barriers as a policy tool, and I calculated measures of input protection. Input tariffs (QRs) faced by industry $i$ are calculated as the weighted sum of tariff (QR) exposure \citep{Amiti2007}: e.g., $\textrm{input-tariff}_i = \sum_j \alpha_{ji}\times\textrm{output-tariff}_j$, where $\alpha_{ji}$ are cost-shares for industry $i$ and input $j$. Cost weights come from the 1970 input-output accounts.

\section{The Main Impacts Heavy and Chemical Industry Drive}\label{sec:directsection}

This section considers the empirical impact of the heavy and chemical industry drive in three parts. First, I introduce the main estimation strategy (Section \ref{sec:directestimation}), which I use to identify how HCI targeting corresponds to industrial development (Section \ref{sec:growthestimates}). Second, I report estimates of the average impact of the policy over time and consider estimates from the double-robust DD estimator (Section \ref{sec:doublerobustestsection}). Finally, I employ a DDD estimation strategy to study the impact of the heavy industry drive using cross-country variation (Section \ref{sec:ddd}).

\subsection{Estimation}\label{sec:directestimation}
To estimate the impact of the industrial policy, I use the temporal and sectoral variation from the heavy and chemical industry drive to employ a differences-in-differences strategy. I take the January 1973 announcement of HCI as the start date and the assassination of President Park in 1979 as the \textit{de facto} end date. I compare differences between the set of targeted manufacturing industries versus the set of non-targeted manufacturing industries relative to 1972. I collect data on the industries until 1986, the year before Korea's formative democratic transition. I consider the following baseline specification:

\begin{equation}\label{eq:mainflexible}
\begin{split}
\ln \left( y_{it} \right)= \alpha_i + \tau_t+ \sum_{j\neq1972} \beta_j \cdot \left( \textrm{Targeted}_i \times \textrm{Year}^j_t \right) + \sum_{j\neq1972} X'_i \times \textrm{Year}^j_t \Omega_j + \epsilon_{it} , 
\end{split}
\end{equation}

\noindent where $y_{it}$ are (log) industrial development outcomes, $i$ indexes each manufacturing industry, and the year is denoted by $t$, which takes the values 1967--1986 for the long panel and 1970--1986 for the short panel. Equation \eqref{eq:mainflexible} is a linear TWFE specification with industry fixed effects $\alpha_i$ and time effects $\tau_t$. I cluster standard errors at the industry level, allowing for within-industry correlation. I also estimate equation \eqref{eq:mainflexible} using pre-treatment variables to control for unobserved productivity correlated with the intervention, including pre-1973 industry averages: total intermediate outlays (material costs), average wage bill (total wage bill per worker), average plant size (employment per plant), and labor productivity (value added per worker). Values are all in real terms and are in logs. Since the pre-treatment averages $X'_i$ are time-invariant, I interact them with year effects to estimate their impact over time.

The impact of the industrial policy drive is estimated using a binary variable $\textrm{Targeted}_{i}$, which is equal to one for a treated industry and zero otherwise (for assignment, see Section \ref{sec:datasection}). The set of $\beta_j$s is the differences between targeted and non-targeted industries for each year $j$ relative to the pre-treatment year $1972$, and coefficients for 1972 are normalized to zero. The binary treatment term allows me to visually assess counterfactual dynamics and pre-trends. I also compare TWFE estimates from equation \eqref{eq:mainflexible} to the double-robust DD estimators below (Section \ref{sec:doublerobustestsection}).

The coefficients of interest, $\beta_j$, convey three aspects of how targeted sectors evolved. First, estimates after 1972 describe the average impact of the targeting for each period after the start of the heavy and chemical industry drive. If the industrial policy is associated with short-term industrial development during the six-year drive, we should observe increasing differences in $y_{it}$ between 1973 and 1979.

Second, estimates after 1979 describe the long-term impacts of the industrial policy drive. In the parlance of the industrial policy literature, their longevity indicates the potential dynamic effects of industrial policy. This evolution may be realized through dynamic economies of scale (Section \ref{sec:theorysection}). Even where differences stabilize in the later period, this may also coincide with a permanent shift in levels of development between the two types of industries.

Third, estimates before 1972 describe average differences between targeted and non-targeted industries before the policy. Thus, they convey information about the common trend assumption of the research design. Before 1972, we should not observe systematic differences between treated and control industries: $\hat{\beta}_{1967} \approx \hat{\beta}_{1968} \approx \hat{\beta}_{1972} \approx 0$. For key analyses, I report the full tables and plotted estimates, including full tests for the joint significance of pre-trends.\footnote{Although null results provide information about DD pre-trend assumptions, they cannot validate the pre-trend assumption alone or do so decisively. This situation is particularly true for more detailed five-digit estimates, which have limited pre-treatment periods.}

Ultimately, the goal of specification \eqref{eq:mainflexible} is to understand the impact of the industrial policy package on treated industries or the ATT. This estimand is particularly relevant for industrial policy, where policymakers are often interested in the impact of a policy on targeted units rather than the average unit in an economy (ATE); as such, the ATT requires different, less stringent assumptions.

Theoretically, sectoral industrial policies often target industries that are most responsive to the relevant legislation or that idiosyncratically gain from interventions. In our setting, targeted industries may be those expected to respond the most to policies, for example, by having stronger dynamic economies of scale. For estimating the ATT, the common trend assumption accounts for this issue under certain assumptions: if selection does not change over time (irrespective of policy), the common trend assumption addresses this form of selection between targeted and non-targeted industries \citep{Heckman1998,Blundell2009}. In other words, if the selection bias remains unchanged between the sectors at the time of treatment, then parallel trends remove unobserved idiosyncratic gains from estimates. This assumption is violated if unobserved factors such as productivity are expected to accelerate in targeted industries, regardless of treatment. Recall, however, that Section \ref{sec:historysection} documented how allies and foreign lenders estimated that South Korea could not cultivate dominance in heavy industries without intervention. Nevertheless, the assumptions above mean that estimating the impact (ATT) of the policy drive requires a proper control group. To this end, the treatment effects literature has emphasized the power of alternative estimators and re-weighting methods \citep{Heckman1998,Smith2005}.

I consider alternative estimation procedures and build on my baseline TWFE estimator for equation \eqref{eq:mainflexible} in two ways. First, I use a double-robust DD estimator---a method that re-weights observations in the control group through their propensity score and adjusts the counterfactual outcome using a linear regression model. Second, I estimate the takeoff of Korean targeted industries using cross-country and cross-industry variation and deploy a triple difference estimation strategy. This DDD strategy attempts to directly address the issues discussed above by comparing Korean industries to similar international industries. However, let us first consider the baseline estimates.

\subsection{Direct Impact on Industrial Development: Results}\label{sec:growthestimates}

\subsubsection{Key patterns and output expansion} \label{sec:outputandkeypatterns} Figure \ref{fig:mainoutputfigure} plots baseline dynamic DD estimates for the impact of HCI on output, measured as real value shipped. Panel A provides estimates for the detailed (`short') five-digit panel, which starts in 1970. Panel B presents estimates for the more aggregated (`long') four-digit panel, which started in 1967. The left columns give estimates from the baseline fixed effect specifications, while the right columns show estimates with controls. The top row of each panel in Figure \ref{fig:mainoutputfigure} presents the average log output for targeted (red) and non-targeted (black) industries using the fit model from equation \eqref{eq:mainflexible}. The bottom row presents the traditional DD plots of the estimated differences between the two industries. 

Figure \ref{fig:mainoutputfigure} delivers three key patterns of industrial development associated with the policy drive, and these patterns reappear across outcomes throughout this study. First, Figure \ref{fig:mainoutputfigure} shows that output from targeted and non-targeted industries evolved similarly over the pre-HCI period (1967--1972). This result is clearest in the longer aggregate four-digit panel, and pre-period coefficients are individually and jointly insignificant (Online Appendix Table \ref{tab:supptablerollingoutput}).

Second, Figure \ref{fig:mainoutputfigure} shows that marked differences between treated and non-treated sectors emerged after the 1973 intervention. These differences widen and become stark over the policy period. This divergence is most pronounced in estimates for the five-digit data in Panel A. Panel B reports a similar, though less precise, divergence in aggregate four-digit data. The top row of Figure \ref{fig:mainoutputfigure} also shows that the estimated differences (bottom) are not driven by the decline in the control industries. This finding is useful since differences between treated and non-treated industries may emerge if policies harm industries in the control group (e.g., \citep{Cerqua2017}), for instance, if policy crowds out investment for other manufacturing industries. I explore this issue in Section \ref{sec:investmentpolicy}.

Third, the impacts of the drive were not transitory. In terms of real output, in Figure \ref{fig:mainoutputfigure}, the gap between treated and non-treated industries persists throughout the post-1979 period. The top row of Figure \ref{fig:mainoutputfigure} also reveals that even though differences stabilize or diminish, the level effects are sticky. The patterns in Figure \ref{fig:mainoutputfigure} are also robust and seen across alternative measures of log output, data sets (four vs. five-digit panels), and specifications in Appendix Figure \ref{fig:appendixoutputrobust}. 

\subsubsection{Industrial development outcomes}\label{sec:otherdevoutcomes} Figure \ref{fig:maindevelopment} presents the impact of the heavy-chemical policy across various industrial development outcomes. Panel A of Figure \ref{fig:maindevelopment} illustrates that the policy drive coincided with a significant increase in simple measures of labor productivity (log real value added per worker) and relatively lower (log) output prices. Like the estimates above, five-digit data estimates are more precise than aggregate four-digit ones. Note that these estimates are not driven by a relative decline in prices for heavy industry. Appendix \ref{sec:appendixprodandprices} shows that heavy industry prices increased less than other industries over the inflationary 1970s.

Panel A (Figure \ref{fig:maindevelopment}) also demonstrates that the policy coincided with a shift in the share of total manufacturing activity toward targeted industries. The log manufacturing share of output and the log employment share both increase for the targeted industry. Moreover, this reallocation of manufacturing activity toward the heavy and chemical industry is durable. Estimates are less precise for aggregate data. Online Appendix Table \ref{tab:supptablerollingdevelopment} jointly rejects pre-trends. Additionally, Figure \ref{fig:maindevelopment} shows a rise in the number of plants operating in HCI markets.

\subsection{Direct Impact on Exports Development}\label{sec:maintradeestimates}

Export performance provides another view of industrial development, and exports were central to the policy program, as was the case for earlier iterations of South Korean industrial policy. For instance, a distinct goal of the HCI drive was that heavy-chemical products would constitute 50\% of exports by 1980 \citep{WorldBank1987,Hong1987}. Figure \ref{fig:maindevelopment}, Panel B reports the impact of industrial policy on export development outcomes, now using SITC (Rev.1) trade flow data, which are substantially more disaggregated than harmonized industry data. 

The analysis in Panel B considers multiple measures of export development. First, I calculate a traditional measure of revealed comparative advantage (e.g., \citep{Balassa1965}) for each industry. The RCA (Balassa) index is defined as the ratio of Korea's export share of good $k$ relative to the world's export share of commodity $k$: 
$$
\textrm{RCA}_k = \left( \dfrac{X^{\textrm{Korea}}_k}{X^{\textrm{Korea}}_{\textrm{Total}}} \middle/ \dfrac{X^{\textrm{World}}_k}{X^{\textrm{World}}_{\textrm{Total}}} \right),
$$
\noindent where $X$ denotes the value of exports. Korea has a comparative advantage in $k$ when $\textrm{RCA}_k$ is larger than one.

Additionally, I estimate the relative export productivity (CDK) using the gravity model methods proposed by \citet{Costinot2012a}. Their CDK estimate provides a theoretically consistent measure of revealed comparative advantage beyond the classic calculations. For industry $k$, I estimate relative export productivity for country $i$, where $\widehat{\textrm{CDK}}_k=\exp(\frac{\delta_{ik}}{\hat{\theta}})$; the trade elasticity $\hat{\theta}$ is taken from \citet{Costinot2012a}. The $\delta_{ik}$ term is the exporter-commodity fixed effect from the bilateral trade regression, $\ln(X_{ijk})= \delta_{ij} + \delta_{jk} + \delta_{ik} + \epsilon_{ijk}$, where $X$ are exports, $i$ is an exporter, $j$ is an importer, and $k$ is a commodity. While the traditional RCA measure accommodates zero trade flows, CDK is estimated from non-zero trade flows and takes positive non-zero values.

Across measures of export development, Panel B in Figure \ref{fig:maindevelopment} depicts a strong positive relationship between industrial policy and treatment. For the classic RCA index, I employ Poisson pseudo-maximum likelihood (PPML), given the prevalence of 0s. I also provide linear estimates using transformed RCA for completeness. Panel B shows a consistent pattern: after 1973, there was a marked rise in the relative RCA and the share of manufacturing exports for targeted SITC industries. Furthermore, the probability of attaining comparative advantage grew markedly after 1973. Second, before 1973, pre-trends were absent across trade outcomes, except for RCA, which trended downward. Third, estimates grew and became highly significant in the post-policy period. Hence, relative comparative advantage emerged during the drive and was fully articulated after the policy period. The ascent of heavy-chemical exports is also shown below (Section \ref{sec:ddd}) using cross-country trade data.

\subsection{Direct Impact: Robustness}\label{sec:directrobustness}

\subsubsection{Total-Factor Productivity: Plant-Level Persistence and Industry-Level Trends}\label{sec:tfpsection}

Above, Section \ref{sec:growthestimates} presented indirect productivity measures. I now turn to total-factor productivity. However, features of the data and the historical context pose constraints for estimating TFP (e.g., microdata availability). Nevertheless, I consider the persistence of plant-level TFP using microdata (which are available after 1979). Specifically, I study the correlation between targeting and plant-level TFP after the termination of HCI in 1979 using a simple pooled panel regression: 

\begin{equation}\label{eq:tfpplantregression}
\textrm{TFP}_{pit} = \alpha_{it}+\beta\textrm{Targeted}_{p}+\epsilon_{pit},
\end{equation}

\noindent where $p$ denotes plant, $t$ are years after 1979. The term $\textrm{Targeted}_{p}$ indicates plants operating in industries targeted by the heavy industry drive. Given that treatment is time-invariant, I include (four-digit) industry-year effects, \(\alpha_{it}\). For completeness, I consider multiple estimates of \(\textrm{TFP}_{pit}\) \citep{Olley1996, Levinsohn2003, Wooldridge2009, Ackerberg2015}. I use two-way clustered standard errors to allow for within-industry and plant correlation.

Table \ref{tab:tfpcrosssection} reports the relationship between plant-level productivity and plants in treated heavy industry. For the period immediately following the HCI drive (1980--1986), treated establishments have significantly higher TFP than non-treated establishments. Across specifications and measures of TFP, estimates in Table \ref{tab:tfpcrosssection} are significant and imply that heavy industry plants in the 1980s have between \resultstfpmin\% and \resultstfpmax\% higher productivity than non-targeted plants. These correlational results are compatible with the industry-level dynamics shown in Section \ref{sec:growthestimates}. 

Next, I turn to industry-level dynamics using aggregate TFP, which I present in Appendix \ref{sec:appendixtfp}. These industry-level estimates also reveal a gentle upward trend in total productivity for targeted industries relative to non-targeted industries. Differences in productivity became significant over the post-1979 period. This upward trajectory is compatible with the relatively high TFP in the cross-section of post-1979 heavy industry plants (Table \ref{tab:tfpcrosssection}). For further robustness, Online Appendix \ref{fig:suppappendixmicrotfp} provides dynamic estimates for plant-level TFP, showing gentle upward trends over the limited post-1979 period. Together, the industry and plant-level estimates appear consistent with the slow emergence of policy effects. Perhaps equally important, I do not find a salient relative \textit{decline} in TFP for the treated industries, which may be commonly associated with poorly performing industrial policy.

\subsubsection{Continuous Treatment and Limited ``Horizontal'' Spillovers}\label{sec:continuousrobust}

For robustness, Online Appendix \ref{sec:supappendixcontinuoustreatment} explores the patterns of industrial development using a more continuous industry-level measure of exposure to HCI. This measure captures the extent to which plants in HCI product markets produce output in other (non-HCI) markets. Dynamic estimates using this continuous measure (Online Appendix Figure \ref{fig:suppcontinuousdev}) track the binary estimates in Section \ref{sec:growthestimates}. Broadly, however, multi-product plants in heavy industry tend not to produce significant output in control industries. Consequently, there is minimal variation in this type of continuous measure and limited potential for this form of horizontal spillover.

\subsection{Direct Impact: Double-Robust DD and Average Effects} \label{sec:doublerobustestsection}

\subsubsection{Double-robust estimator}\label{sec:doublerobustestimator} I now employ the double-robust DD estimator proposed by \citet{SantAnna2020a} and \citet{Callaway2020}. Doing so allows me to consider the policy bundle's overall effect (ATT) and provides a robustness check on the TWFE estimates. In particular, this procedure relaxes some of the constraints of the traditional DD estimators and coherently adjusts counterfactuals. I consider the following specification:

\begin{equation}\label{eq:semi}
\begin{split}
\textrm{ATT}_t =\mathbf{E} \left[ \frac{\textrm{Targeted}}{\mathbf{E}\left[ \textrm{Targeted} \right] } - \frac{ \frac{\pi(X)(1-\textrm{Targeted})}{1-\pi(X)} }{\mathbf{E} \left[ \frac{\pi(X)(1-\textrm{Targeted})}{1-\pi(X)} \right] } \right] \left( Y_t - Y_{1972} \right) - f_{0,Y_t-Y_{1972}}\left(X\right) ,
\end{split}
\end{equation}

\noindent where equation \eqref{eq:semi} refers to the weighted average differences in industry outcomes. More precisely, equation \eqref{eq:semi} is the difference in outcomes between targeted industries ($\textrm{Targeted}$) and non-targeted industries ($1-\textrm{Targeted}$). Weights in equation \eqref{eq:semi} are defined as follows \citep{SantAnna2020a,Callaway2020}: the term $\pi(X) \equiv \mathbf{E}[\textrm{Targeted|X}]$ is the propensity score for the treated industries. The term $f_{0,Y_t-Y_{1972}}(X) \equiv \mathbf{E} [Y_t-Y_{1972}|\textrm{Targeted=0,X}]$ is a regression for the change in outcomes for non-treated industries between post-period $t$ and the baseline, pre-treatment period, $t=1972$. Propensity scores $\pi(X)$ and regression $f_{0,Y_t-Y_{1972}}(X)$ are estimated by logit and OLS, respectively. The estimator \eqref{eq:semi} is doubly robust in that if either component is correctly specified, it provides a consistent estimate of the ATT.

The double-robust estimator ensures a balance between targeted and non-targeted industries. The two-step procedure relaxes some of the functional-form assumptions of the evolution of potential outcomes. The pre-trend assumptions are also less stringent than those of other DD estimators. The average effects in equation \eqref{eq:semi} do not rely on zero pre-trends over \emph{all} pre-treatment periods, instead using a long-difference (between post-period $t$ and the last pre-treatment period, 1972). Confidence intervals for equation \eqref{eq:semi} are calculated using a bootstrap procedure, which allows industry-level clustering \citep{Callaway2020}. I use the same controls as the TWFE estimates above. Note that equation \eqref{eq:semi} requires a binary treatment and is not used for cases of continuous treatment, such as the indirect analysis in Section \ref{sec:networksection}.

\subsubsection{Results: average impacts}\label{sec:resultsatt} I first consider the overall average impact of the policy before and after 1972. Table \ref{tab:semitable} reports the ATTs, comparing double-robust and OLS estimates. Columns (1) and (3) list the doubly robust results, and columns (2) and (4) list the linear TWFE results. Because the double-robust estimator uses controls, I compare them only to TWFE estimates using controls. Panels A and B present estimates for the five-digit and four-digit panels, respectively. 

The estimates in Table \ref{tab:semitable} reveal that the overall average impact of HCI targeting was meaningful and significant. The preferred estimates in Panel A, column (1) indicate \resultssemiship\% growth in output for targeted manufacturers relative to non-targeted manufacturers.\footnote{Calculated using $100\times \left( \exp \left( \hat{\beta}-.5\times(\textrm{SE})^2 \right) -1 \right)$.} Similarly, linear TWFE estimates in Panel A (col. 2) suggest \resultsolsship\% output growth, significant at the 1\% level. The average impact on labor productivity (column 1) translates into a \resultslaborprodmax\% increase in value added per worker for targeted industries after 1973. Labor productivity growth ranges from \resultslaborprodmin--\resultslaborprodmax\% across four-digit and five-digit panels. Table \ref{tab:semitable} shows relatively lower prices, implying prices were \resultssemiprice\% (Panel A, column 1) lower relative to other industries over the period. The average employment effects of HCI in Table \ref{tab:semitable} are also substantial. Preferred double-robust DD estimates imply a \resultssemilabor\% increase in employment in Panel A, column (1), or a \resultssemilaborfour\% increase for four-digit data in Panel B, column (3). The reallocation of labor share is also positive and significant across specifications.

Table \ref{tab:semitablerca} shows substantial development in the heavy export industry. These results are significant and similar across measures of export development. Before 1973, the mean RCA index for targeted sectors was \resultsrcahcimean, while the average RCA for non-targeted Korea was \resultsrcanonhcimean (refer to Appendix Table \ref{tab:appendixtabledescriptive}). Table \ref{tab:semitablerca}, column (1) reports a significant increase in (log) RCA. These estimates translate into a \resultslogrca\% rise in RCA for targeted industry products. Column (1) implies that targeted industries saw a \resultsprobrca percentage point increase in the probability of attaining comparative advantage, or, alternatively, a \resultsexportshare\% increase in the (log) share of manufacturing exports (over non-targeted sectors).\footnote{The World Bank calculated that for HCI industries, the export share of output tripled during the period (\citep{Kim1993,Cho1995}).} The grand export target of the original heavy industry plan (50\% of manufacturing exports) was surpassed by 1983 \citep{Kim1993,Cho1995}.

\subsubsection{Results: Robust dynamic estimates}\label{sec:robustnessdreventstudy} Doubly robust event-study estimates show similar patterns to the direct effects discussed in Section \ref{sec:growthestimates}. Appendix \ref{sec:appendixdreventstudy} records and provides the dynamic DD estimates using the re-weighting estimator above. Here, the patterns (Appendix Figures \ref{fig:semiplot}--\ref{fig:semitrade}) are qualitatively similar to the linear TWFE estimates (Sec. \ref{sec:growthestimates}), although the double-robust DD relaxes some assumptions relative to the traditional TWFE DD. The general dynamic pattern associated with HCI is robust across estimators. Do these same patterns hold when using cross-country variation? I turn to this next using a triple difference estimation strategy.

\subsection{Direct Impact on Trade Development: Cross-Country Evidence}\label{sec:ddd}

\subsubsection{Cross-country variation and triple difference estimation}\label{sec:dddestimationtext} How did heavy-chemical industries in South Korea fare relative to the rest of the world? Cross-country data allows me to move beyond the within-country comparisons above. I use a DDD estimation strategy to expand on the DD analysis---intuitively, I compare the original DD estimates between HCI and control manufacturers in Korea to placebo DDs across international markets. I start with the following baseline specification:

\begin{subequations}
\renewcommand{\theequation}{\arabic{parentequation}\alph{equation}}

\begin{equation}\label{eq:ddd}
\begin{split}
Y_{ict} = \alpha_{i}+ \tau_{t} + \sigma_{c}+ \sum_{j\neq1972} \beta_{1j} \cdot \left( \textrm{HCI}_i \times \textrm{Year}^j_t \right) + \\
          \sum_{j\neq1972} \beta_{2j} \cdot \left( \textrm{Korea}_c \times \textrm{Year}^j_t \right) + \sum_{j\neq1972} \beta_{3j} \cdot \left( \textrm{Korea}_c  \times \textrm{HCI}_i \times \textrm{Year}^j_t \right) + \epsilon_{ict} , 
\end{split}
\end{equation}

\noindent where $c$ denotes country, $i$ denotes industry, and $t$ denotes time. I estimate equation \eqref{eq:ddd} using cross-country trade data (based on the SITC four-digit level). I focus on the triple interaction, $\textrm{Korea}_c \times \textrm{HCI}_i \times \textrm{Year}_t$, where $\textrm{Korea}_c$ is a dummy indicator for Korean observations. The simplest specification \eqref{eq:ddd} includes industry, time, and country effects: $\alpha_{i}$, $\tau_{t}$, and $\sigma_{c}$, respectively. However, using cross-country trade data allows me to control for a rich set of higher-dimensional fixed effects. Hence, I also consider a more stringent specification:

\begin{equation}\label{eq:ddd2}
\begin{split}
Y_{ict} = \alpha_{it}+ \tau_{ct} + \sigma_{ci}+ \sum_{j\neq1972} \beta_{3j} \cdot \left( \textrm{Korea}_c  \times \textrm{HCI}_i \times \textrm{Year}^j_t \right) + \epsilon_{ict} ,  
\end{split}
\end{equation}

\end{subequations}

\noindent where equation \eqref{eq:ddd2} controls for aggregate industry-year shocks ($\alpha_{it}$), aggregate country-year shocks ($\tau_{ct}$), and time-invariant country-by-industry factors ($\sigma_{ic}$). The effects in equation \eqref{eq:ddd2} thus subsume the $\textrm{Korea}_c \times \textrm{Year}_t$ and $\textrm{HCI}_i \times \textrm{Year}_t$ interactions from equation \eqref{eq:ddd}. 

Triple difference estimates (eq. \ref{eq:ddd}-\ref{eq:ddd2}) capture the impact of Korea's industrial policy on industrial development. The coefficients of interest are $\beta_{3j}$, estimated from the three-way interaction term: $\textrm{Korea}_c \times \textrm{HCI}_i \times \textrm{Year}_t$. In effect, I compare the conventional DD for Korea to placebo DDs over the same period. The identifying assumptions of DDD require differences in targeted and non-targeted outcomes for Korea to have trended similarly to differences in targeted and non-targeted industries (elsewhere) before the intervention.\footnote{Note that the difference between two biased DD estimators is considered unbiased when the bias is similar in both \citep{Olden2022}.} Triple difference estimates use two-way standard errors clustered at the industry and country level. I follow the empirical trade literature and estimate DDD specifications using PPML \citep{Silva2006} for RCA outcomes, given the preponderance of zeros. I also show alternative transformations and estimators for completeness.

\subsubsection{Results: cross-country trade development} \label{sec:dddresults} Figure \ref{fig:dddtrade} presents the triple differences estimates for the impact of Korean HCI on comparative advantage. The panels in Figure \ref{fig:dddtrade} plot the coefficient from the interaction: $\textrm{Korea}_c \times \textrm{HCI}_i \times \textrm{Year}_t$. I present multiple specifications: one using individual country, year, and industry effects; one using industry--year and country--year effects; and one that adds additional country--industry effects.

The DDD estimates in Figure \ref{fig:dddtrade} show a substantial policy impact on Korean heavy industry exports across trade outcomes. These outcomes include revealed comparative advantage (RCA, standard and normalized inverse hyperbolic sine), measures of relative export productivity (CDK), and the probability of achieving comparative advantage. RCA estimates use PPML to accommodate zeros; all others use OLS. The four plots in Figure \ref{fig:dddtrade} show similar patterns across the export development measures: muted differences before 1973 and a post-1973 shift in comparative advantage for the targeted Korean industry, which continued to ascend after the end of the drive. 

These cross-country patterns are robust even when using alternative, aggregate industry data, which I show in Online Appendix \ref{sec:suppunidorobustness}. The Online Appendix Figure \ref{fig:suppddaltrca} also shows a DD version of Figure \ref{fig:dddtrade}, comparing only targeted Korean industries to targeted placebo industries in ``non-treated'' counties. These results show a qualitatively similar pattern to those in Figure \ref{fig:dddtrade}.

How unusual is it for a country to cultivate an export advantage in targeted industries? Perhaps it is inevitable that countries naturally cultivate a comparative advantage in heavy or chemical industries. I assess the probability that an heavy-chemical industry achieves comparative advantage on the world market, $p(\textrm{RCA}>1)$, in Korea versus foreign controls in Appendix \ref{sec:appendixprobrca}. Appendix Table \ref{tab:appendixprobrca} shows that Korea had a significantly higher probability (between \resultsprobrcamin and \resultsprobrcamax\%) of achieving comparative advantage in HCI products after 1972 when compared to countries with similar levels of development (OLS estimates in Appendix Table \ref{tab:appendixprobrca}).

\subsection{Direct Impact: Discussion}\label{sec:directdiscussion}

The empirical relationship between industrial policies and industrial development is not a foregone conclusion. For many reasons, we may anticipate a negative relationship between an industrial policy and development outcomes (see \citep{Harrison2010, Lane2019, Juhasz2023}). Historically, there is no shortage of failures \citep{Pack2000}. Above, I showed a positive relationship between an industrial policy and industrial development outcomes from output growth to export development. The impact of these policies is seen throughout the HCI period (1973--1979) and continued through the 1980s. These results are robust across data sets (short-term and long-term panels), the type of estimator (TWFE vs. double-robust), and cross-country variation (DDD). Next, I turn to the forces underlying these results.

\section{Policy and Mechanisms}\label{sec:policysection}

\subsection{Policy: Credit Expansion, Investment, and Input Use}\label{sec:investmentpolicy}

The heavy and chemical industry drive aimed to promote investment and expand the sector through directed credit and investment incentives. This section examines the role and impacts of these policies. However, observing an explicit effect of these policy levers is challenging. Industrial statistics rarely capture such policy details, and these issues of observability are common in studies of industrial policy \citep{Kalouptsidi2018, Juhasz2022}. The HCI drive is no exception. Given these limitations, I examine indirect outcomes related to investment policy, following approaches in the credit policy literature \citep{Banerjee2014,Manova2015}. First, I demonstrate that intermediate outlays and investments responded differentially for treated industries. Then, drawing on \citep{Bau2021}, I show that input use responded in ways consistent with credit policy, specifically in treated sectors.

\subsubsection{Baseline results: input use and policy variations} Figure \ref{fig:maincapital} presents baseline DD estimates (eq. \ref{eq:mainflexible}) for input-related outcomes at the five-digit level. Plots A and B in Figure \ref{fig:maincapital} show that this divergence is starkest for intermediate outlays (total and per worker) beginning in 1973 and widening throughout the drive. Plots C--E report estimates related to investment and capital formation. These estimates grow significantly different between the treated and non-treated industries soon after the start of the industrial policy drive. Additionally, I observe similar investment patterns across asset classes, especially those targeted by investment incentives (e.g., machinery equipment; see Online Appendix Table \ref{tab:supptablerollingcapital2}).

The increase in intermediate outlays and investment for heavy industry was substantial. Table \ref{tab:semitableinvest} provides double-robust and TWFE DD estimates of the total average impact (ATT). Preferred estimates (Panel A) in column (1) translate into a \resultssemicosts\% relative increase in total intermediate outlays for treated over non-treated manufacturers. DD estimates for materials are highly significant for five-digit data and noisily estimated in four-digit panels. Similarly, for investment, double-robust DD estimates translate into a \resultssemiinvest\% increase in total investment for treated over non-treated industries, as shown in column (1) of Table \ref{tab:semitableinvest}. Investment estimates for four-digit data are imprecise, and investment per worker is negative, given the high employment growth. In light of industrial policy history, it is not obvious that we should expect positive effects on investment and input outlays (e.g., if lending leads to crowding out; see Appendix \ref{sec:appendixcapitaldiscussion}). 

\subsubsection{Policy mechanisms: changes in investment and input wedges}\label{sec:mrpksection} In theory, directed credit should reduce wedges on inputs for the treated industry during the heavy industry drive. Capital market policies that expand credit should disproportionately impact firms with high wedges, and these wedges can be captured through the pre-treatment marginal revenue product of capital (MRPK) \citep{Bau2021}. In other words, a policy should disproportionately impact investment in high-MRPK industries and increase the marginal revenue of other inputs. 

I find evidence consistent with investment policy operating in heavy-chemical sectors, reducing wedges for high-MRPK industries, specifically among targeted producers. I analyze the differential impact of industrial policy targeting on high-MRPK versus low-MRPK industries, using a basic measure of industry-level MRPK \textit{à la} \citet{Bau2021}. I present these calculations and details in Appendix \ref{sec:appendixmrpkanalysis} and show a marked increase in input use across intermediate inputs (intermediate materials, investment, and labor) for high-MRPK industries relative to low-MRPK industries (see: Appendix Figure \ref{fig:appendixmrpkpolicy}). Importantly, this change is only seen in targeted industries---no such effect is seen for non-treated industries. These results suggest that credit expansion differentially impacted the heavy-chemical sector.

\subsubsection{Robustness: investment and crowding out}\label{sec:crowdingout} Was the policy deleterious for investment in non-targeted industries? Although higher in treated industries, investment did not decline for non-HCI industries. Before formal tests, it is worth considering the evidence thus far in favor of crowding out. Section \ref{sec:historysection} demonstrated that, although biased toward heavy-chemical industries, lending continued for the non-treated sectors throughout the drive. Recall that commercial banks continued to lend to non-targeted industries and remained a major source of financial support during the period (see Appendix \ref{sec:appendixcommercialbanks} for details). Trends in non-targeted industry growth support this finding. Recall, Section \ref{sec:growthestimates} suggested that the relative ascent of heavy industry was not driven by the absolute contraction of non-targeted industries.

I now consider whether investment dynamics were different across treated versus untreated industries. I do so by regressing investment outcomes on time effects separately for each class of industry, as shown in Appendix \ref{sec:capitalcrowdingoutappendix}. I find that investment was high in targeted relative to non-targeted industries during the period, though it generally increased across both sectors. These results are consistent with the patterns of lending and growth described above.

Furthermore, although the heavy industry drive altered investment patterns, it did not reduce absolute investment in non-targeted industries. The analysis in Appendix Section \ref{sec:capitalcrowdingoutappendix} shows that investment was not crowded out in untreated, capital-intensive industries. Although less striking than investment in heavy industry, investment in the light industry sector continued, as non-heavy industry companies had access to domestic commercial credit and credit from countries like Japan \citep{Castley1997}.

\subsection{Policy: Trade Policy and the Weak Case for Nominal Protection}\label{sec:tradepolicysection}

Scholars have emphasized the role of trade policy, and some have characterized it as overtly protectionist \citep{Lall1997}. Evidence for the latter claim is weak, as shown in Appendix \ref{sec:appendixtradepolicy}. Instead, I posit that targeted cuts on import duties for intermediates could have been advantageous. Using a simple fixed effects regression and five periods of disaggregated protection data, I find that the average level of output protection was significantly lower for treated versus non-treated industries during the policy drive (see Appendix Table \ref{tab:appendixtradepolicy}, Panel A). Appendix Table \ref{tab:appendixtradepolicy} also suggests that nominal output protection fell more for non-treated than treated industries. Similarly, assuming that trade policy allowed for discounts on imported inputs (Section \ref{sec:policyvariation}), many heavy industrial producers enjoyed significantly lower duties on foreign inputs (Appendix Table \ref{tab:appendixtradepolicy}, Panel B). Together, the evidence does not suggest a surge in overt, nominal protectionism over the period.

\subsection{Mechanisms: Targeted Industry and Learning}\label{sec:mechanisms}

Did the industrial policy promote industries with strong learning-by-doing forces? I now examine the potential learning-by-doing effects in treated industries. If learning-by-doing forces were at work, we would expect increased cumulative experience to correspond to higher productivity or a lower unit cost. To assess whether learning was particularly strong in treated sectors, I employ a simple, reduced-form regression for the post-1972 period. Specifically, I consider the following equation:

\begin{equation}\label{eq:lbdindustry}
Y_{it} = \beta_1 \textrm{Experience}_{it} + \beta_2 \left( \textrm{Experience}_{it} \times \textrm{Targeted}_{i} \right) + \theta \textrm{Size}_{it} + \alpha_i+ \tau_t + X'_{it}\Omega + \epsilon_{it},
\end{equation}

\noindent where $Y_{it}$ represents industry (or plant) log prices, log unit cost, or TFP, following \citet{Gruber1998,Barrios2004,Fernandes2005}. I measure unit cost as total intermediate costs per unit of real gross output. Equation \eqref{eq:lbdindustry} examines a reduced-form relationship between these outcomes and log $\textrm{Experience}_{it}$, measured as real cumulative gross output up to time $t$. All baseline regressions control for measures of plant size ($\textrm{Size}_{it}$) to account for conventional scale effects. Additionally, I control for the effects of technological progress embodied in input use, $X_{it}$ (e.g., total input outlays, capital intensity). These (log) covariates are normalized by the number of workers to further account for scale effects. I include year effects ($\tau_t$) and industry effects ($\alpha_i$) in industry-level regressions or plant effects in micro-level regressions.

The correlations in equation \eqref{eq:lbdindustry} indicate potential learning externalities over the policy period. The coefficient $\beta_1$ is the general impact of cumulative output ($\textrm{Experience}_{it}$), and $\beta_2$ is the differential impact of $\textrm{Experience}_{it}$ for the treated industries. Hence, estimates from equation \eqref{eq:lbdindustry} examine whether dynamic externalities are present in targeted industries and their strength in treated sectors relative to non-treated sectors (see \citep{Beason1996} and \citep{Pons-Benaiges2017}). These estimates are indicative and not causal.

First, consider the industry-level estimates of equation \eqref{eq:lbdindustry}. Table \ref{tab:lbdindustry}, columns (1)--(4) demonstrate that experience is positively related to reductions in prices and unit costs, with the effect being significantly stronger for targeted sectors. Estimates for the interaction $\textrm{Targeted}_i \times \textrm{Experience}_{it}$ are negative and highly significant. Similarly, columns (5)--(10) show a positive relationship between experience and productivity using three measures of TFP. In these cases, the correlation between experience and TFP is stronger for targeted industries, with interactions being significant for Levinsohn-Petrin (LP) measures of TFP. Furthermore, the combined effect of experience for treated industries (shown at the bottom of Table \ref{tab:lbdindustry}) is strong and significant across all specifications. Appendix Table \ref{tab:appendixindustrylbd} confirms that these results are robust to alternative measures of experience, unit cost, and TFP.

Second, I analyze microdata to investigate the correlation between learning and targeting after 1979, when microdata first became available. Expanding on equation \eqref{eq:lbdindustry}, I regress plant TFP and log unit cost on two types of log cumulative experience: (i) plant-level and (ii) industry-level (four-digit), both measured from the beginning of the sample period. All regressions include plant and industry-level fixed effects to account for time-invariant factors (micro and sectoral) that influence learning. As before, I control for year effects. I employ two-way clustered standard errors to allow for sectoral and plant-level correlation.

The plant-level estimates in Table \ref{tab:lbdmicro} provide evidence of learning, even in the period following infant industry policy. Similar to the industry estimates (Table \ref{tab:lbdindustry}), columns (1)--(3) of Table \ref{tab:lbdmicro} show a negative relationship between experience and unit cost reduction, now decomposing learning into plant and industry levels. The estimates for plant-level experience are differentially stronger among targeted establishments (cols. 1--3). Moreover, the estimates for industry-level experience are also significant---and significantly stronger---for targeted industries (cols. 2--3). Similarly, industry-level experience has a positive impact on TFP (cols. 5--6). Including the industry learning reduces estimates for the $\textrm{Targeted}\times\textrm{(Plant Experience)}$ interaction; however, the effect of $\textrm{Targeted}\times\textrm{(Industry Experience)}$ remains significant.

These micro estimates indicate that plant and industry-level learning may be more pronounced for treated establishments. The combined effects of plant and industry-level estimates are substantial for heavy-chemical industry establishments, as shown at the bottom of Table \ref{tab:lbdmicro}. Predictably, plant-level experience generally exerts a larger effect than industry-level experience. Appendix Table \ref{tab:appendixmicrolbd} demonstrates the robustness of these results across alternative measures of experience, unit costs, and TFP. However, it is important to note that the plant-level estimates only cover the post-1979 period, thereby excluding potentially steep learning curves during the earlier stages of the industrial drive.

The correlational results in this section indicate that learning externalities plausibly impact targeted industries. Of course, this correlational analysis cannot definitively identify the strength of the externalities or whether they originate from plant-level learning or industry-wide learning. Nevertheless, taken together, the industry and plant-level analyses suggest the potential for learning-by-doing spillovers operating in heavy and chemical industries.

\section{Indirect Impact of Industrial Policy}\label{sec:networksection}

I now consider how the industrial policy drive may have impacted industries outside of the targeted sectors through linkages. I use the terms ``backward'' and ``forward'' from the vantage point of the targeted industry. When the impact of industrial policy propagates from the treated heavy-chemical industry to upstream suppliers, suppliers are impacted through backward linkages. When the impact of industrial policy propagates downstream to users of heavy-chemical industry products, buyers are impacted through forward linkages. I refer to both as linkage effects. The following analysis draws on the empirical study of foreign direct investment (FDI) spillovers, particularly \citet{Javorcik2004}, and empirical work on the propagation of policy shocks (e.g., \citep{Acemoglu2015}).

I measure an industry's linkage exposure to industrial policy using South Korea's 1970 input-output accounts, which predate the HCI drive. Specifically, I calculate industry $i$'s exposure to industrial policy through backward and forward linkages as follows:

\begin{subequations}
\begin{align}
    \textrm{Backward Linkage}_i &= {\displaystyle\sum_{j \in \textrm{HCI}} \alpha_{ij}} \label{eq:backwardlinkage}, \\[1em]
    \textrm{Forward Linkage}_i &= {\displaystyle\sum_{j \in \textrm{HCI}} \alpha_{ji}} \label{eq:forwardlinkage},
\end{align}
\end{subequations}

\noindent where $j$ represents the treated heavy-chemical industries. For industry $i$, its $\textrm{Backward Linkage}_i$ \eqref{eq:backwardlinkage} equals the weighted sum of output supplied to treated industries $j$. The weight $\alpha_{ij}$ denotes the value of $i$'s output used by $j$ as a share of $j$'s total output and comes from the IO accounts. For industry $i$, $\textrm{Forward Linkage}_i$ \eqref{eq:forwardlinkage} equals the weighted sum of inputs sourced from treated industries $j$. The weights $\alpha_{ji}$ denote the value of $j$'s output sold to $i$ as a share of $i$'s total value of output in the input-output accounts. For further details on these calculations, refer to Appendix \ref{sec:appendixlinkcalculations}.

The measures above (\ref{eq:backwardlinkage}-\ref{eq:forwardlinkage}) capture direct spillovers to industries one degree away from the heavy-chemical sector. To account for both direct and indirect effects, I extend this analysis using the Leontief inverse, which captures the full network of linkages (first, second, ..., and \textit{n}-degree) between Korean industries. For industry $i$, I construct $\textrm{Total Backward Linkages}_i$ and $\textrm{Total Forward Linkages}_i$ using a method analogous to equations \ref{eq:backwardlinkage}-\ref{eq:forwardlinkage}, but now employing weights derived from the Leontief inverse calculated from the 1970 IO accounts. For example, $\textrm{Total Backward Linkages}_i = {\sum_{j \in \textrm{HCI}} \ell_{ij}}$, where $\ell_{ij}$ is an element of the Leontief inverse matrix. See Appendix \ref{sec:appendixlinkcalculations} for details.

To study the impact of linkages, I compare outcomes across industries with strong versus weak linkages to treated industries relative to 1972. In the spirit of the main DD analysis (eq. \ref{eq:mainflexible}), I consider the following specification:

\begin{equation}\label{eq:networkflexible}
\begin{split}
\ln\left(y_{it}\right) = \alpha_i + \tau_t+ 
\sum_{j\neq1972} \gamma_j \cdot \left( \textrm{Backward Linkage}_i \times \textrm{Year}^j_t \right) + \\
\sum_{j\neq1972} \delta_j \cdot \left( \textrm{Forward Linkage}_i \times \textrm{Year}^j_t \right) + \epsilon_{it}, 
\end{split}
\end{equation}

\noindent where $Y_{it}$ is an outcome and $i$ indexes each five-digit (or four-digit) industry. Subscript $t$ denotes the years, which are 1967--1986 for the four-digit panel and 1970--1986 for the five-digit panel. As before, equation \eqref{eq:networkflexible} uses two-way fixed effects for time $\tau_t$ and industry $\alpha_i$. I first estimate equation \eqref{eq:networkflexible} using only non-treated industries. I show these estimates alongside estimates from the full sample, which provide additional power. For the full-sample estimation, I control separately for the direct impact of policy using the interaction term $\textrm{Targeted}_i\times\textrm{Year}_t$.

The coefficients of interest, $\gamma_j$ and $\delta_j$, reflect the differential evolution of industries with strong versus weak exposure to treated industries, measured by $\textrm{Backward Linkage}_i$ and $\textrm{Forward Linkage}_i$. The set of estimates, $\widehat{\gamma_j}$ ($\widehat{\delta_j}$), captures the differential development of industries with strong backward (forward) linkages to targeted industries relative to those with weaker linkages. Note that specification \eqref{eq:networkflexible} uses a continuous treatment, whereas estimates in the first part of this paper (Section \ref{sec:directsection}) used a binary treatment. I estimate the model using the baseline linear TWFE estimator.

Before 1972, the set of coefficients should be zero, reflecting no prior differences between industries with stronger linkages. Estimates over the policy period suggest the potential strength and direction of linkage spillovers to non-treated industries. For instance, if the industrial policy increases the cost of key inputs over the policy period, we may expect negative estimates for $\widehat{\delta}_{1973},...,\widehat{\delta}_{1979}$. Estimates for the post-1979 period indicate, among other things, longer-term spillovers from the policy. The identifying assumption is that differences in industrial development between stronger or weaker backward (forward) linked industries would have evolved similarly in the absence of the heavy and chemical industry policy.

\subsection{Indirect Effects: Results}\label{sec:networkresults}

Below, I find that industries with relatively strong forward linkages with targeted industries developed more robustly over the policy period. Specifically, downstream industries that were more dependent on inputs from targeted sectors showed greater industrial development. In contrast, the impact of backward linkages---where industries supply inputs to targeted sectors---appears to have been more limited. 

\subsubsection{Downstream industrial development}\label{sec:forwarddevresults} Figure \ref{fig:mainlinkage} plots the relationship between the strength of forward linkages and downstream output (eq. \ref{eq:networkflexible}). Rows in Figure \ref{fig:mainlinkage} correspond to estimates for real value added (top) and output prices (bottom). For this analysis, I consider output measured in terms of the value added, given different stages of production and input intensity. The columns in Figure \ref{fig:mainlinkage} present estimates across different data sets (four-digit versus five-digit) and samples (full sample versus only non-treated). Panels B and D restrict the sample to non-targeted industries, which significantly reduces the sample size and power, especially in aggregate four-digit data. Alternatively, Panels A and C provide estimates using the entire sample of industries and flexibly control for targeted industries $(\textrm{Targeted}_i \times\textrm{Year}_t)$.

Figure \ref{fig:mainlinkage} shows that industries with stronger forward linkage exposure expanded more often following the policy drive. Before 1973, differences among the industries were noisy, trending upwards in the 1960s, and centered on zero.\footnote{Online Appendix Table \ref{tab:supprollingforwardoutput} rejects pre-trends across specifications, except those for the non-HCI sample in the four-digit data.} Table \ref{tab:prepostlinkoutput} reports the average pre-post version of equation \eqref{eq:networkflexible} and presents both forward linkage estimates and backward linkage estimates. For output, average forward linkage estimates imply that a 1\% rise in the share of links (between 0 and 1) from a treated industry is associated with \resultforwardoutnonhci\% more output (col. 2) for the non-treated industry; estimates for the full sample imply a semi-elasticity of \resultforwardoutall (col.~1). Estimates across specifications are positive and significant for direct forward linkages. A similar pattern also holds for the total forward linkages (see Appendix \ref{sec:appendixforwardmoredev} and Table \ref{tab:appendixprepostlfoutput}).

Similarly, Table \ref{tab:prepostlinkprices} shows that greater exposure to forward linkages is associated with reduced output prices. Panel A, column (2) implies that a 1\% rise in the share of direct HCI linkages is associated with \resultforwardpricesnonhci\% lower output prices of non-HCI industry (\resultforwardpricesall for the full sample, col. 1). Appendix Table \ref{tab:appendixprepostlfprices} shows a similar strong negative relationship for total forward linkages. Dynamic estimates plotted in Figure \ref{fig:mainlinkage} demonstrate that industries using more treated inputs had relatively low output prices during and after the drive. However, prices were relatively higher and began converging before the policy introduction. Thus, the price effects in Figure \ref{fig:mainlinkage} may have already been in motion before HCI. Nevertheless, the policy is associated with declining output prices in the downstream industry. This result contrasts with arguments that similar industrial policies are associated with increased prices for downstream firms \citep{Blonigen2015}.

There is also a positive relationship between forward linkage exposure and development outcomes. This relationship is particularly strong and significant for relative employment and plant entry in downstream industries using large shares of treated inputs. This finding holds across datasets and applies to direct and total linkage measures (see average DD estimates in Appendix Tables \ref{tab:appprepostlinkmoredev}--\ref{tab:appprepostlfmoredev}). I provide a more detailed analysis of these effects in Appendix \ref{sec:appendixforwardmoredev}. The results show weakly positive estimates between forward linkages and labor productivity, wages, and TFP.

\subsubsection{Evolution of downstream comparative advantage}\label{sec:resultsforwardrca} What was the relationship between forward linkages and trade development? To explore this question, I combine information on linkages with the SITC-level trade data and consider the same regressions as above. As before, I employ a PPML estimator for trade flow outcomes.

Like output, Figure \ref{fig:rcalinkage} shows a positive relationship between the strength of forward linkage exposure and improved export development in downstream industries (Online Appendix Table \ref{tab:supprcaforwardlink} shows full estimates). Prior to 1973, forward-linked sectors did not demonstrate a relative export advantage or export productivity over other downstream sectors. Post-1973, Figure \ref{fig:rcalinkage} shows a shift in comparative advantage that emerged over the 1973--1979 period. However, it took time for a comparative advantage to manifest, and estimates appear strongest in the 1980s. These patterns hold across traditional and modern measures of RCA. These effects are also seen in measures of total forward linkages, reported in Appendix Figure \ref{fig:appendixtradetotallinkage}.

The previous section presented evidence of positive, contemporaneous spillovers from industrial policy. However, other spillovers may take time to materialize. Furthermore, the positive relationship between forward linkages and export development further supports the direct main effects of policy shown in Section \ref{sec:directsection}. Had the policy been unsuccessful, it may well have harmed downstream exports.

\subsubsection{Downstream linkages: mechanisms, investment, and intermediates}\label{sec:resultslinkmech} Where the industrial policy affected downstream industries, it likely did so by supplying domestic inputs for their benefit. I analyze this in Appendix \ref{sec:appendixforwardmechanism} and find that material outlays expanded relatively more for downstream users (both direct and indirect) of heavy industrial goods. This finding is illustrated in Appendix Figure \ref{fig:appendixmechanismlinkage} (Panels A and B).

\subsubsection{Backward linkage: weak relationship with industrial development}\label{sec:backwardlink} The expansion of a targeted sector may promote upstream suppliers by increasing the demand for their goods. However, for this episode, the spillovers from the heavy industry drive to upstream suppliers appear to have been limited. This minimal effect may be because policy planners (Section \ref{sec:historysection}) chose relatively upstream industries \citep{Liu2019}, potentially constraining the extent of spillovers through backward linkages.

For instance, Table \ref{tab:prepostlinkoutput} shows that an upstream industry with high backward linkage exposure is not associated with a differential increase in output, unlike the positive impact of forward linkage exposure. The same pattern is also seen for similar DD estimates using total backward linkage measures (Appendix Table \ref{tab:appendixprepostlfoutput}). Similarly, the relationship between backward linkage exposure is undetectable for employment, plant entry, and other development outcomes, as seen in Appendix Table \ref{tab:appprepostlinkmoredev}. I discuss the muted estimates of backward linkages further in Appendix \ref{sec:appendixbackwards}.

\subsection{Robustness and the Stable Unit Treatment Value Assumption}\label{sec:sutvaresults}

The indirect effects above (Section \ref{sec:networkresults}) pose a dilemma in light of the direct effects of the policy highlighted in Section \ref{sec:directestimation}. That is, the network effects of the policy may contaminate the control group by virtue of linkage spillovers, violating the stable unit treatment value assumption (SUTVA). For robustness, I demonstrate that the pattern of direct effects largely survives after accounting for the indirect effects in three analyses: 

First, I examine how the main effects change when limiting the control group to industries with lower exposure to forward linkages from treated industries. Specifically, I restrict control sectors to industries with below-median linkage measures. For both output and labor productivity, estimates using the ``limited exposure" group (for both direct and total linkages) do not significantly alter the main policy effect (Appendix Figure \ref{fig:appendixsutvalinkexposure}).

Second, I report the main effects while controlling for linkage exposure in the control group. Appendix \ref{sec:appendixsutvacontrollink} shows that after controlling for positive downstream spillovers in non-treated industries, the main impact of HCI becomes more pronounced. This finding is intuitive, as positive spillovers may cause the control group to benefit, slightly biasing estimates downward. While controlling for linkages increases the standard errors, the main effects persist. 

Third, Appendix \ref{sec:appendixcrowdingoutio} provides additional evidence that investment is not crowded out when accounting for linkages.

\subsection{Indirect Impact: Discussion}\label{sec:networksdiscussion}

The analysis above demonstrates policy spillovers through linkages to and from treated heavy-chemical industries. I find that non-treated industries with high exposure to policy through forward linkages are associated with higher development outcomes and increased use of intermediates. This positive relationship extends to the later export development of downstream sectors. However, the impact of backward linkages appears to have been limited and ambiguous, possibly because treated sectors were, by design, upstream. While indirect effects, even if weak, may influence the control group, Section \ref{sec:sutvaresults} shows that these linkage effects do not significantly alter the qualitative pattern observed in the main policy effects.

\section{Conclusion}\label{sec:conclusionsection}

This paper shows that Korea's heavy and chemical industry drive promoted industrial development in the manufacturing sectors targeted by the policy. I find that this intervention had wide ramifications. First, the drive created positive effects in treated industries long after its major elements had been retrenched. In the case of export performance, policy effects took time and fully materialized after the policy had ended. I provide cursory evidence that the dynamic effects may correspond to learning mechanisms. Moreover, the regime's policy likely impacted the development of industries not targeted by the policy, both in the short and long run. Thus, this study takes a multidimensional view of industrial development, demonstrating that HCI targeting corresponded to improvements across an array of outcomes, from export performance to the labor market.

Aspects of these findings correspond to arguments proposed by \citet{Wade1990} and \citet{Amsden1992}, mainly that active policy may have contributed to Korea's industrialization and its shift in comparative advantage to more advanced industries. My results, however, emphasize conventional policy forces rather than miraculous ones. These included the use of directed credit to facilitate investment, the purchase of key intermediates, and the promotion of sectors with dynamic economies and linkage spillovers.

History is not a clean laboratory, and South Korea's experience is no exception. Like many transformations, South Korea's was tumultuous and multifaceted. Nevertheless, this study attempts to decipher a key episode of industrial policy using the contemporary econometric toolbox. The goal is to structure coherent insights around a key historical case of industrial transformation. By doing so, I hope to extract more coherent workings of the policy---those that are useful more broadly---and emphasize a more empirically grounded narrative around East Asian interventions. The findings here are not final, nor could they be. Instead, they point to a potential direction for further empirical work. 

The limitations of this study are manifold and show the necessity of further study. Although heavy industrial policy may have promoted forms of industrial development, it did so at a cost that cannot be accounted for within the scope of this study. Nor have I examined the aggregate or allocative consequences of the episode. I leave those questions to future quantitative and empirical work. Importantly, the context of this study suggests that successful industrial policy likely hinges on bureaucratic capacity and political incentive compatibility \citep{Haggard1990,Evans1995,Robinson2009,Juhasz2024}. Such factors highlight the importance of future research on the political economy of industrial policy.

\ifx\undefined\bysame
\newcommand{\bysame}{\leavevmode\hbox to\leftmargin{\hrulefill\,\,}}
\fi

\clearpage

\setcounter{figure}{0}

\begin{figure}[h]

\begin{center}
\includegraphics[width=0.9\columnwidth]{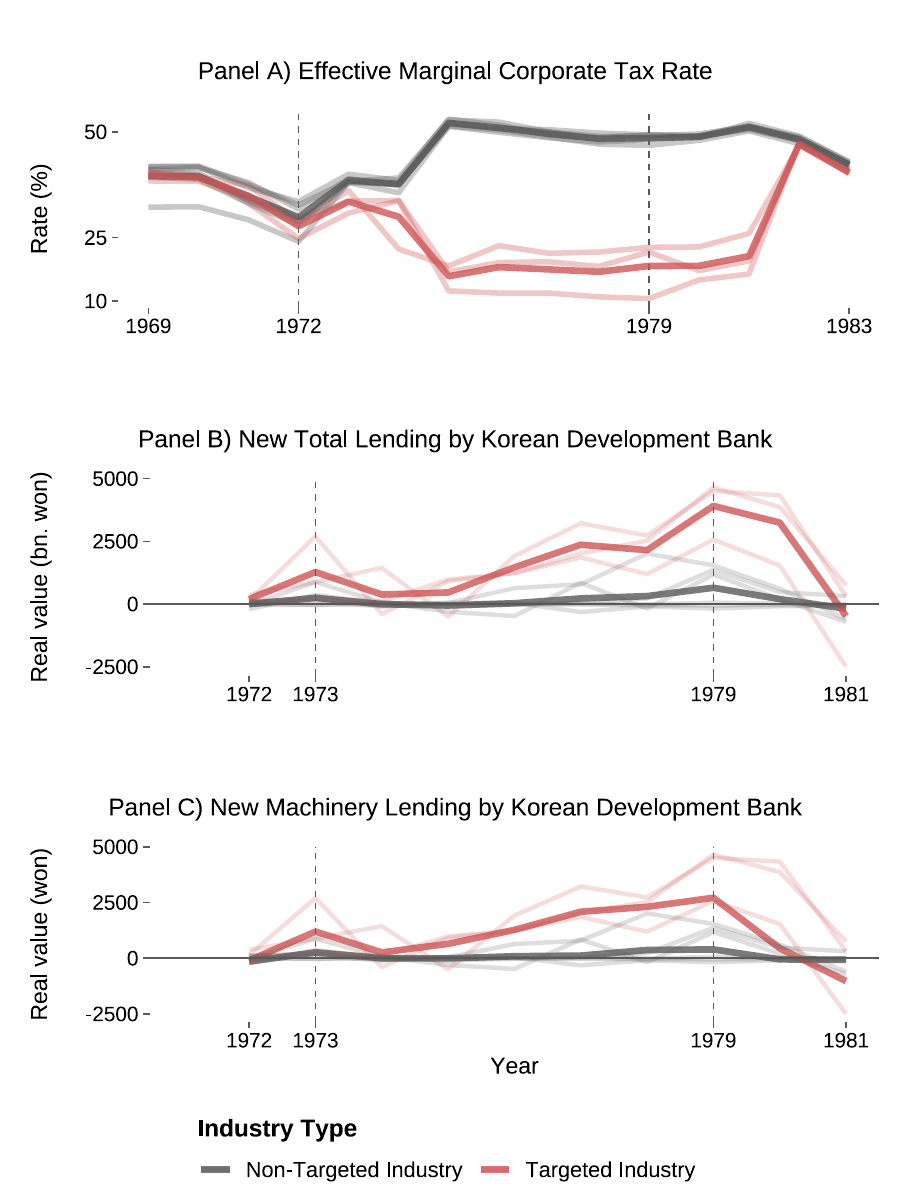}    
\end{center}

\caption{Investment Policy and the Heavy and Chemical Industry Drive}\label{fig:lendingpolicyplot}

\scriptsize
{
\setlength{\parindent}{2em}
    \input{policyplot_note.tex}
}
\end{figure}

\FloatBarrier

\blandscape

\begin{figure}[h]
\begin{center}
\includegraphics[width=\columnwidth]{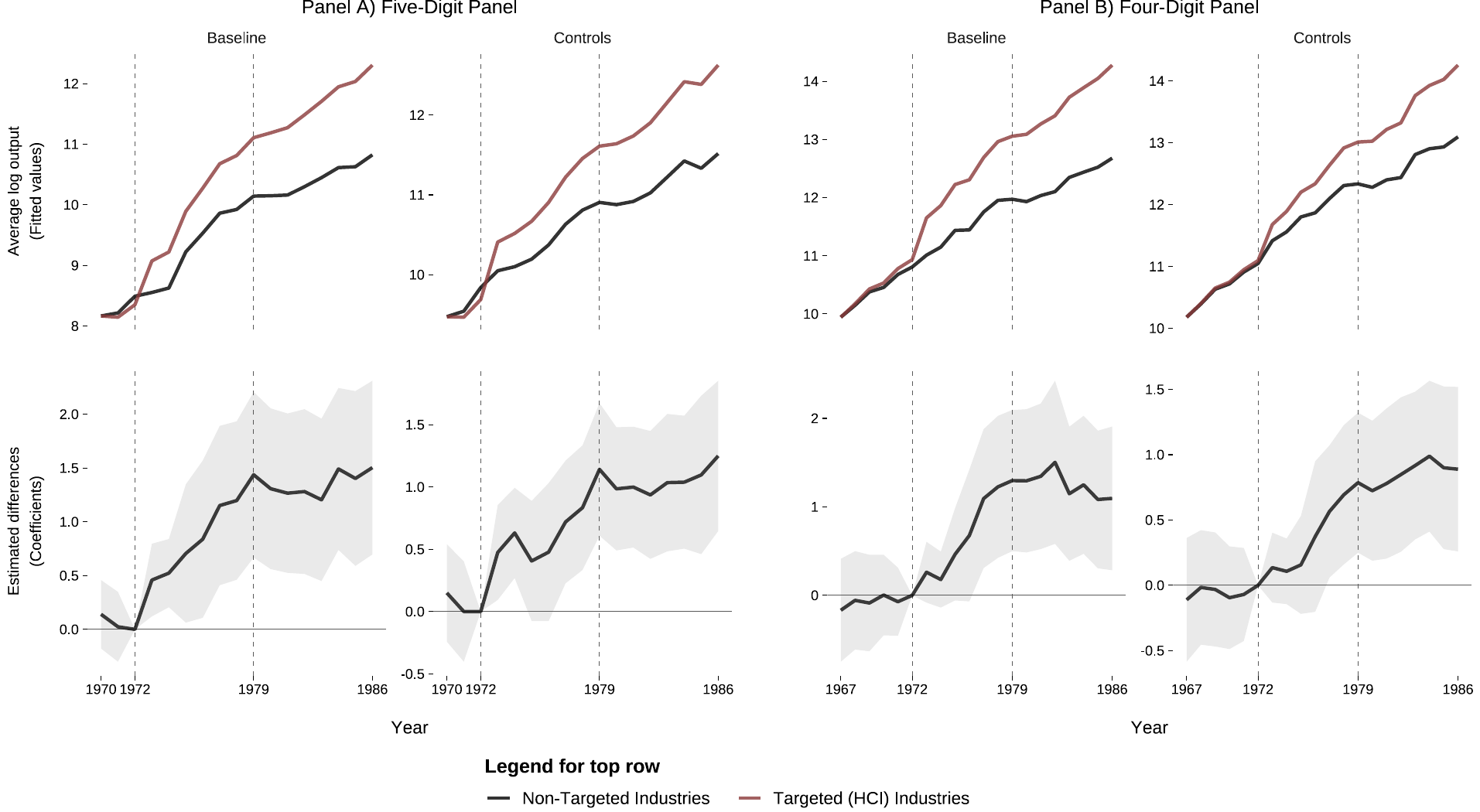}
\end{center}

\caption{Industrial Policy and Industry Output}\label{fig:mainoutputfigure}

\scriptsize
{
\setlength{\parindent}{2em}
    \input{mainoutputplot_note.tex}
}
\end{figure}
\elandscape

\pagebreak

\blandscape
\begin{figure}[h]
\begin{center}
    \vspace*{-1.5cm} 
    \includegraphics[height=\textwidth]{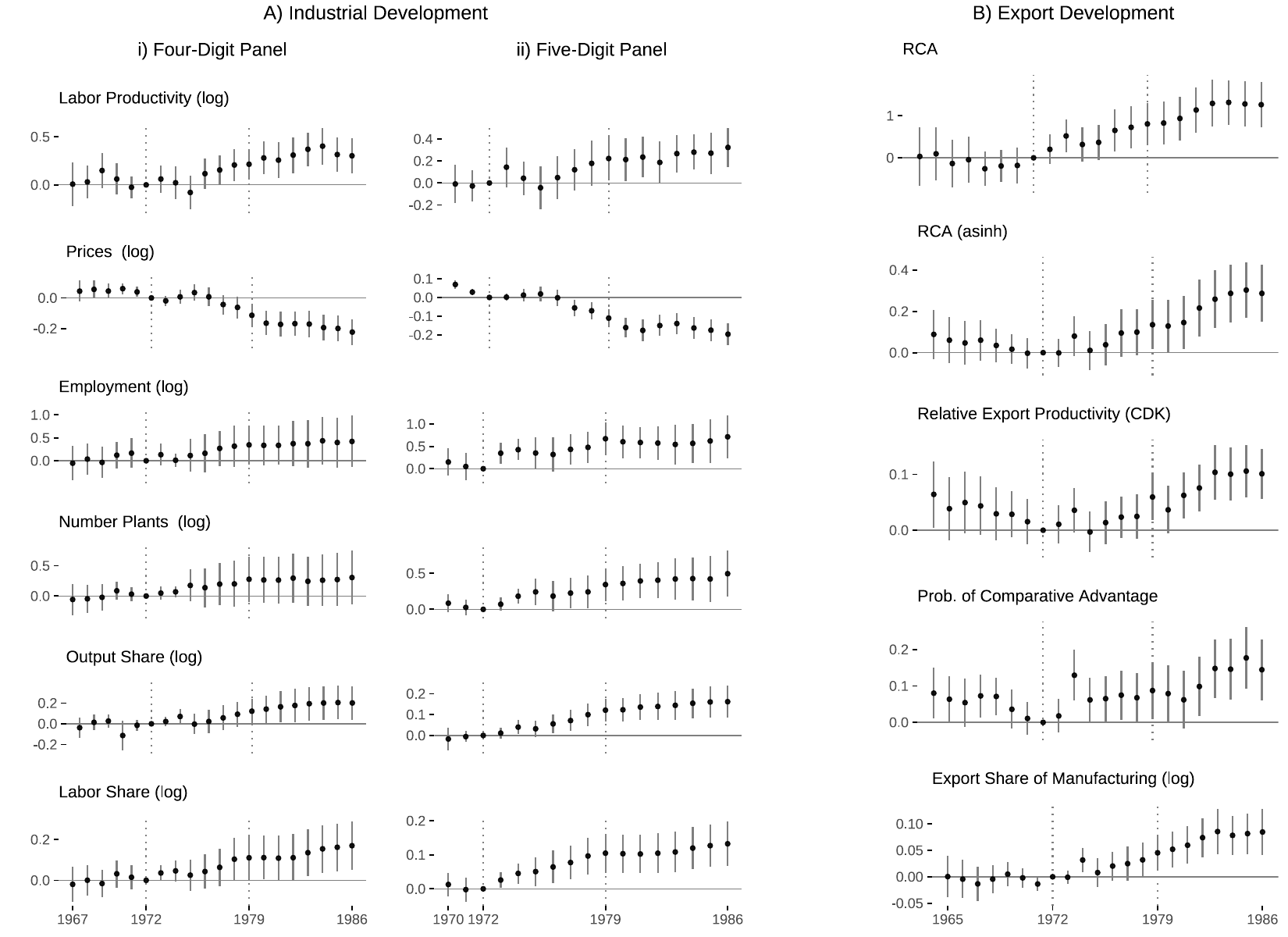}    
\end{center}

\caption{Industrial Policy: Industrial and Export Development}\label{fig:maindevelopment}

\scriptsize
{
\setlength{\parindent}{2em}
    \input{devcombinedplot_note.tex}
}
\end{figure}
\elandscape
\FloatBarrier

\blandscape

\begin{figure}[h]
\begin{center}
    \vspace*{-1.25cm} 
    \includegraphics[height=.85\textwidth]{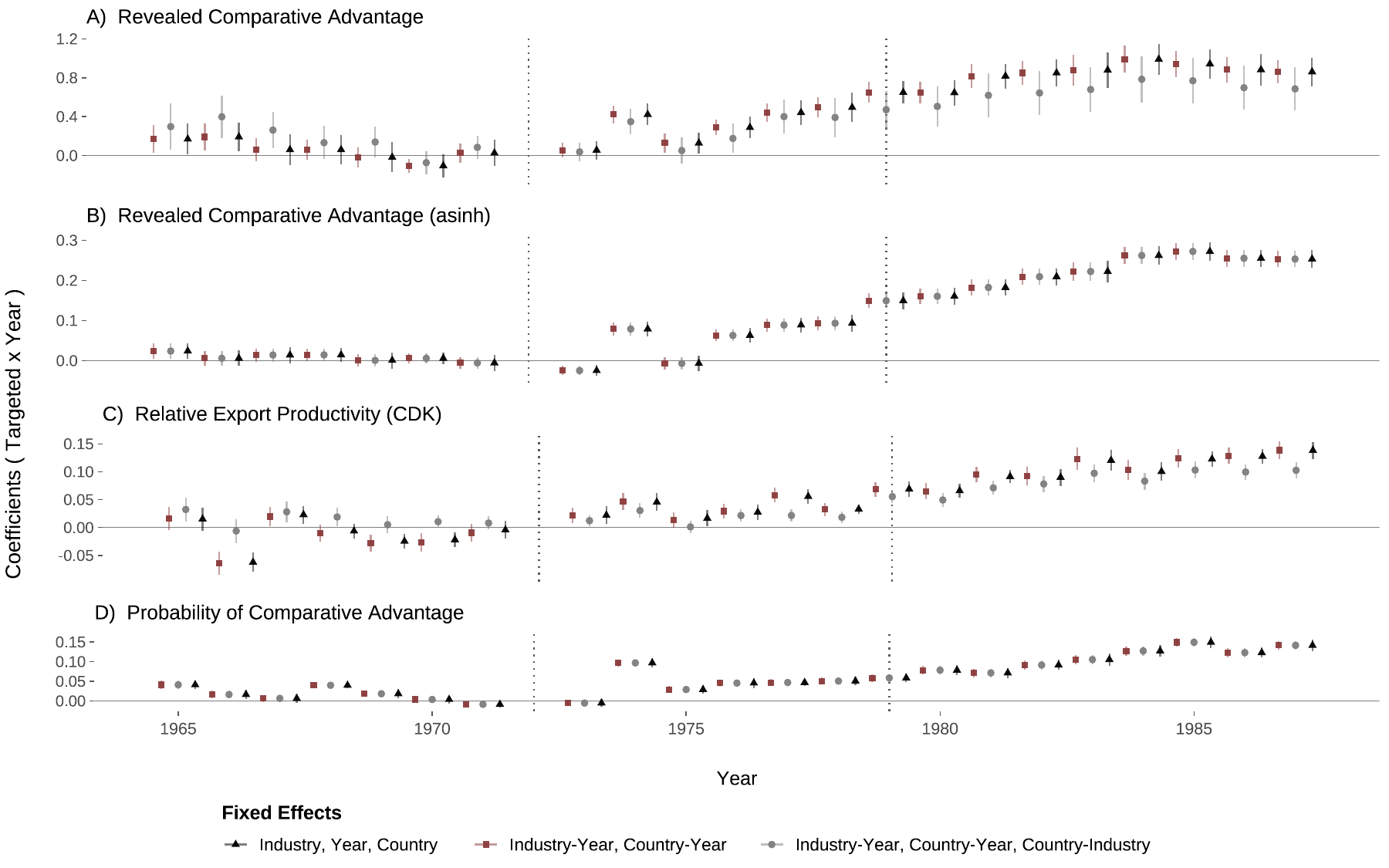}
\end{center}

\caption{Cross-Country Triple Differences and Export Development}\label{fig:dddtrade}

\scriptsize
{
\setlength{\parindent}{2em}
    \input{dddtradeplot_note.tex}
}

\end{figure}

\elandscape

\FloatBarrier

\begin{figure}[h]
\begin{center}
\includegraphics[width=\textwidth]{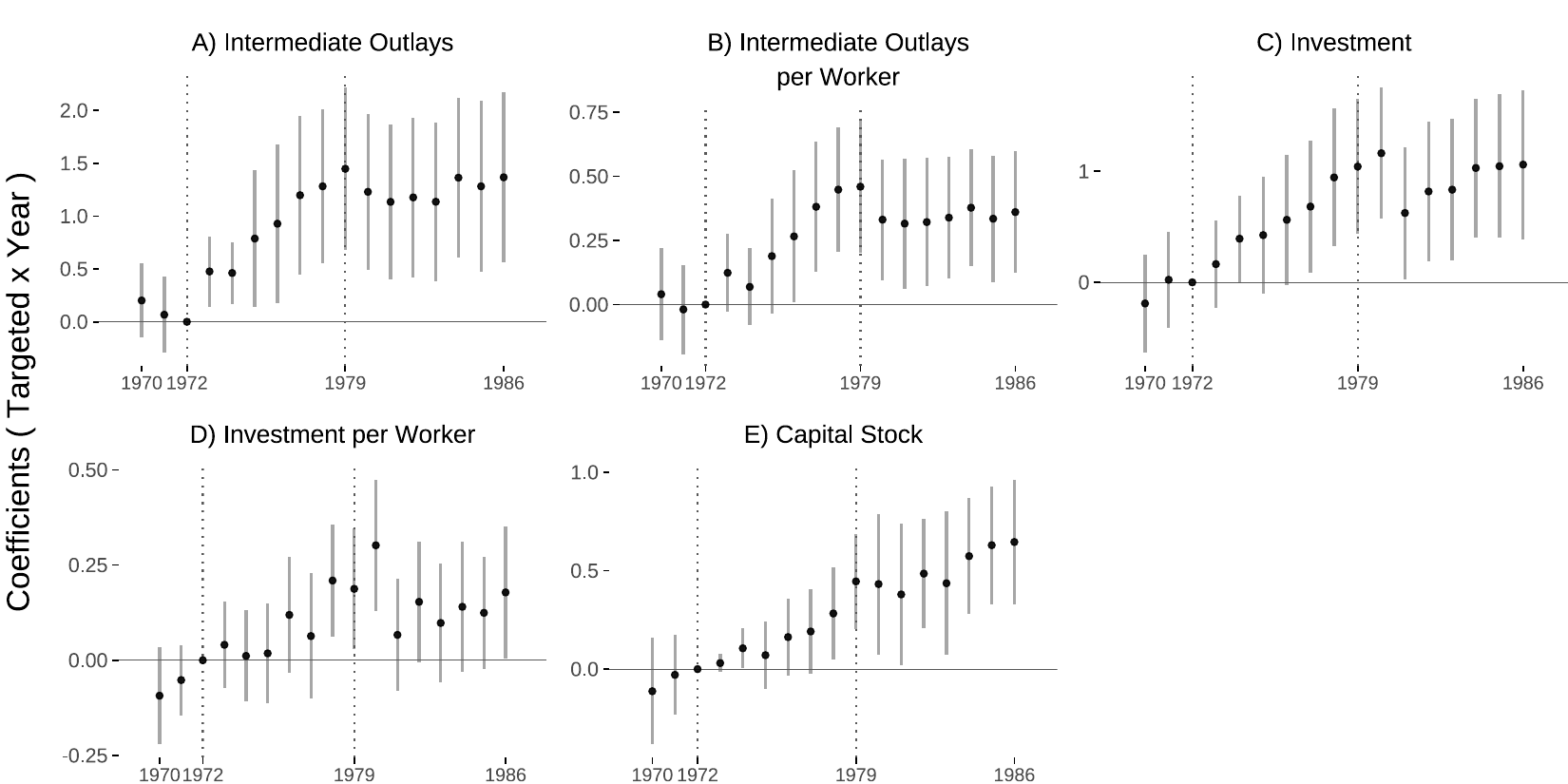}
\end{center}

\caption{Changes in Input Use and Investment}\label{fig:maincapital}

\scriptsize
{
\setlength{\parindent}{2em}
    \input{capitalplot_note.tex}
}

\end{figure}

\FloatBarrier

\blandscape

\begin{figure}[h]
\begin{center}
\includegraphics[height=.70\textwidth]{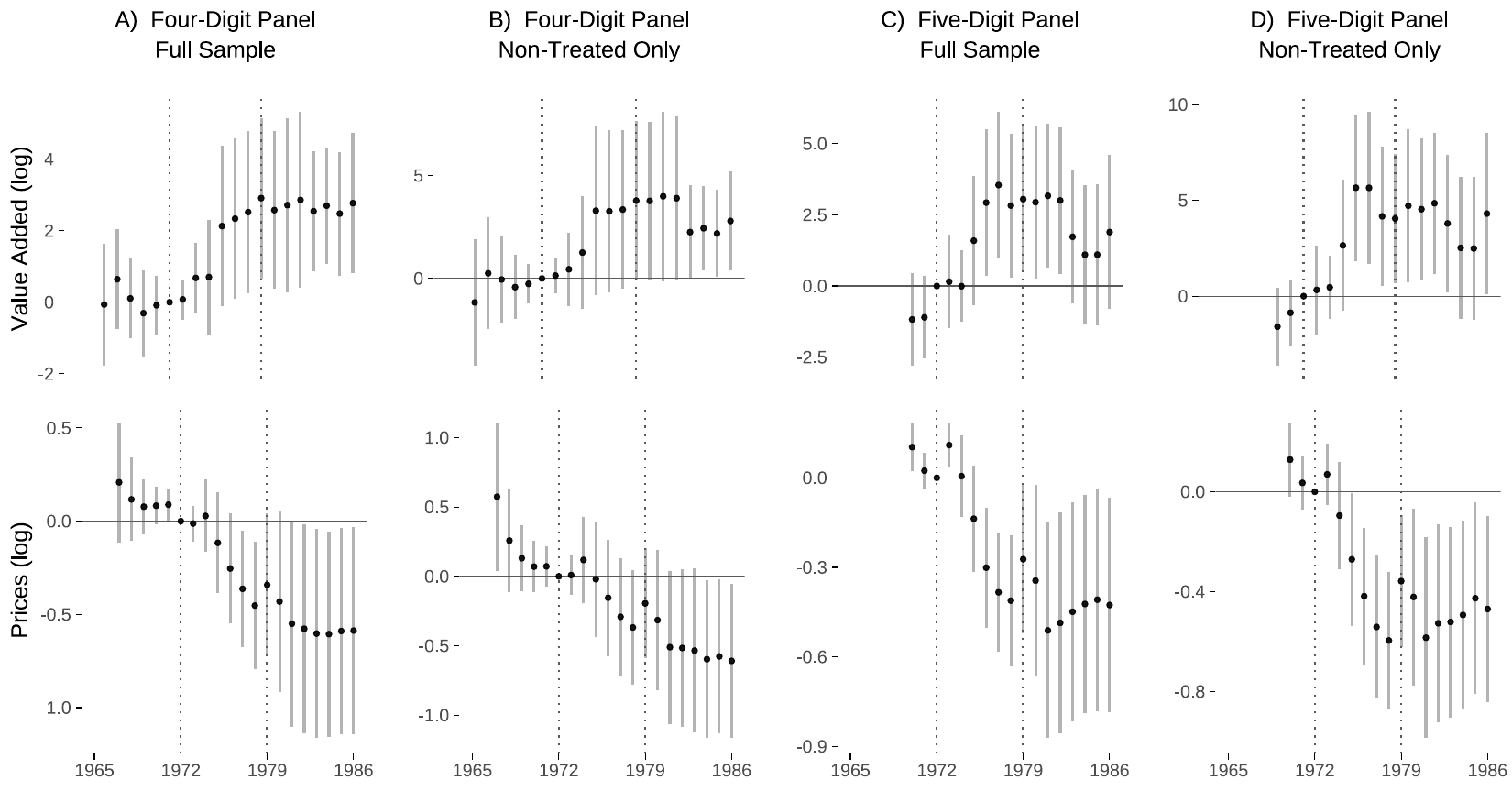}    
\end{center}

\caption{Forward Linkages Exposure: Value Added and Output Prices}\label{fig:mainlinkage}

\scriptsize
{
\setlength{\parindent}{2em}
    \input{forwardlinkageplot_note.tex}
}
\end{figure}

\elandscape
\FloatBarrier

\begin{figure}[h]
\begin{center}
\includegraphics[width=\textwidth]{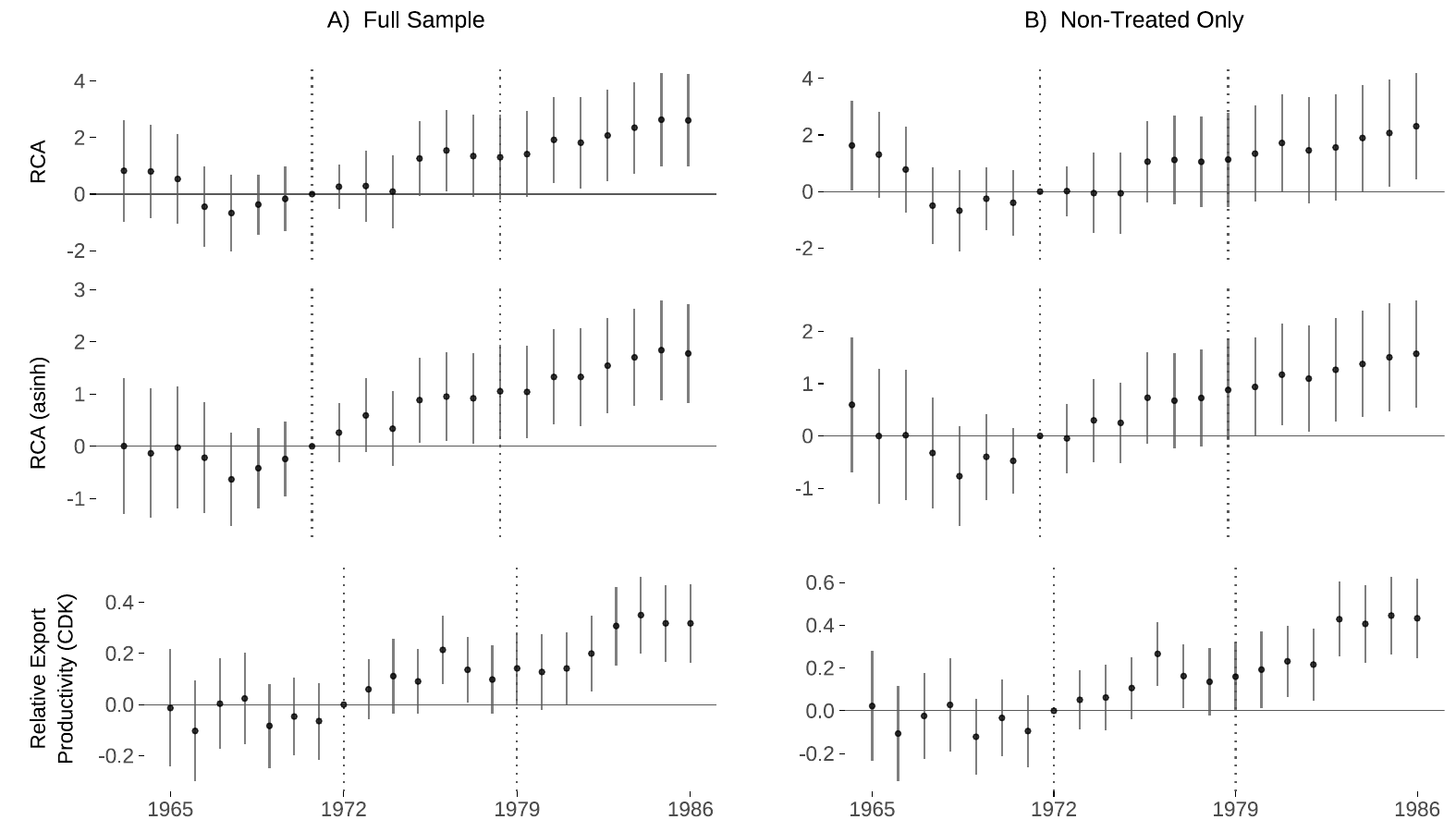}
\end{center}

\caption{Forward Linkages Exposure: Export Development}\label{fig:rcalinkage}

\scriptsize
{
\setlength{\parindent}{2em}
    \input{forwardlinkagetrade_note.tex}
}
\end{figure}

\FloatBarrier

\pagebreak

\clearpage

\setcounter{table}{0}

\begin{table}
\centering
\caption{\label{tab:tfpcrosssection}Differences in Plant-Level Total Factor Productivity, By Treatment Status(1980-1986)}
\centering
\fontsize{10}{12}\selectfont
\begin{threeparttable}
\begin{tabular}[t]{lccccc}
\toprule
\multicolumn{1}{c}{\bgroup\fontsize{11}{13}\selectfont \egroup{}} & \multicolumn{5}{c}{\bgroup\fontsize{11}{13}\selectfont Outcomes: Total Factor Productivity (TFP)\egroup{}} \\
\cmidrule(l{3pt}r{3pt}){2-6}
 & (1) & (2) & (3) & (4) & (5)\\
\midrule
Targeted & 0.043*** & 0.010 & 0.054*** & 0.042*** & 0.003\\
 & (0.015) & (0.014) & (0.015) & (0.014) & (0.013)\\
\addlinespace[0.1em]
\multicolumn{6}{l}{\textbf{ }}\\
\hspace{1em}Industry \(\times\) Year & Yes & Yes & Yes & Yes & Yes\\
\hspace{1em}\(R^2\) & 0.368 & 0.446 & 0.342 & 0.234 & 0.117\\
\hspace{1em}Observations & 272150 & 272150 & 272150 & 272150 & 272150\\
\hspace{1em}Two-way Cluster (Industry Plant) & 488 x 91094 & 488 x 91094 & 488 x 91094 & 488 x 91094 & 488 x 91094\\
\addlinespace[0.3em]
\hline
\multicolumn{6}{l}{\textbf{ }}\\
\hspace{1em}Estimation Type (TFP) & W & ACF & LP & OP & OLS\\
\bottomrule
\end{tabular}
\begin{tablenotes}[para]
\item \textit{\hspace{1em}\textit{Notes.}} 
\item This table shows the relationship between plant-level TFP  and HCI (targeted industries) for the post-HCI period (1980-1986), using equation  \eqref{eq:tfpplantregression}. TFP is estimated using Ackerberg-Caves-Frazer  (ACF), Levinsohn-Petrin (LP), Olley-Pakes (OP), and Wooldridge (W) methods. I also  include TFP estimated using OLS as a baseline estimate. The table reports estimates  of plant-level TFP regressed on the Targeted indicator and year-by-industry  (4-digit level) effects. Plant-level TFP is estimated using a log-transformed,  value added production function. The Targeted indicator is defined by the plant's  main industry. Regressions use two-way clustered standard errors at the plant and  industry levels. * Significant at the 10 percent level. ** Significant at the 5  percent level. *** Significant at the 1 percent level. \label{tab:tfpcrossection}
\end{tablenotes}
\end{threeparttable}
\end{table}

\FloatBarrier

\begin{table}
\centering
\caption{\label{tab:semitable}Average Impact of Industrial Policy: Industrial Development}
\centering
\fontsize{10}{12}\selectfont
\begin{threeparttable}
\begin{tabular}[t]{lcccc}
\toprule
\multicolumn{1}{c}{} & \multicolumn{2}{c}{A) Five-Digit Panel} & \multicolumn{2}{c}{B) Four-Digit Panel} \\
\cmidrule(l{3pt}r{3pt}){2-3} \cmidrule(l{3pt}r{3pt}){4-5}
\multicolumn{1}{c}{} & \multicolumn{1}{c}{Double Robust} & \multicolumn{1}{c}{TWFE} & \multicolumn{1}{c}{Double Robust} & \multicolumn{1}{c}{TWFE} \\
\cmidrule(l{3pt}r{3pt}){2-2} \cmidrule(l{3pt}r{3pt}){3-3} \cmidrule(l{3pt}r{3pt}){4-4} \cmidrule(l{3pt}r{3pt}){5-5}
\textit{Outcomes (log)} & (1) & (2) & (3) & (4)\\
\midrule
 & 0.8378*** & 0.8235*** & 0.5923*** & 0.5452**\\

\multirow[t]{-2}{*}{\raggedright\arraybackslash Output (Shipm.)} & (0.1734) & (0.1846) & (0.2063) & (0.2223)\\

 & 0.7426*** & 0.7292*** & 0.5063** & 0.4586**\\

\multirow[t]{-2}{*}{\raggedright\arraybackslash Value Added} & (0.1702) & (0.1742) & (0.1966) & (0.209)\\

 & 0.8383*** & 0.8236*** & 0.5962*** & 0.5481**\\

\multirow[t]{-2}{*}{\raggedright\arraybackslash Gross Output} & (0.1843) & (0.1852) & (0.2107) & (0.2217)\\

 & 0.504*** & 0.4972*** & 0.2941 & 0.2679\\

\multirow[t]{-2}{*}{\raggedright\arraybackslash Employment} & (0.1479) & (0.1509) & (0.2012) & (0.1915)\\

 & -0.1002*** & -0.1012*** & -0.1154*** & -0.1152***\\

\multirow[t]{-2}{*}{\raggedright\arraybackslash Prices} & (0.0212) & (0.0205) & (0.0319) & (0.0304)\\

 & 0.1608** & 0.1548** & 0.1602** & 0.1371*\\

\multirow[t]{-2}{*}{\raggedright\arraybackslash Labor Prod.} & (0.067) & (0.068) & (0.0733) & (0.0829)\\

 & 0.0996*** & 0.0993*** & 0.1072* & 0.097\\

\multirow[t]{-2}{*}{\raggedright\arraybackslash Output Share} & (0.026) & (0.0261) & (0.0555) & (0.0599)\\

 & 0.0979*** & 0.0967*** & 0.1254** & 0.116**\\

\multirow[t]{-2}{*}{\raggedright\arraybackslash Labor Share} & (0.0307) & (0.028) & (0.0528) & (0.0495)\\

 & 0.297*** & 0.2908*** & 0.1986 & 0.1831\\

\multirow[t]{-2}{*}{\raggedright\arraybackslash Num. Plants} & (0.0975) & (0.1018) & (0.1458) & (0.1549)\\
\bottomrule
\end{tabular}
\begin{tablenotes}[para]
\item \textit{\hspace{1em}\textit{Notes.}} 
\item This table shows the average treatment effect on the treated  (ATT) for industrial policy. Average DD estimates are shown for double robust  and TWFE estimators. Outcomes are log: output is the real value of gross  output shipped (shipments), alongside other measures of real output: value  added and gross output. Employment is the total number of workers. Prices  are industry output prices. Labor Prod. is real value added per employee.  Output Share is the manufacturing share of industry output. Labor Share is  the manufacturing share of industry employment. Specifications include controls for pre-1973 industry averages  (log): avg. wages, avg. plant size, intermediate outlays, and labor productivity.  Standard errors are clustered at the industry level. Double robust DD estimates come from \eqref{eq:semi}. Double robust estimators use bootstrapped  standard errors (10,000 iterations) and are adjusted to allow for  within-industry correlation. * Significant at the 10 percent level.  ** Significant at the 5 percent level. *** Significant at the 1 percent level.
\end{tablenotes}
\end{threeparttable}
\end{table}

\FloatBarrier

\begin{table}
\centering
\caption{\label{tab:semitablerca}Average Impact of Industrial Policy: Export Development}
\centering
\fontsize{10}{12}\selectfont
\begin{threeparttable}
\begin{tabular}[t]{lccc}
\toprule
\multicolumn{1}{c}{} & \multicolumn{3}{c}{Type of Estimator} \\
\cmidrule(l{3pt}r{3pt}){2-4}
\multicolumn{1}{c}{} & \multicolumn{1}{c}{Double Robust} & \multicolumn{1}{c}{TWFE} & \multicolumn{1}{c}{PPML} \\
\cmidrule(l{3pt}r{3pt}){2-2} \cmidrule(l{3pt}r{3pt}){3-3} \cmidrule(l{3pt}r{3pt}){4-4}
\textit{Outcomes} & (1) & (2) & (3)\\
\midrule
 & 0.4853*** & 0.4701*** & 0.9142***\\

\multirow[t]{-2}{*}{\raggedright\arraybackslash RCA} & (0.1803) & (0.1806) & (0.26)\\

 & 0.1251*** & 0.1192*** & 0.5939***\\

\multirow[t]{-2}{*}{\raggedright\arraybackslash RCA (log)} & (0.0416) & (0.042) & (0.1535)\\

 & 0.1633*** & 0.1557*** & 0.6059***\\

\multirow[t]{-2}{*}{\raggedright\arraybackslash RCA (asinh)} & (0.0533) & (0.0537) & (0.1577)\\

 & 0.0502*** & 0.0498*** & 0.0302\\

\multirow[t]{-2}{*}{\raggedright\arraybackslash RCA (CDK)} & (0.0186) & (0.0176) & (0.0189)\\

 & 0.1057*** & 0.1021*** & 0.6486***\\

\multirow[t]{-2}{*}{\raggedright\arraybackslash Prob. Comparative Adv.} & (0.0311) & (0.0307) & (0.1945)\\

 & 0.071** & 0.0727** & 0.8346***\\

\multirow[t]{-2}{*}{\raggedright\arraybackslash Export Share} & (0.0297) & (0.0293) & (0.2582)\\

 & 0.0481*** & 0.048*** & 0.7658***\\

\multirow[t]{-2}{*}{\raggedright\arraybackslash Export Share (log)} & (0.0146) & (0.0147) & (0.1991)\\

 & 0.0596*** & 0.0599*** & 0.7998***\\

\multirow[t]{-2}{*}{\raggedright\arraybackslash Export Share (asinh)} & (0.0191) & (0.0191) & (0.2166)\\
\bottomrule
\end{tabular}
\begin{tablenotes}[para]
\item \textit{\hspace{1em}\textit{Notes.}} 
\item This table shows the average treatment effect on the treated (ATT)  for industrial policy. Average DD estimates are shown for double robust, PPML   TWFE, and linear TWFE estimators. RCA is the standard Balassa index measure of   revealed comparative advantage. RCA (CDK) is relative productivity estimated   using CDK. See text for their calculation. The indicator I[RCA>1] is a binary   dummy variable equal to 1 when an industry has achieved comparative advantage,   0 otherwise. I also show transformed versions of RCA (asinh and log). Specifications include controls for pre-1973 industry averages  (log): avg. wages, avg. plant size, intermediate outlays, and labor productivity.  Standard errors are clustered at the industry level. Double robust DD estimates come from \eqref{eq:semi}. Double robust estimators use bootstrapped  standard errors (10,000 iterations) and are adjusted to allow for  within-industry correlation. * Significant at the 10 percent level.  ** Significant at the 5 percent level. *** Significant at the 1 percent level.
\end{tablenotes}
\end{threeparttable}
\end{table}

\FloatBarrier

\begin{table}
\centering
\caption{\label{tab:semitableinvest}Average Impact of Industrial Policy: Input Use and Investment}
\centering
\fontsize{10}{12}\selectfont
\begin{threeparttable}
\begin{tabular}[t]{>{\raggedright\arraybackslash}p{17.5em}cccc}
\toprule
\multicolumn{1}{c}{} & \multicolumn{2}{c}{A) Five-Digit Panel} & \multicolumn{2}{c}{B) Four-Digit Panel} \\
\cmidrule(l{3pt}r{3pt}){2-3} \cmidrule(l{3pt}r{3pt}){4-5}
\multicolumn{1}{c}{} & \multicolumn{1}{c}{Double Robust} & \multicolumn{1}{c}{TWFE} & \multicolumn{1}{c}{Double Robust} & \multicolumn{1}{c}{TWFE} \\
\cmidrule(l{3pt}r{3pt}){2-2} \cmidrule(l{3pt}r{3pt}){3-3} \cmidrule(l{3pt}r{3pt}){4-4} \cmidrule(l{3pt}r{3pt}){5-5}
\textit{Outcomes (log)} & (1) & (2) & (3) & (4)\\
\midrule
 & 0.7544*** & 0.7408*** & 0.5606** & 0.5147**\\

\multirow[t]{-2}{17.5em}{\raggedright\arraybackslash Intermediate Outlays} & (0.1847) & (0.1894) & (0.2319) & (0.2428)\\

 & 0.213*** & 0.2074*** & 0.2823*** & 0.2659***\\

\multirow[t]{-2}{17.5em}{\raggedright\arraybackslash Intermediate Outlays (Per Worker)} & (0.0715) & (0.0719) & (0.0932) & (0.0975)\\

 & 0.6198*** & 0.6089*** & 0.2445 & 0.21\\

\multirow[t]{-2}{17.5em}{\raggedright\arraybackslash Investment} & (0.2237) & (0.2205) & (0.2322) & (0.2186)\\

 & 0.134** & 0.1316** & 0.1124 & 0.1048\\

\multirow[t]{-2}{17.5em}{\raggedright\arraybackslash Investment (Per Worker)} & (0.0622) & (0.0612) & (0.0938) & (0.0901)\\
\bottomrule
\end{tabular}
\begin{tablenotes}[para]
\item \textit{\hspace{1em}\textit{Notes.}} 
\item This table shows the average treatment effect on the treated (ATT)  for industrial policy. Average DD estimates are shown for double robust and   TWFE estimators. Intermediate outlays (log) is real intermediate input costs.   Investment Total (log) is real total gross capital formation. I also show   outcomes in per worker terms. Specifications include controls for pre-1973 industry averages  (log): avg. wages, avg. plant size, intermediate outlays, and labor productivity.  Standard errors are clustered at the industry level. Double robust DD estimates come from \eqref{eq:semi}. Double robust estimators use bootstrapped  standard errors (10,000 iterations) and are adjusted to allow for  within-industry correlation. * Significant at the 10 percent level.  ** Significant at the 5 percent level. *** Significant at the 1 percent level.
\end{tablenotes}
\end{threeparttable}
\end{table}

\FloatBarrier

\begin{table}
\centering
\caption{\label{tab:lbdindustry}Industry-Level Learning by Treatment Status}
\centering
\resizebox{\ifdim\width>\linewidth\linewidth\else\width\fi}{!}{
\fontsize{7}{9}\selectfont
\begin{threeparttable}
\begin{tabular}[t]{lcccccccccc}
\toprule
\multicolumn{1}{c}{\bgroup\fontsize{8}{10}\selectfont  \egroup{}} & \multicolumn{4}{c}{\bgroup\fontsize{8}{10}\selectfont  \egroup{}} & \multicolumn{6}{c}{\bgroup\fontsize{8}{10}\selectfont Total Factor Productivity\egroup{}} \\
\cmidrule(l{3pt}r{3pt}){6-11}
\multicolumn{1}{c}{\bgroup\fontsize{8}{10}\selectfont  \egroup{}} & \multicolumn{2}{c}{\bgroup\fontsize{8}{10}\selectfont Prices (log)\egroup{}} & \multicolumn{2}{c}{\bgroup\fontsize{8}{10}\selectfont Unit cost (log)\egroup{}} & \multicolumn{2}{c}{\bgroup\fontsize{8}{10}\selectfont (ACF)\egroup{}} & \multicolumn{2}{c}{\bgroup\fontsize{8}{10}\selectfont (LP)\egroup{}} & \multicolumn{2}{c}{\bgroup\fontsize{8}{10}\selectfont (W)\egroup{}} \\
\cmidrule(l{3pt}r{3pt}){2-3} \cmidrule(l{3pt}r{3pt}){4-5} \cmidrule(l{3pt}r{3pt}){6-7} \cmidrule(l{3pt}r{3pt}){8-9} \cmidrule(l{3pt}r{3pt}){10-11}
 & (1) & (2) & (3) & (4) & (5) & (6) & (7) & (8) & (9) & (10)\\
\midrule
 &  &  &  &  &  &  &  &  &  & \\
Experience & -0.011 & -0.195*** & 0.015*** & -0.109*** & 0.182*** & 0.369*** & 0.053 & 0.360*** & 0.060 & 0.348***\\
 & (0.008) & (0.029) & (0.004) & (0.014) & (0.065) & (0.059) & (0.083) & (0.060) & (0.081) & (0.064)\\
Targeted \(\times\) Experience & -0.042*** & -0.044*** & 0.005 & -0.035*** & 0.014 & 0.033 & 0.087** & 0.120*** & 0.094*** & 0.125***\\
 & (0.008) & (0.012) & (0.007) & (0.009) & (0.031) & (0.022) & (0.039) & (0.025) & (0.036) & (0.024)\\
\addlinespace[0.5em]
\multicolumn{11}{>{\centering\arraybackslash}p{\linewidth}}{\textbf{ }}\\
\hspace{1em}Controls for Size-Scale & Yes & Yes & Yes & Yes & Yes & Yes & Yes & Yes & Yes & Yes\\
\hspace{1em}Controls for Capital Intensity & No & Yes & No & Yes & No & Yes & No & Yes & No & Yes\\
\hspace{1em}Controls for Intermediates & No & Yes & No & Yes & No & Yes & No & Yes & No & Yes\\
\hspace{1em}Controls for Investment & No & Yes & No & Yes & No & Yes & No & Yes & No & Yes\\
\addlinespace[0.5em]
\multicolumn{11}{l}{\textbf{ }}\\
\hspace{1em}Industry Effects & Yes & Yes & Yes & Yes & Yes & Yes & Yes & Yes & Yes & Yes\\
\hspace{1em}Year Effects & Yes & Yes & Yes & Yes & Yes & Yes & Yes & Yes & Yes & Yes\\
\hspace{1em}R2 & 0.951 & 0.961 & 0.845 & 0.900 & 0.823 & 0.877 & 0.976 & 0.985 & 0.983 & 0.990\\
\hspace{1em}Observations & 3890 & 3429 & 3890 & 3429 & 3512 & 3428 & 3512 & 3428 & 3512 & 3428\\
\hspace{1em}Clusters & 278 & 263 & 278 & 263 & 264 & 263 & 264 & 263 & 264 & 263\\
\addlinespace[0.5em]
\hline
\multicolumn{11}{p{\linewidth}}{\textbf{ }}\\
\hspace{1em}Linear Combination & -0.053 & -0.239 & 0.020 & -0.143 & 0.196 & 0.402 & 0.140 & 0.480 & 0.154 & 0.474\\
\hspace{1em}(St.Err.) & (0.009) & (0.029) & (0.007) & (0.016) & (0.060) & (0.054) & (0.073) & (0.059) & (0.072) & (0.062)\\
\bottomrule
\end{tabular}
\begin{tablenotes}[para]
\item \textit{\hspace{1em}\textit{Notes.}} 
\item This table shows the industry-level relationship between  industrial outcomes and (log) Experience in targeted vs. non-targeted industries.  Estimates come from equation \eqref{eq:lbdindustry}. The analysis is for the  post-1972 period, using the 5-digit industry panel. The outcomes are log Unit Cost  (total intermediate costs per unit of real gross output) and TFP, estimated using  Ackerberg-Caves-Frazer (ACF), Levinsohn-Petrin (LP), and Wooldridge (W) methods.  (log) Experience is measured as cumulative output (the sum of real gross output  until the current year). All equations control for size/scale, measured as (log)  industry employment and (log) average plant size. Additional controls include  log: capital intensity, investment per worker, and intermediate input intensity  per worker. Linear Combination, at the bottom, gives the combined effects. All  specifications are estimated using industry and year fixed effects. Standard errors  are clustered at the industry level. * Significant at the 10 percent level.  ** Significant at the 5 percent level. *** Significant at the 1 percent level.
\end{tablenotes}
\end{threeparttable}}
\end{table}

\FloatBarrier

\begin{table}
\centering
\caption{\label{tab:lbdmicro}Plant and Industry-Level Learning by Treatment Status}
\centering
\resizebox{\ifdim\width>\linewidth\linewidth\else\width\fi}{!}{
\fontsize{9}{11}\selectfont
\begin{threeparttable}
\begin{tabular}[t]{lcccccc}
\toprule
\multicolumn{1}{c}{\bgroup\fontsize{10}{12}\selectfont  \egroup{}} & \multicolumn{3}{c}{\bgroup\fontsize{10}{12}\selectfont Unit cost (log)\egroup{}} & \multicolumn{3}{c}{\bgroup\fontsize{10}{12}\selectfont TFP\egroup{}} \\
\cmidrule(l{3pt}r{3pt}){2-4} \cmidrule(l{3pt}r{3pt}){5-7}
 & (1) & (2) & (3) & (4) & (5) & (6)\\
\midrule
Plant Experience & -0.072*** & -0.072*** & -0.069*** & 0.479*** & 0.485*** & 0.484***\\
 & (0.002) & (0.002) & (0.002) & (0.011) & (0.011) & (0.011)\\
Targeted × Plant Experience & -0.009*** & -0.008*** & -0.005*** & 0.032*** & 0.014* & 0.018**\\
 & (0.001) & (0.001) & (0.001) & (0.007) & (0.008) & (0.007)\\
Industry Experience &  & -0.008*** & -0.006** &  & 0.022 & 0.026*\\
 &  & (0.003) & (0.003) &  & (0.014) & (0.015)\\
Targeted × Industry Experience &  & -0.002** & -0.003*** &  & 0.030*** & 0.030***\\
 &  & (0.001) & (0.001) &  & (0.009) & (0.009)\\
\addlinespace[0.5em]
\multicolumn{7}{p{\linewidth}}{\textbf{ }}\\
\hspace{1em}Control for Plant Size & Yes & Yes & Yes & Yes & Yes & Yes\\
\hspace{1em}Control for Capital & Yes & Yes & Yes & Yes & Yes & Yes\\
\hspace{1em}Control for Skill Ratio & Yes & Yes & Yes & Yes & Yes & Yes\\
\hspace{1em}Control for Investment & Yes & Yes & Yes & Yes & Yes & Yes\\
\hspace{1em}Control for Intermediates & Yes & Yes & Yes & Yes & Yes & Yes\\
\hspace{1em}Polynomial Controls & No & No & Yes & No & No & Yes\\
\addlinespace[0.5em]
\multicolumn{7}{p{\linewidth}}{\textbf{ }}\\
\hspace{1em}Plant Effect & Yes & Yes & Yes & Yes & Yes & Yes\\
\hspace{1em}Year Effect & Yes & Yes & Yes & Yes & Yes & Yes\\
\hspace{1em}Industry Effect & Yes & Yes & Yes & Yes & Yes & Yes\\
\hspace{1em}R2 & 0.882 & 0.882 & 0.890 & 0.691 & 0.691 & 0.694\\
\hspace{1em}Observations & 250989 & 250989 & 250989 & 235940 & 235940 & 235940\\
\hspace{1em}Clusters (Industry and Plant) & 489 x 60009 & 489 x 60009 & 489 x 60009 & 489 x 57942 & 489 x 57942 & 489 x 57942\\
\addlinespace[0.5em]
\hline
\multicolumn{7}{p{\linewidth}}{\textbf{ }}\\
\hspace{1em}Linear Combination (Plant-Level) & -0.081 & -0.080 & -0.073 & 0.510 & 0.499 & 0.502\\
\hspace{1em}(St.Err.) & (0.002) & (0.002) & (0.002) & (0.013) & (0.013) & (0.013)\\
\hspace{1em}Linear Combination (Industry-Level) &  & -0.010 & -0.009 &  & 0.052 & 0.056\\
\hspace{1em}(St.Err.) &  & (0.003) & (0.003) &  & (0.014) & (0.015)\\
\bottomrule
\end{tabular}
\begin{tablenotes}[para]
\item \textit{\hspace{1em}\textit{Notes.}} 
\item This table shows the plant-level relationship between industrial development and (log) Experience in targeted vs. non-targeted industries. Estimates come from a plant-level version of equation \eqref{eq:lbdindustry}.  Outcomes are the following: log Unit Cost (total intermediate costs per unit  of real gross output) and TFP (estimated using Ackerberg-Caves-Frazer).  Experience is measured as cumulative output (the sum of real gross output  until the current year). 'Plant Experience' refers to plant-level cumulative  learning, and 'Industry Experience' refers to industry-level learning,  calculated at the 4-digit industry level. All equations control for log  plant size (employment). Additional controls include log: capital intensity,  skill ratio, investment per worker, and intermediate input intensity per  worker. Linear Combination, at the bottom, gives the combined effects. All  specifications are estimated using plant, industry, and year fixed effects.  'Polynomial Controls' adds cubic polynomials in the control variables. Two-way  standard errors are clustered at the industry and plant levels. * Significant  at the 10 percent level. ** Significant at the 5 percent level. *** Significant  at the 1 percent level.
\end{tablenotes}
\end{threeparttable}}
\end{table}

\FloatBarrier

\begin{table}
\centering
\caption{\label{tab:prepostlinkoutput}Linkage Exposure and Value Added, Before and After 1973}
\centering
\resizebox{\ifdim\width>\linewidth\linewidth\else\width\fi}{!}{
\fontsize{9}{11}\selectfont
\begin{threeparttable}
\begin{tabular}[t]{lcccc}
\toprule
\multicolumn{1}{c}{\bgroup\fontsize{10}{12}\selectfont  \egroup{}} & \multicolumn{4}{c}{\bgroup\fontsize{10}{12}\selectfont Outcome: Value Added (log)\egroup{}} \\
\cmidrule(l{3pt}r{3pt}){2-5}
\multicolumn{1}{c}{\bgroup\fontsize{10}{12}\selectfont  \egroup{}} & \multicolumn{2}{c}{\bgroup\fontsize{10}{12}\selectfont A) Five-Digit Panel (1970-1986)\egroup{}} & \multicolumn{2}{c}{\bgroup\fontsize{10}{12}\selectfont B) Four-Digit Panel (1967-1986)\egroup{}} \\
\cmidrule(l{1pt}r{1pt}){2-3} \cmidrule(l{1pt}r{1pt}){4-5}
\multicolumn{1}{c}{\bgroup\fontsize{9}{11}\selectfont \em{}\egroup{}} & \multicolumn{1}{c}{\bgroup\fontsize{9}{11}\selectfont \em{\makecell[c]{Full\\Sample}}\egroup{}} & \multicolumn{1}{c}{\bgroup\fontsize{9}{11}\selectfont \em{\makecell[c]{Non-HCI\\Sample}}\egroup{}} & \multicolumn{1}{c}{\bgroup\fontsize{9}{11}\selectfont \em{\makecell[c]{Full\\Sample}}\egroup{}} & \multicolumn{1}{c}{\bgroup\fontsize{9}{11}\selectfont \em{\makecell[c]{Non-HCI\\Sample}}\egroup{}} \\
\cmidrule(l{3pt}r{3pt}){2-2} \cmidrule(l{3pt}r{3pt}){3-3} \cmidrule(l{3pt}r{3pt}){4-4} \cmidrule(l{3pt}r{3pt}){5-5}
 & (1) & (2) & (3) & (4)\\
\midrule
Post \(\times\) Forward Linkage & 2.832*** & 4.405*** & 2.095** & 2.906**\\
 & (0.914) & (1.504) & (0.802) & (1.174)\\
Post \(\times\) Backward Linkage & -0.0167 & 0.176 & -0.693 & -2.163*\\
 & (0.334) & (0.375) & (0.559) & (1.279)\\
\addlinespace[0.25em]
\multicolumn{5}{l}{\textbf{ }}\\
\hspace{1em}Industry Effects & Yes & Yes & Yes & Yes\\
\hspace{1em}Year Effects & Yes & Yes & Yes & Yes\\
\hspace{1em}Targeted \(\times\) Year & Yes & No & Yes & No\\
\hspace{1em}\(R^2\) & 0.776 & 0.763 & 0.847 & 0.819\\
\hspace{1em}Observations & 4720 & 2986 & 1750 & 1096\\
\hspace{1em}Clusters & 278 & 176 & 88 & 55\\
\bottomrule
\end{tabular}
\begin{tablenotes}[para]
\item \textit{\hspace{1em}\textit{Notes.}} 
\item Average differences-in-differences estimates, before and  after 1973. Estimates correspond to equation \eqref{eq:networkflexible}.  Regressions interact linkage measures with a Post indicator. The outcome is real  log value added. Both linkage interactions (forward and backward) are shown.  Analysis is performed for the sample of i) only non-treated industries and ii)  the full sample of industries. Estimates for the full sample separately control  for the Targeted \(\times\) Year effects to account for the main impact  of policy. Standard errors are clustered at the industry level. * Significant at  the 10 percent level. ** Significant at the 5 percent level. *** Significant at  the 1 percent level.
\end{tablenotes}
\end{threeparttable}}
\end{table}

\begin{table}
\centering
\caption{\label{tab:prepostlinkprices}Linkage Exposure and Output Prices, Before and After 1973}
\centering
\resizebox{\ifdim\width>\linewidth\linewidth\else\width\fi}{!}{
\fontsize{9}{11}\selectfont
\begin{threeparttable}
\begin{tabular}[t]{lcccc}
\toprule
\multicolumn{1}{c}{\bgroup\fontsize{10}{12}\selectfont  \egroup{}} & \multicolumn{4}{c}{\bgroup\fontsize{10}{12}\selectfont Outcome: Output Prices (log)\egroup{}} \\
\cmidrule(l{3pt}r{3pt}){2-5}
\multicolumn{1}{c}{\bgroup\fontsize{10}{12}\selectfont  \egroup{}} & \multicolumn{2}{c}{\bgroup\fontsize{10}{12}\selectfont A) Five-Digit Panel (1970-1986)\egroup{}} & \multicolumn{2}{c}{\bgroup\fontsize{10}{12}\selectfont B) Four-Digit Panel (1967-1986)\egroup{}} \\
\cmidrule(l{1pt}r{1pt}){2-3} \cmidrule(l{1pt}r{1pt}){4-5}
\multicolumn{1}{c}{\bgroup\fontsize{9}{11}\selectfont \em{}\egroup{}} & \multicolumn{1}{c}{\bgroup\fontsize{9}{11}\selectfont \em{\makecell[c]{Full\\Sample}}\egroup{}} & \multicolumn{1}{c}{\bgroup\fontsize{9}{11}\selectfont \em{\makecell[c]{Non-HCI\\Sample}}\egroup{}} & \multicolumn{1}{c}{\bgroup\fontsize{9}{11}\selectfont \em{\makecell[c]{Full\\Sample}}\egroup{}} & \multicolumn{1}{c}{\bgroup\fontsize{9}{11}\selectfont \em{\makecell[c]{Non-HCI\\Sample}}\egroup{}} \\
\cmidrule(l{3pt}r{3pt}){2-2} \cmidrule(l{3pt}r{3pt}){3-3} \cmidrule(l{3pt}r{3pt}){4-4} \cmidrule(l{3pt}r{3pt}){5-5}
 & (1) & (2) & (3) & (4)\\
\midrule
Post \(\times\) Forward Linkage & -0.359*** & -0.459*** & -0.483** & -0.510***\\
 & (0.128) & (0.144) & (0.184) & (0.176)\\
Post \(\times\) Backward Linkage & 0.103*** & 0.0880*** & 0.251 & 0.673***\\
 & (0.0213) & (0.0142) & (0.154) & (0.226)\\
\addlinespace[0.25em]
\multicolumn{5}{l}{\textbf{ }}\\
\hspace{1em}Industry Effects & Yes & Yes & Yes & Yes\\
\hspace{1em}Year Effects & Yes & Yes & Yes & Yes\\
\hspace{1em}Targeted \(\times\) Year & Yes & No & Yes & No\\
\hspace{1em}\(R^2\) & 0.957 & 0.942 & 0.962 & 0.956\\
\hspace{1em}Observations & 4721 & 2987 & 1751 & 1097\\
\hspace{1em}Clusters & 278 & 176 & 88 & 55\\
\bottomrule
\end{tabular}
\begin{tablenotes}[para]
\item \textit{\hspace{1em}\textit{Notes.}} 
\item Average differences-in-differences estimates, before and  after 1973. Regressions interact linkage measures with a Post indicator.  Estimates correspond to equation \eqref{eq:networkflexible}. The outcome  variable is log output price. Both linkage interactions (forward and backward)  are shown. Analysis is performed for the sample of i) only non-treated  industries and ii) the full sample of industries. Estimates for the full sample  separately control for the Targeted \(\times\) Year effects to account  for the main impact of policy. Standard errors are clustered at the industry  level. * Significant at the 10 percent level. ** Significant at the 5 percent  level. *** Significant at the 1 percent level.
\end{tablenotes}
\end{threeparttable}}
\end{table}

\clearpage
\appendix
\setcounter{section}{0}
\renewcommand{\thesection}{\Alph{section}}
\part*{Online Appendix}

\setcounter{secnumdepth}{4}

\setcounter{section}{0}
\renewcommand{\thesection}{\Alph{section}}
\titleformat{\section}
  {\centering\normalfont\large\scshape}
  {\thesection.\linebreak}
  {1em}
  {} 

\titlespacing{\subsection}
  {0pt} 
  {3.5ex} 
  {1ex} 
  [1em] 

\renewcommand{\thesubsection}{\thesection.\arabic{subsection}}
\titleformat{\subsection}
  {\normalfont\itshape}
  {\textit{\thesubsection}}
  {1em}
  {}
\titlespacing{\subsection}
  {0pt} 
  {2ex plus 1ex minus .2ex} 
  {1ex plus .25ex minus 0ex} 
  [1em] 

\renewcommand{\thesubsubsection}{\arabic{subsubsection}}
\titleformat{\subsubsection}[runin]
  {\normalfont\itshape}
  {\thesubsubsection}
  {1em}
  {\titlecap}
  [.~~] 
\titlespacing{\subsubsection}
  {2em} 
  {3ex plus 1ex minus .2ex} 
  {0pt} 

\renewcommand{\theparagraph}{\roman{paragraph}}
\titleformat{\paragraph}[runin]
  {\normalfont\itshape}
  {\theparagraph}
  {1em}
  {}
  [.~~] 

\titlespacing{\paragraph}
  {2em} 
  {0pt} 
  {0pt} 

\section{History Appendix}\label{sec:historyappendix}

\subsection{Troop Withdrawal Threat and the Nixon Shock
}\label{sec:troophistorysection}

In 1969, President Richard Nixon declared that the United States would no longer provide direct military support to its allies in the Asia-Pacific region, creating the risk of a complete American troop withdrawal from the Korean Peninsula \citep{Nixon2010, Kim1970, Kwak2003}. Panel B of Appendix Figure \ref{fig:appendixtrooprobust} shows American press coverage of the troop withdrawal, measured by the share of \textit{New York Times} articles containing "South Korea" and "troop withdrawal." The first peak appeared around 1970 when the United States confirmed its withdrawal from the Peninsula. Coverage increased during the 1971 pullout of 24,000 troops and three air force battalions.

This confirmation "shocked" the South Korean leadership, who had expected exemptions from Nixon's doctrine \citep[p.34]{Kwak2003,Rogers2010,Trager1972}. The second jump coincided with the 1976 U.S. presidential contest and Jimmy Carter's election, which further committed to an American pullout \citep{Han1978, Taylor1990}. This goal was later complicated by the fall of the Park regime during President Carter's administration. See Online Supplemental History Appendix \ref{sec:supphistoryappendix} for further details.

The United States' pivot coincided with growing antagonism from North Korea. Panel A of Appendix Figure \ref{fig:appendixtrooprobust} illustrates North Korea's increasing hostility during the U.S. policy shift, using the full-text archives of two major Korean newspapers, \emph{Dong-A Ilbo} and \emph{Kyunghyang Shinmun}. The Online Supplemental Data Appendix \ref{sec:suppappendixattacks} describes the data construction. The data shows the number of articles covering military antagonism, counted using a dictionary of Korean-language keywords related to military hostility. Panel A traces a series of high-profile security emergencies that tipped the Park regime into crisis \citep{Scobell2007,Kim2001}. Additionally, Online Supplemental Data Appendix \ref{sec:suppappendixattacks} demonstrates that these patterns are robust to alternative data sources.

\subsection{Commercial Banks and Heavy-Chemical Drive Lending}
\label{sec:appendixcommercialbanks}

\begin{figure}[h]

{
\centering 
\includegraphics[width=.8\textwidth]{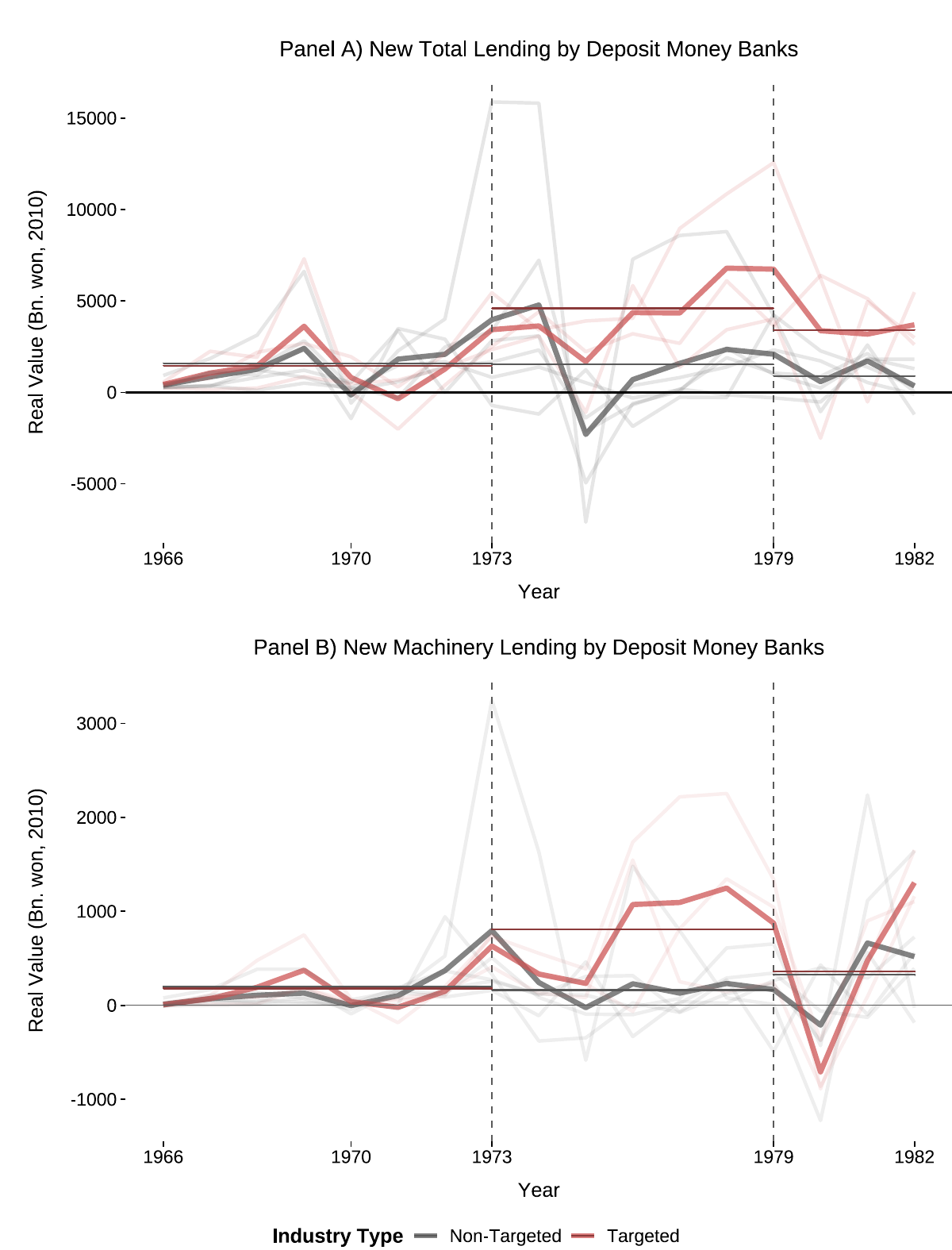} 
}

\caption{New Loans Issued By Commercial Deposit Money Banks}\label{fig:appendixloans}

\scriptsize
{
\setlength{\parindent}{2em}
    \input{appendixpolicyplot_note.tex}
}
\end{figure}

Appendix Figure \ref{fig:appendixloans} illustrates commercial bank loans during the heavy-chemical drive period. Although technically private, the commercial banking sector was deeply intertwined with the state throughout the Park era. Commercial deposit banks played a significant role in this period, distributing 60\% of policy loans during the 1970s \citep{Cho1995, TheWorldBank1993}.

Panel A of Appendix Figure \ref{fig:appendixloans} reveals that before the heavy industry drive, the value of new loans from commercial deposit banks was similar across sectors. However, it rose sharply for targeted sectors after 1973. After 1979, new total heavy industry lending declined. In contrast to the Korean Development Bank (in the main paper), total private lending continued. These post-1979 policy loans were qualitatively different; liberalization removed preferential rates and equalized borrowing costs across industries \citep[pp.443-444]{Lee1991,Woo1991}. For more information about the liberalization of the banking sector, refer to Online Supplemental Appendix \ref{sec:supplementalliberalization}.

\section{Direct Impact Appendix}\label{sec:sectionb}

\subsection{Labor Productivity and Prices}\label{sec:appendixprodandprices}

An initial interpretation of the event study estimates in the main paper might suggest that prices declined for targeted versus non-targeted industries and that pre-1973 pre-trends indicate a literal downward trend in prices for targeted industries. However, the top row of Appendix Figure \ref{fig:appendixproductivity} Panel A (five-digit panel) reveals that the trends between the two industries are similar throughout the mid-1970s and diverge over the policy period.

Labor productivity rises through the HCI period, which is notable in the five-digit data and less precisely estimated in the four-digit data. The top row of Appendix Figure \ref{fig:appendixproductivity} also demonstrates that the effects stem from increased labor productivity for treated industries rather than a decline in non-treated industries.

Appendix Figure \ref{fig:appendixproductivity} (Panel B) shows that average prices increased during the inflationary 1970s. However, HCI prices diverged from the control industry averages and did not increase as sharply over this inflationary period. These price effects, shown in Appendix Figure \ref{fig:appendixproductivity}, contrast with industrial policy experiences elsewhere, where inefficient industrial policy has typically increased the prices for targeted outputs.

A positive relationship between prices and industrial policy may be the norm rather than the exception. For example, heavy industrial policy in Egypt, India, and Turkey may have effectively increased the relative price of capital and intermediate goods \citep{SchmitzJr2001}. For a case study on steel, see \citep{Blonigen2015}, which shows how heavy industrial policies can raise output prices to the detriment of downstream exporters.

\subsection{Robustness: Direct Impacts}\label{sec:appendixdirectrobustness}

\subsubsection{Robustness: Industry-Level TFP}\label{sec:appendixtfp}
This section explores the relationship between the heavy-chemical industrial policy package and estimated total factor productivity (TFP) using the more granular (five-digit) industry-level panel (1970–1986).\footnote{The short, five-digit data contains capital stock data and is subject to less harmonization/aggregation. See the Data Section in the main paper. For aggregation and harmonization of the four-digit data, refer to Online Supplemental Data Appendix \ref{sec:suppappendixdataappendix}.} Although I estimate industry-level TFP over the study period, I emphasize caution. Modern best practices for estimating TFP focus on micro-econometric estimation strategies and corrections modeled by micro-level behavior \citep{VanBeveren2012}. For this reason and more, the following industry-level estimates may have limitations.

Practically, aggregate data can limit the power to estimate production function parameters and may exacerbate measurement issues that confound TFP estimation (e.g., \citep{Diewert2000}). Market imperfections may further complicate TFP estimation, especially in distorted miracle economies \citep{Felipe1999, Fernald2011}. Aggregate data precludes some micro-level corrections. Nevertheless, I estimate industry-level TFP using standard micro-econometric estimators.

I estimate (log-linearized value added) production function parameters at approximately the two-digit level. To improve power, I combine sectors with sparse observations to properly estimate production function parameters when additional power is required.\footnote{For example, some mining and minerals processing sectors contain limited five-digit industries, so a broader two-digit category is used.} Following the empirical TFP literature, I Winsorize estimates for extreme values.

Figure \ref{fig:appendixindustrytfp} estimates HCI's impact on industry-level TFP using five common measures. To be conservative, I use 1970 as the baseline for regression estimates of total factor productivity. Figure \ref{fig:appendixindustrytfp} demonstrates that 1972 was a particularly low year for heavy-chemical industry TFP. Hence, using 1972 as the baseline can overstate post-1972 TFP differences.

Figure \ref{fig:appendixindustrytfp} reveals a slow upward trend in TFP for targeted industries relative to non-targeted industries over the study period. Although estimates are noisy and vary across TFP outcomes, they show a slight increase in TFP for heavy-chemical industries. For the (limited) pre-1973 period, TFP in the targeted industries seemed stagnant, perhaps even declining. After 1973, this trend reversed, and estimates gained momentum through the later 1970s. TFP measures became significant post-1979 across the board.

Earlier studies stress that treated industries experienced low productivity \citep{Dollar1990}, yet early work did not consider the \textit{relative} trends in TFP before and after the intervention.\footnote{The issues of cross-sectional variation versus panel variation appear in \citet{Harrison1994,Lucas1984}.} Limited relative growth (in TFP) over the period matched an earlier analysis \citep{Felipe1999}. Moreover, a subtle upward trajectory seems compatible with a story of industrial learning taking time.

\subsubsection{Robustness: Dynamic Double Robust Results}\label{sec:appendixdreventstudy}

For robustness, I demonstrate that the patterns observed in \textit{dynamic} (or event study) two-way fixed effects (TWFE) difference-in-differences (DD) estimates are robust to using the doubly robust estimator of \citet{SantAnna2020a,Callaway2020}. I use the same log outcomes and controls as in the standard TWFE estimates for equation (1). The adjustments performed by the doubly robust estimator rely on pre-treatment controls, so only specifications with controls are used.\footnote{Note that without controls, the estimator package defaults to a standard TWFE method.} I provide bootstrap confidence intervals (95\%), which allows for clustering at the industry level.

Appendix Figures \ref{fig:semiplot}--\ref{fig:semitrade} present estimates from the doubly robust estimator. Figure \ref{fig:semitrade} reports estimates for export development outcomes aggregated to the four-digit industry level. The patterns in Appendix Figures \ref{fig:semiplot}--\ref{fig:semitrade} are qualitatively similar to the linear TWFE estimates.

Consider first the relationship between the industrial policy drive and industrial development given by Appendix Figures \ref{fig:semiplot} (four-digit panels) and \ref{fig:semiplot2} (five-digit panels). Although the doubly robust DD relaxes some assumptions related to the traditional TWFE DD, the general dynamic pattern is robust. This finding is particularly important because this estimator re-weights the treatment and control groups. In other words, the same dynamics shown in the OLS TWFE estimates are present in the semi-parametric DD estimates in the main paper. See the main paper for comparisons between the average estimates.

\section{Direct Impact on Trade Appendix}\label{sec:appendixprobrca}

In Table \ref{tab:appendixprobrca}, I examine the probability of achieving a comparative advantage in heavy-chemical goods using cross-country data. To do so, I compare Korea to control countries. I restrict the data to the post-1972 period and focus on HCI products only, using the following regression:

\begin{equation}\label{eq:appendixprodrca}
  Y_{ict}=\alpha_{kt}+\beta_1 \text{Korea}_{i}+ \beta_2 \ln \left( \text{Income}_{i1972}\right)+\epsilon_{ict}.
\end{equation}

For completeness, I present both PPML and linear probability estimates. The linear probability estimates in columns (1)–(4) provide a more straightforward interpretation.

For 1972–1986, the average country had a comparative advantage in \resultsmeanprobrca percent of HCI products, as shown by the mean in column (1). Estimates in Appendix Table \ref{tab:appendixprobrca} show that, across samples and estimates, Korea had a significantly higher probability of achieving comparative advantage in heavy-chemical industry goods. The effect of the Korea indicator is highly significant across specifications, including after controlling for 1972 income per capita, PPP adjusted in 2010 dollars, in columns (2) and (6).

Additional estimates in Appendix Table \ref{tab:appendixprobrca} demonstrate that Korea had a significantly higher probability of achieving comparative advantage when we limit estimates to specific subsets. This holds for sample countries in the same pre-treatment income decile, as shown in columns (4) and (8). It also applies to countries in similar income deciles, defined as those in the same decile \textit{and} those in the immediate deciles above and below Korea's 1972 income group, as presented in columns (3) and (7).

\section{Policy and Mechanisms Appendix}\label{appendixmechanisms}

\subsection{Investment and Industrial Policy Discussion}\label{sec:appendixcapitaldiscussion}

Is it obvious that we would observe responses to investment or production incentives from industrial policy? Based on the history of industrial policy, the answer is no. If financial policies are redundant, they may not create new investments (or outlays)---investments that would have occurred without policy shifts. In many contexts, \emph{de jure} investment policy may not bind.

Work by \citet{Lazzarini2015} shows that in Brazil, capital from a major national development bank did not translate into increased investment and was allocated to politically connected firms where investments would otherwise have occurred. For East Asia, \citet{Yang1993} argues that investment subsidies in Taiwan did not contribute to capital formation, echoing a common criticism that investment would have occurred without industrial investment schemes. 

\subsection{Policy Mechanisms: The Impact of Directed Credit and Marginal Revenue Product of Capital (MRPK)}\label{sec:appendixmrpkanalysis}

I explore the relationship between high-MRPK and low-MRPK industries and input use. Specifically, I test (i) whether input use increases differentially for industries with a high marginal revenue product of capital and (ii) whether this increase occurs specifically for treated industries.

The MRPK calculation is constrained by industry-level (as opposed to micro) data and is calculated for the most disaggregated five-digit panel. The marginal revenue product of capital for industry \(i\) is \(\textrm{MRPK}_i=\alpha^k_i \times ( \textrm{Revenue}_{i}/K_{i})\). I calculate a version of the measure proposed by \citet{Bau2021}, using total sales (real shipments) divided by total tangible capital stock. I estimate capital coefficients \(\alpha^k_i\) at the two-digit level. Capital shares are calculated using pre-policy drive shares. Industries are then split into high-MRPK or low-MRPK groups based on whether they are above or below the median level of MRPK.

I consider the following regression equation:

\begin{equation}\label{eq:appmrpkregression}
  \ln \left( \textrm{input}_{it} \right) = \alpha_i+\alpha_t+\sum_{j\neq 1972} \beta_j \left(\textrm{High-MRPK}_{i} \times \textrm{Year}^j_t \right)+\epsilon_{it},
\end{equation}

\noindent where the outcome \(\ln\left(\textrm{input}_{it}\right)\) is investment or intermediate input use for industry \(i\) at year \(t\). I estimate equation \eqref{eq:appmrpkregression} separately for targeted and non-targeted industries. The set of coefficients \(\beta_j\) conveys differences in input use between high-MRPK and low-MRPK industries relative to 1972. In other words, the estimates in \eqref{eq:appmrpkregression} reveal whether inputs respond for those sectors most exposed to HCI credit policies (see the History Section of the main text). Specifically, I assess whether this relationship is observed for targeted industries during the drive period.

Appendix Figure \ref{fig:appendixmrpkpolicy} illustrates the relationship between MRPK and the increase in input use. I estimate regressions separately for targeted and non-targeted industries. Panels A–B show estimates for (log) total material outlays and real total investment. Panels A–B demonstrate that inputs increased in high-MRPK industries relative to low-MRPK industries after 1973, but only for targeted industries. Similarly, high-MRPK industries show increases in (log) labor (Panel C) and, consequently, output, which is measured as the log real output shipped (Panel D).

Thus, the estimates in Figure \ref{fig:appendixmrpkpolicy} suggest that policy differentially relaxed constraints for high-MRPK industries, increasing input use. Note that these results do not imply MRPK convergence or reduced misallocation due to the policy. Instead, they provide indirect evidence that credit expansion operated differentially for targeted industries.

The expansion of credit to targeted industries during the policy drive shares similarities with the directed credit literature and the macroeconomics literature on credit booms and instability \citep{Gorton2020,Mendoza2008}. While this literature has emphasized the aggregate correlates of credit booms, the sectors receiving credit may also have significant implications for the impact of credit booms on industrializing economies.

\subsection{Robustness: Testing Investment Crowding Out}\label{sec:capitalcrowdingoutappendix}

To explore crowding out, I compare investment patterns in targeted and non-targeted sectors using a simple regression analysis. Specifically, I first regress (log) investment outcomes on year effects, controlling for five-digit industry fixed effects: 

\begin{equation}\label{eq:appinvestmentestimates}
  \ln\left(\textrm{investment}_{it}\right) =\alpha_i + \sum_{j\neq1972} \beta_j \cdot \text{Year}^j_t + \epsilon_{it}. 
\end{equation}

I report the estimates for equation \eqref{eq:appinvestmentestimates} separately in Panel A, Appendix Figure \ref{fig:appendixcrowdingout}. Panel A shows investment patterns for each sector relative to 1972, revealing no evidence of crowding out during the drive. Instead, it illustrates a relative increase in investment for both manufacturing sectors, with targeted heavy industry experiencing a more substantial increase.

To examine potential crowding out in capital-intensive, non-targeted industries, Panel B of Appendix Figure \ref{fig:appendixcrowdingout} illustrates the impact of pre-treatment capital intensity on investment during the HCI period. It plots coefficients from the interaction \(\textrm{Year}_t \times \ln (\textrm{Capital Intensity})_{i0}\), where capital intensity is measured using pre-treatment capital stock per employee. Like Panel A, estimates in Panel B are presented separately for targeted and non-targeted samples.

Panel B of Appendix Figure \ref{fig:appendixcrowdingout} shows no relative decline in investment for capital-intensive, non-treated sectors during the drive. The relationship between capital intensity and investment is noisy across the treatment period and slightly positive for non-treated capital-intensive industries (i.e., not crowded out). However, the relationship between capital intensity and investment is neutral in targeted heavy-chemical sectors, which typically have higher capital intensity. Hence, treated industry capital is not necessarily higher for (ex-ante) capital-intensive industries. Additionally, recall that Panel B of Figure \ref{fig:appendixmrpkpolicy} demonstrated that investment did not differentially change for non-targeted (high-MRPK versus low-MRPK) industries during the drive.

\subsection{HCI, Trade Policy, and Nominal Protectionism}\label{sec:appendixtradepolicy}

The following analysis considers evidence of overt nominal protectionism of targeted heavy industry. Before considering quantitative evidence, I first turn to the conceptual and historical context for South Korean trade policy over the period.

\subsubsection{Historical Context: HCI as ISI?}

Although the HCI period has been associated with rising protectionism and import substitution-style industrialization (ISI) policies \citep{Kim1990, Yoo1990, Lee1992}, the qualitative pattern of policymaking is more complex. Since the 1960s, South Korea has undergone a "continuous process of tariff reform" under Park Chung-hee \citep[p.52]{Gatt1992}, including multiple rounds of tariff cuts during the HCI period \citep{Gatt1978,Young1988}.\footnote{The 1978 GATT Consultation reports tariff reductions in 1973, 1974, July 1975, December 1976, January 1977, January–November 1977, and April–July 1978 \citep[p.6]{Gatt1978}.} Average import liberalization ratios gradually increased from 1973 to 1979.\footnote{Economic instability in 1979--1980 postponed further import liberalization, planned in 1978, until the post-HCI era \citep{Kim1988}. See Online Supplemental Appendix \ref{sec:supplementalliberalization} for more information.} Exemptions from trade policy were widely used in the 1960s and during the heavy industry drive. Consequently, reported tariffs and quantitative restrictions may represent a theoretical upper bound for an industry's effective protection \citep{Yoo1993}.\footnote{For example, income from customs duties accounted for less than 14 percent of total tax revenue in 1975.}

\subsubsection{Trade Policy Analysis}

Having established the qualitative patterns above, I now study the role of trade policy during the heavy industry drive period quantitatively.

Before conducting a regression analysis, however, it is worth considering the aggregate data presented in Appendix Figure \ref{fig:appendixaggregatepolicy}. Panels C and D of Appendix Figure \ref{fig:appendixaggregatepolicy} show two simple aggregate measures of market protection across targeted and non-targeted industries for five periods: 1968, 1974, 1978, 1980, and 1982.\footnote{Trade policy data is limited to these periods. Refer to the Data Section of the main text.} Panel D reports the average tariff rates (percent), and Panel C presents measures of quantitative restriction (QR) coverage. These panels demonstrate that output protection, measured in terms of tariffs and QR coverage, is lower in targeted sectors compared to non-targeted sectors. Panel D shows that average measures of nominal tariff protection fell continually throughout the period. QRs in Panel C rose slightly in the 1970s but fell by 1982.

Next, consider the distribution of trade policy by sector. Figure \ref{fig:appendixtraderidgeplot} plots the (kernel density) distribution of protection by treatment status for the same period. The histograms in Figure \ref{fig:appendixtraderidgeplot} show a steady convergence in the distribution of nominal (output) protection between targeted and non-targeted sectors from 1968 to 1982. Liberalization will proceed fully after 1982. For details of liberalization, refer to the History Section of the main text and Online Supplemental Appendix \ref{sec:supplementalliberalization}. Additionally, Appendix Figure \ref{fig:appendixtraderidgeplot} shows a mass of low tariff and QR protection for targeted industries.

Next, I turn to regression analysis and consider the following specification,

\begin{equation}\label{eq:tradepanel}
  Y_{it} = \alpha + \beta \cdot \left( \text{Targeted}_i \right) + \tau_t + X'_{i}\Omega + \epsilon_{it},
\end{equation}

\noindent where \(i\) represents industries and \(t\) represents the five periods. Specification \eqref{eq:tradepanel} controls for period effects, \(\tau_t\), and includes baseline controls (log avg. wages, material outlays, avg. plant size, and labor productivity). I estimate this relationship in terms of levels \textit{and} differences, \(Y_{it}\) and \(\Delta Y_{it}\). The coefficient of interest, \(\beta\), provides the difference in the average level—or change—in policy between heavy and non-heavy industries from 1968 to 1982.

Appendix Table \ref{tab:appendixtradepolicy} Panel A first considers differences in output protection between treated and non-treated sectors. Panel A reports that the \textit{level} of output protection is, on average, significantly lower for targeted heavy industries: columns (1--4) show this for log tariffs, and columns (5--8) for QR coverage. Estimates (cols. 3--4) imply that the level of tariffs is significantly lower for targeted industries, even during HCI. Quantitative restrictions are also lower (cols. 7--8). Panel A, columns (9--12) of Table \ref{tab:appendixtradepolicy} report estimated changes in output protection between 1968 and 1982. Estimates are positive, though imprecisely estimated. However, the level of output protection is significantly lower for the targeted industry.

imported inputs (see History Section). Appendix Table \ref{tab:appendixtradepolicy} Panel B shows differential exposure to input protection using industry-level measures of input protection built from input-output tables. These measures account for potential exemptions afforded to targeted industries during the drive. Panel B shows that the targeted industry has significantly lower levels of input protection (cols. 1--8) than the non-targeted industry. Likewise, the targeted industry sees significant \textit{reductions} in input exposure for tariffs (cols. 9--10) and QRs (cols. 11--12).

In sum, the analysis above does not provide strong evidence that the heavy-chemical industry drive means an appreciable rise in conventional means of market protection. The findings comport with general trends in liberalization and South Korea's incorporation into multilateral institutions during the Park era.

\section{Linkage Appendix}\label{sec:appendixinkages}

\subsection{Linkage Measurement}\label{sec:appendixlinkcalculations}

The linkage measures in this study capture exposure to HCI industrial policy through backward and forward linkages. Note that the measures below do not model the causal relations. Rather, they are proxies capturing the extent to which an industry is exposed to policy indirectly through inter-industry linkages. A long literature in input-output economics has considered far more complicated means of measuring and decomposing linkage effects. The following is a simple baseline implementation of backward and forward linkage measures.

\subsubsection{Direct Linkages}

First, consider exposure to industrial policy through backward linkages: this is when the impact of industrial policy propagates to upstream suppliers (through the backward linkages with treated sectors). Let $i$ be a non-targeted industry that sells its output to a treated industry $j$. Industry $i$'s exposure to the industrial policy through backward linkages is equal to

\begin{equation}\label{eq:hcilinkage}
  \text{Backward Linkage}_i = {\displaystyle\sum_{j \in \textrm{HCI}} \alpha_{ij}} \quad \text{with} \quad \alpha_{ij} = \frac{\textrm{x}_{ij}}{\textrm{x}_{j}},
\end{equation}

\noindent where $\alpha_{ij}$ represents the share of $i$'s sales to treated heavy-chemical industries $j$ ($j \in\textrm{HCI}$). Specifically, $\alpha_{ij}$ is the proportion of $i$ used to produce one unit of output $j$, calculated as the value of $i$'s sales to industry $j$, $x_{ij}$, divided by the value of $j$'s total output: $x_j = \sum_{i=1}^n x_{ij}$.\footnote{The denominator $x_j$ includes $j$'s output sold to all sectors, including manufacturing, services, and final output. I follow the literature and do not count $i$'s sales to itself and exclude diagonal elements $\alpha_{ii}$ in the input-output matrix (e.g., $\alpha_{11}=0$).} The coefficients $\alpha_{ij}$ come from technical matrix:\footnote{I calculate the matrix $A$ manually for 1970 from the table of inter-industry flows $X = [x_{ij}]_{n \times n}$. The vector $x$ is a vector of the total output sold by each sector. I compute $A = X\left[{\text{diag}(x)}\right]^{-1}$, and each element is $\alpha_{ij}=x_{ij} / x_j$.
}

\begin{equation}
\vspace{1em}
A = \left[
\begin{array}{cccc}
  \alpha_{11} & \alpha_{12} & \cdots & \alpha_{1n} \\[0.8em]
  \alpha_{21} & \alpha_{22} & \cdots & \alpha_{2n} \\[0.8em]
  \vdots & \vdots & \ddots & \vdots \\[0.8em]
  \alpha_{n1} & \alpha_{n2} & \cdots & \alpha_{nn}
\end{array}
\right].
\label{eq:technical_coefficient_matrix}
\vspace{1em}
\end{equation}

Practically, to calculate the backward linkage measure \eqref{eq:hcilinkage}, I take the row-wise sum of elements from the technical coefficient table \eqref{eq:technical_coefficient_matrix}. This means that for each row $i$, I add the coefficients across columns $j$ that correspond to HCI sectors.

Second, consider exposure to industrial policy through forward linkages. In this case, industrial policy propagates \textit{downstream} to purchasers (through \textit{forward linkages} from HCI sectors). In this case, let $i$ be an untreated sector that purchases inputs from a treated sector $j$. The forward linkage analog of equation \eqref{eq:hcilinkage} is the following:

\begin{equation}\label{eq:hcilinkage2}
  \text{Forward Linkage}_i = {\displaystyle\sum_{j \in \textrm{HCI}} \alpha_{ji}} \quad \text{with} \quad \alpha_{ji} = \frac{\textrm{x}_{ji}}{\textrm{x}_{i}},
\end{equation}

\noindent where $\alpha_{ji}$ denotes the sales from treated industry $j$ to downstream industry $i$ ($x_{ji}$), per unit of $i$'s total output ($x_i$). Practically, to calculate the exposure to policy through forward linkages \eqref{eq:hcilinkage}, I take the column-wise sum of elements from table \eqref{eq:technical_coefficient_matrix}. That is, for each column $i$, I add coefficients across rows $j$ corresponding to HCI sectors. As with backward linkages, I exclude diagonal elements.

\subsubsection{Total Linkages}

In addition, I also calculate the exposure of non-treated industries to HCI policy through total---direct and indirect---links with treated industries. Equations \eqref{eq:hcilinkage}-\eqref{eq:hcilinkage2} above capture the extent to which HCI policy propagates through \emph{direct}, or first-degree, connections. I now consider the total $n$-degree effects; I calculate Total Backward Linkages and Total Forward Linkage Measures using a method analogous to the direct linkages described above. Instead of using coefficients $\alpha_{ij}$ from the coefficient matrix $A$, I use coefficients $\ell_{ij}$ from the Leontief matrix: 

\begin{equation}\label{eq:leontief_inverse_matrix}
\small
L = \begin{bmatrix}
  \ell_{11} & \ell_{12} & \cdots & \ell_{1n} \\[0.8em]
  \ell_{21} & \ell_{22} & \cdots & \ell_{2n} \\[0.8em]
  \vdots & \vdots & \ddots & \vdots \\[0.8em]
  \ell_{n1} & \ell_{n2} & \cdots & \ell_{nn}
\end{bmatrix}.
\vspace{1em}
\end{equation}

The Leontief inverse matrix in equation \eqref{eq:leontief_inverse_matrix} is calculated from the technical coefficient matrix $A$ (eq. \ref{eq:technical_coefficient_matrix}). More precisely, $L = (I - A)^{-1}$, where $I$ is the identity of matrix $A$. The matrix $L$, or the Leontief inverse, captures the full chain of inter-industry relationships between sectors.

I calculate the total exposure between treated HCI industry $j$ and non-treated industry $i$ using elements from table $L$. Formally, the two measures are

\begin{subequations}
\begin{align}
 \text{Total Backward Linkages}_i &= {\displaystyle\sum_{j \in \textrm{HCI}} \ell_{ij}} \label{eq:totalhcilinkage} \\[1em]
 \text{Total Forward Linkages}_i &= {\displaystyle\sum_{j \in \textrm{HCI}} \ell_{ji}} \label{eq:totalhcilinkage2}.
\end{align}
\end{subequations}

Industry $i$'s total exposure to policy through backward linkages is given by equation \eqref{eq:totalhcilinkage}, which equals the sum of coefficients between supplier $i$ to each HCI purchaser $j$. To compute \eqref{eq:totalhcilinkage}, I perform row-wise calculations over matrix $L$: for each row $i$, I sum across columns $j$ that correspond to HCI industries. Similarly, $i$'s total exposure to policy through forward linkages is given by \eqref{eq:totalhcilinkage2}, which equals the sum of coefficients between HCI supplier $j$ and the purchasing industry $i$. To compute \eqref{eq:totalhcilinkage2}, I perform column-wise calculations over elements of matrix $L$: for each column $i$, I sum across rows $j$ that correspond to HCI industries.

\subsection{Forward Linkage Appendix: Developmental Effects }\label{sec:appendixforwardmoredev}

\paragraph{Total Forward Linkages, Output, and Prices.} This section considers the total linkage effects of policy in more detail. Appendix Table \ref{tab:appendixprepostlfoutput} reports pre-post estimates for total forward linkages, those accounting for \textit{n}-degree linkages between downstream industries and HCI suppliers. Like the direct linkages, Appendix Table \ref{tab:appendixprepostlfoutput} reports a robust relationship between total forward linkage exposure and the change in downstream value added. These total effects are strongest in the non-HCI sample. Likewise, Appendix Table \ref{tab:appendixprepostlfprices} shows the average pre-post impact of total linkages on output prices. The estimates for total forward linkages are negative across specifications in Appendix Table \ref{tab:appendixprepostlfprices}.

\paragraph{Forward Linkages and Other Development Outcomes.} I now consider the impact of forward linkages on outcomes besides log output (value added) and log prices. These results are provided in Appendix Figure \ref{fig:appendixmoredevlinks}. Beyond these core outcomes, I observe similar patterns across outcomes, such as entry into and higher employment in downstream sectors with stronger connections. Likewise, I find a weak relationship between forward linkages and productivity outcomes. Appendix Tables \ref{tab:appprepostlinkmoredev} and \ref{tab:appprepostlfmoredev} show pre-post estimates for direct and total linkages, respectively.

\subsection{Forward Linkage Appendix: Mechanisms and Intermediate Input Use}\label{sec:appendixforwardmechanism}

Appendix Figure \ref{fig:appendixmechanismlinkage} examines input use and investment among industries with more versus less exposure to HCI suppliers. Pre-1973, differences in (log) total intermediate outlays and (log) investment were closing for sectors with differential forward links to HCI suppliers. After 1973, the trend reversed; Appendix Figure \ref{fig:appendixmechanismlinkage}, Panel A shows a jump in material outlays (Panel A, top row) and total investment (Panel A, bottom row). The post-1973 divergence is seen in both non-HCI and full samples, as well as across data sets. Likewise, these estimates are strong when limited to non-targeted industries. Joint F-tests reject pre-trends across most specifications, shown in the Online Supplemental Appendix \ref{tab:supptablelinkmechanism}, except four-digit panels, where inputs trended upward and converged before 1973.

Additionally, Panel B of Appendix Figure \ref{fig:appendixmechanismlinkage} shows qualitatively similar effects for the total forward linkage exposure. However, the effects are less precisely estimated for the total linkage effects (see Online Supplemental Table \ref{tab:supptablelinkmechanism} for the full regression table). Thus, during the HCI period, direct downstream users of HCI inputs expanded outlays and inputs during the drive.

\section{Backward Linkage Results}\label{sec:appendixbackwards}

Although estimates for forward linkages correspond with the industrial development of downstream industries, backward linkages do not. Broadly, the effects are weak and quite limited. This is seen in estimates for direct linkages exposure in the main text and Appendix Table \ref{tab:appendixprepostlfoutput} for total linkage exposure. The impact of forward linkage exposure is consistently stronger than noisy backward linkage effects. 

In the case of output, the higher backward linkage exposure---direct or total (Appendix Table \ref{tab:appendixprepostlfoutput})---is negatively related to log value added in an upstream industry. Yet, estimates are mostly imprecise. The indeterminate impact of backward linkages is also seen in Appendix Figures \ref{fig:appendixbacklinks}-\ref{fig:appendixbacklinkstotal}, which show dynamic estimates for output. The figures show the ambiguous, weak relationship between backward linkage exposure and upstream output---both for direct and total backward linkage exposure.

\section{SUTVA and Linkage Appendix}\label{sec:appendixsutvalinkages}

\subsection{Main Effects, Restricting Estimates to Low-Linkage Control Industries}\label{sec:appendixsutvalowlink}

Appendix Figure \ref{fig:appendixsutvalinkexposure} shows the TWFE event study estimates (eq. 1) for output and labor productivity, but with alternative control groups. Specifically, I restrict the control groups to \textit{only} industries with low downstream linkages (triangles) or low upstream linkages (squares). To do so, I split non-targeted industries into those with low and high linkage exposure to HCI sectors. Specifically, I base these categories on whether they are below or above $\text{Forward Linkage}_i$ ($\text{Backward Linkage}_i$). I then re-run baseline DD specifications with these truncated control groups.

For both output and labor productivity, estimates using a "low forward linkage" control group increase slightly, and the baseline pattern is preserved. Intuitively, it would make sense that the main effects of HCI increase after I remove the control industries most likely to benefit from positive policy spillovers (e.g., those with high forward linkage exposure). Standard errors increase, which is not surprising given the truncated sample.

Across outcomes, Figure \ref{fig:appendixsutvalinkexposure} shows that limiting control industries to those with low upstream connections has a minimal impact on point estimates for the main, direct impact of HCI (e.g.~\(\textrm{Targeted}_{i}\times\textrm{Year}_{t}\)). This is expected, as the upstream linkage effects of the policy were more muted than the downstream effects (and slightly negative). In sum, limiting the impact of the strongest first-order linkage effects on the control group is insufficient to overcome the main direct impact of HCI.

\subsection{Main Effects, Controlling for Linkages}\label{sec:appendixsutvacontrollink}
I now test whether the main DD estimates survive after including these effects. I do so by re-running the main regression equation (1), now saturated with linkage controls. That is, I specifically control for linkage exposure for non-treated industries. Linkages are multiplied by an indicator equal to one for non-treated industries and zero for treated industries.

Appendix Figure \ref{fig:appendixsutvapluslinks} Panel A shows baseline results for the main effect, \(\textrm{Targeted}_{i} \times \textrm{Year}_{t}\), versus estimates that include varieties of linkage controls. These results are given for both direct linkages (left) and total (Leontief) linkages (right). The baseline estimates are in red, and those controlling for linkages are in dark gray. I control for linkages using the interaction \(\textrm{Forward Linkage}_{i} \times \textrm{Post}_{t}\), which controls for the linkages more parsimoniously. (Controlling flexibly for linkages, \(\textrm{Forward Backward Linkage}_{i} \times \textrm{Year}_{t}\), significantly increases the number of parameters.)

Once I control for the positive downstream spillovers in non-treated industries, Panel A (Fig. \ref{fig:appendixsutvapluslinks}) shows that the main direct effect \(\textrm{Targeted}_{i} \times \textrm{Year}_{t}\) becomes larger. Furthermore, estimates are more prominent after controlling for the total linkage effects. This is intuitive, as positive spillovers may also benefit the control group and thus bias baseline estimates downward. Recall that I have demonstrated in the main text that there may have been weak negative spillovers into backward-linked industries (direct linkages). This is seen specifically for five-digit panel estimates, which more precisely capture the linkage effects. 

Panel B in Appendix Figure \ref{fig:appendixsutvapluslinks} builds off the regressions in Panel A but now includes controls for \emph{both} backward and forward linkage exposure. Panel B shows that including both linkages maintains the main pattern while increasing the standard errors. The main effect estimates are now less positive than those in Panel A. Including backward linkages means we now control for the negative upstream spillovers. The main pattern is preserved in Appendix Figure \ref{fig:appendixsutvapluslinks} Panels A and B, although slightly increased (along with standard errors), once we control for the most prominent linkage effects.

\subsection{SUTVA: Investment Crowding Out and Linkages}\label{sec:appendixcrowdingoutio}

The crowding out of investment is another way the SUTVA assumption is violated. The policy estimates in the main paper demonstrated that investment, however higher in targeted industries, was not diminishing in non-targeted sectors, nor was this the case in capital-intensive non-HCI sectors. I now consider whether crowding out may occur after controlling for linkage intensity. Appendix Figure \ref{fig:appendixcrowdingoutio} shows the relationship between investment and capital intensity (log, pre-1973 capital stock divided by employment) controlling for linkages. Estimates are shown separately for HCI and non-HCI industry samples. The left panel plots the estimates with controls for linkages using \(\textrm{Forward or Backward Linkage}_{i} \times \textrm{Post}_{t}\). The right panel plots the estimates using the more intensive \(\textrm{Forward or Backward Linkage}_{i} \times \textrm{Year}_{t}\) control.

After controlling for linkages, I do not identify a negative relationship between measures of capital intensity and investment. Broadly, the relationship between capital intensity and investment in non-treated sectors is similar to the robustness estimates that did not account for linkages in Appendix Figure \ref{fig:appendixcrowdingout}. The relationship between capital intensity and investment---now controlling for linkages---is similar in both industries during the drive. There is a positive relationship between capital intensity and investment after 1973 for both industries, although the relationship is zero during the HCI period. After capital market liberalization (see Online Supplemental Appendix \ref{sec:supphistoryappendix}), the relationship becomes more pronounced in both industries, with a stronger relationship among non-treated industries.

\pagebreak

\clearpage

\begina

\blandscape
\begin{figure}[h]

{
\centering
\includegraphics{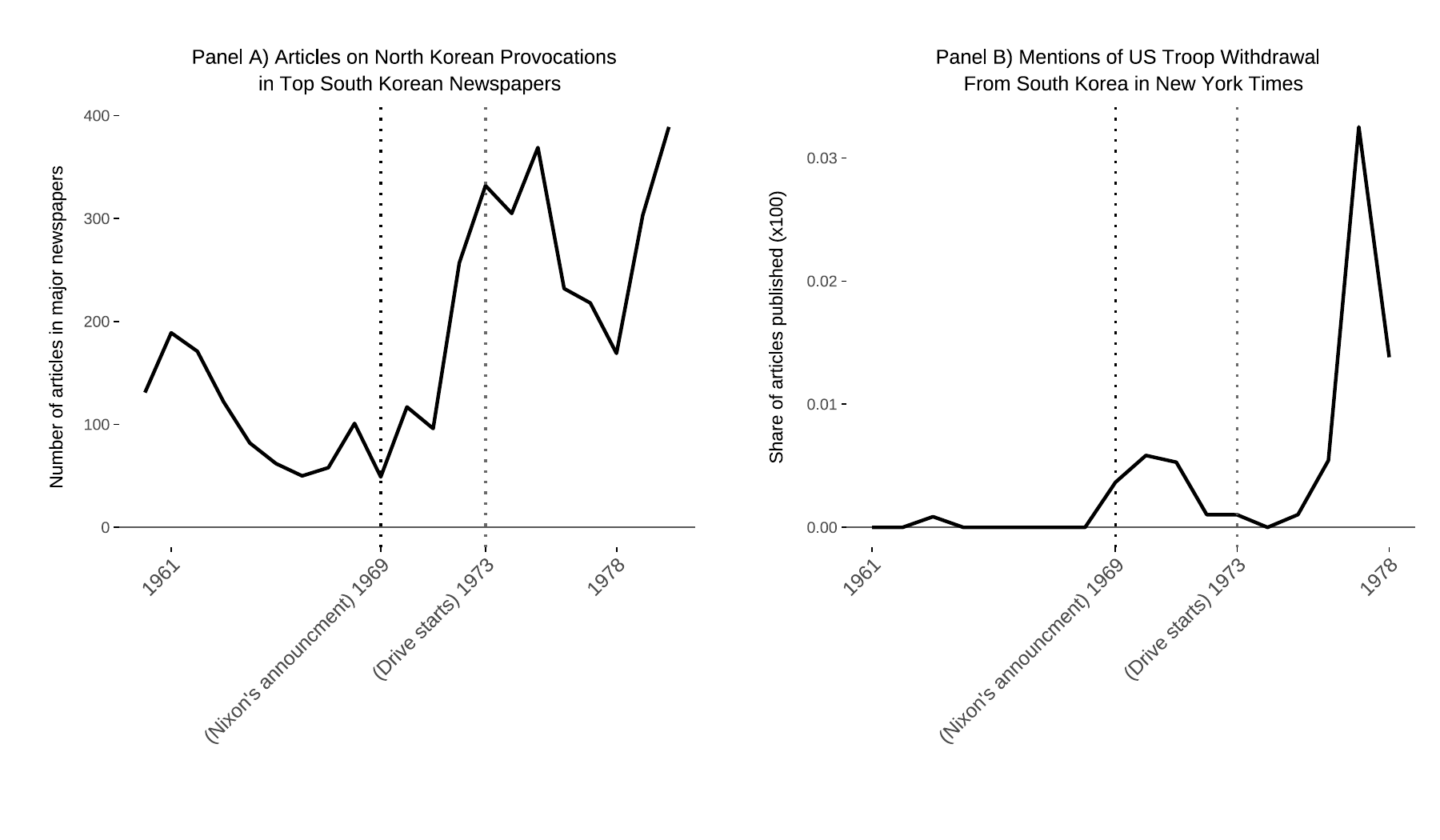}
}

\caption{Political Events Surrounding Heavy and Chemical Industry Drive} \label{fig:appendixtrooprobust}

\scriptsize
{
\setlength{\parindent}{2em}
    \input{appendixnewsplotrobust_note.tex}
}

\end{figure}
\elandscape

\FloatBarrier

\FloatBarrier

\beginb

\begin{figure}[h]

{
\centering
\includegraphics[width=.95\textwidth]{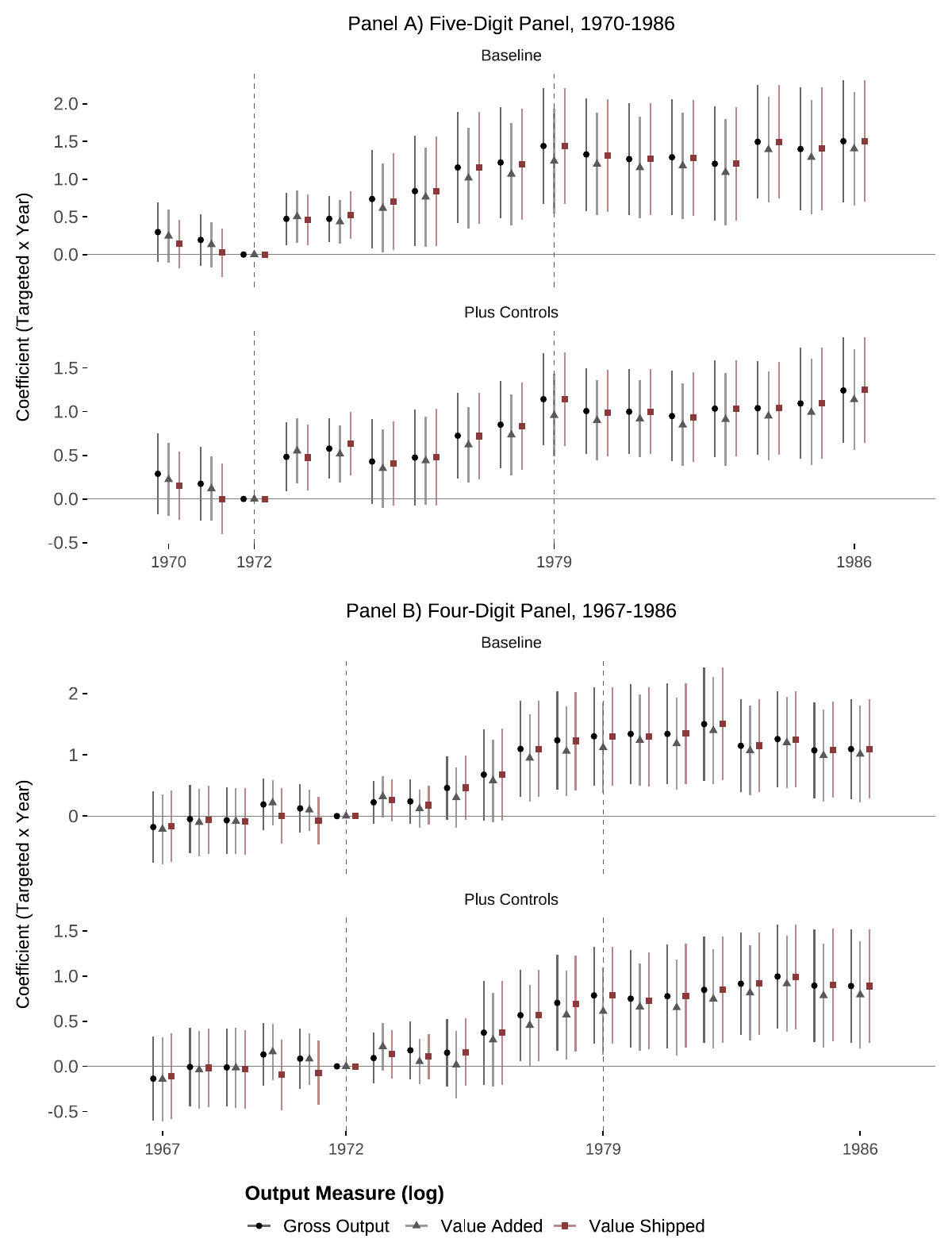}
}

\caption{Robustness: Industrial Policy and Measures of Output}\label{fig:appendixoutputrobust}

\scriptsize
{
\setlength{\parindent}{2em}
\input{appendixrobustoutput_note.tex}
}
\end{figure}

\blandscape
\vspace*{\fill} 
{
  \centering 
  \includegraphics[height=.8\textwidth]{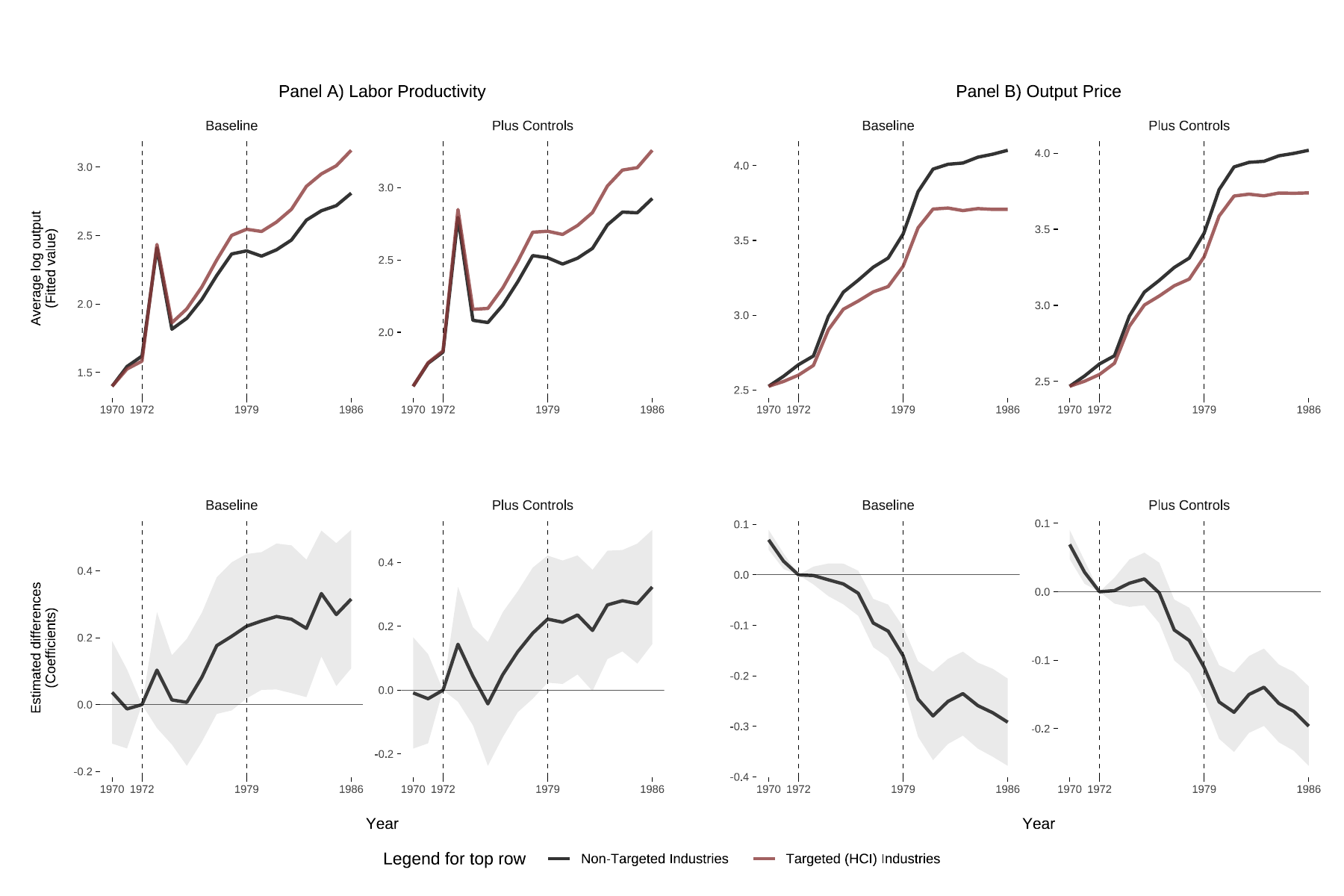}

  \captionof{figure}{Differences in Value Added Per Worker and Output Prices}
  \label{fig:appendixproductivity}
}

\scriptsize
\setlength{\parindent}{2em}
\noindent 
\indent This figure plots dynamic differences-in-differences  estimates for the relationship between HCI and labor productivity (value added  per worker) in Panel A and output prices in Panel B. Estimates come from equation (1) The top row shows the average outcomes for targeted (red) and non-targeted industries (black) using the fitted model. For specifications with controls, the model is evaluated using  means of the controls. The bottom row plots the differences-in-differences estimates.  All estimates are relative to 1972, the year before the HCI policy. The line at 1979 demarcates  the end of the Park regime. Standard errors are clustered at the industry level.  95 percent confidence intervals are shown in gray.
\vspace*{\fill} 
\elandscape
\FloatBarrier

\begin{figure}[h]

{
\centering 
\includegraphics{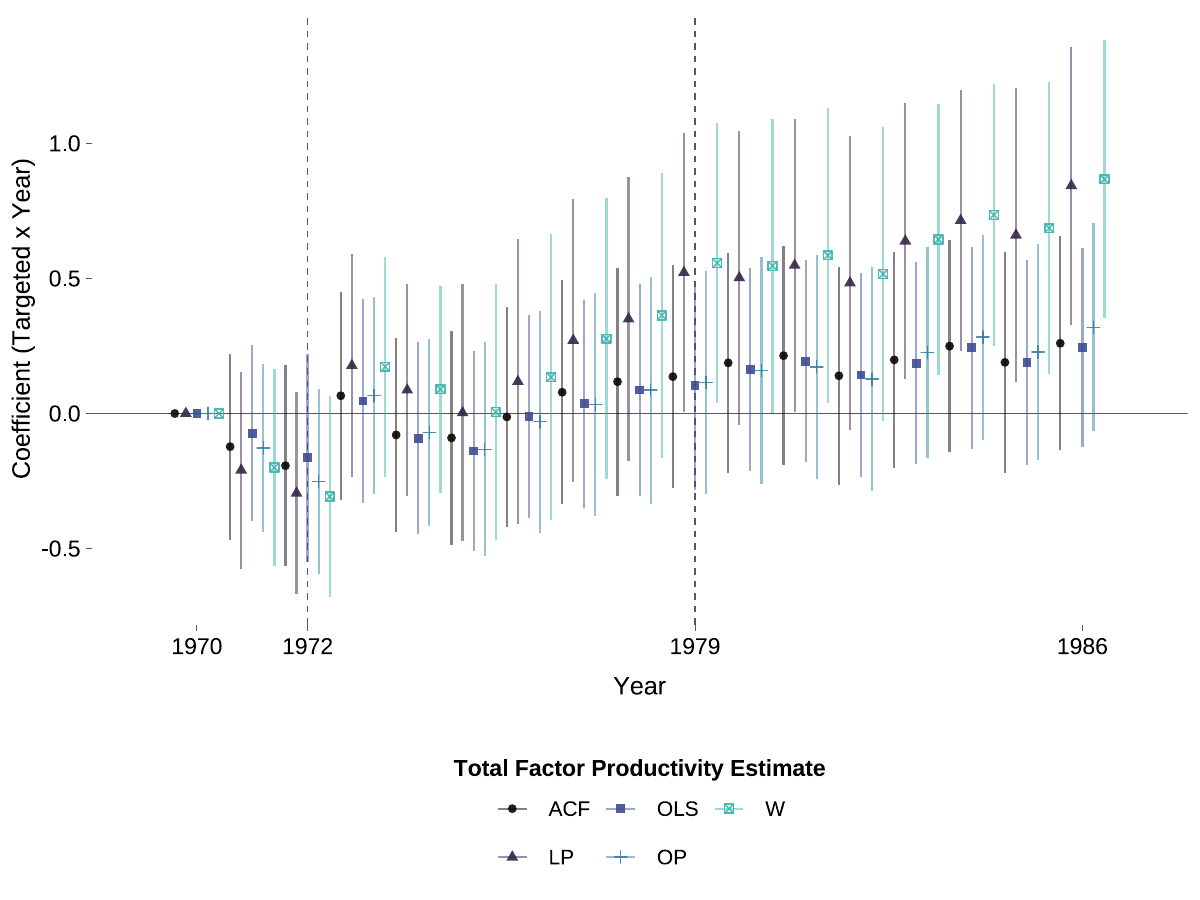}
}

\caption{Robustness: Industry Policy and Industry-Level Total Factor Productivity}\label{fig:appendixindustrytfp}

\scriptsize
{
\setlength{\parindent}{2em}
    \input{industrytfpfigure_note.tex}
}

\end{figure}

\FloatBarrier

\blandscape
\begin{figure}[p]

{
\centering 
\includegraphics[height=.85\textwidth]{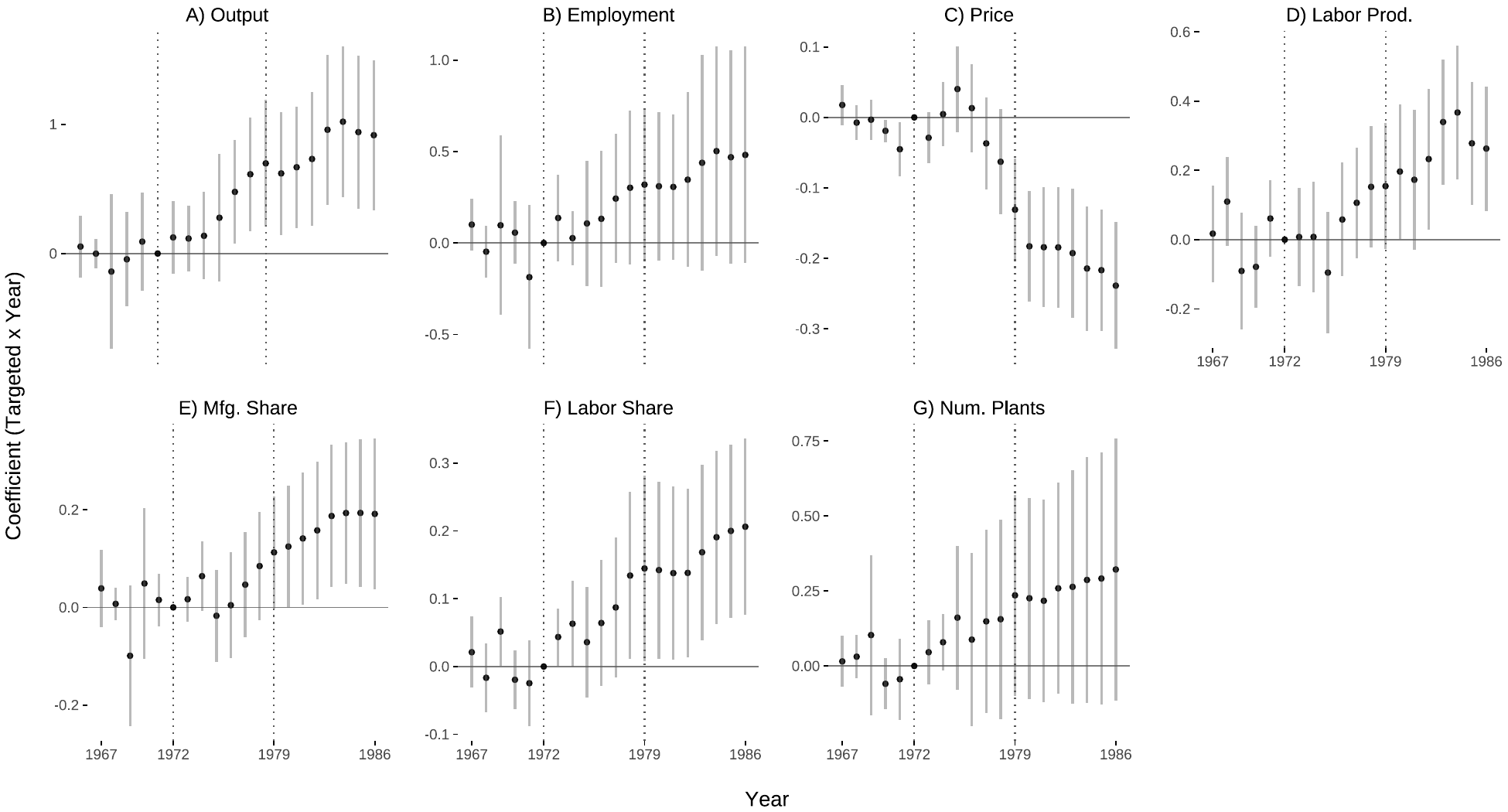}
}

\caption{Double Robust Estimates: Industrial Policy and Industrial Development, Four-Digit Panel}\label{fig:semiplot}

\scriptsize
{
\setlength{\parindent}{2em}
    \input{semidid_4digit_plot_note.tex}
}

\end{figure}
\elandscape

\FloatBarrier

\blandscape
\begin{figure}[p]

{
\centering 
\includegraphics[height=.85\textwidth]{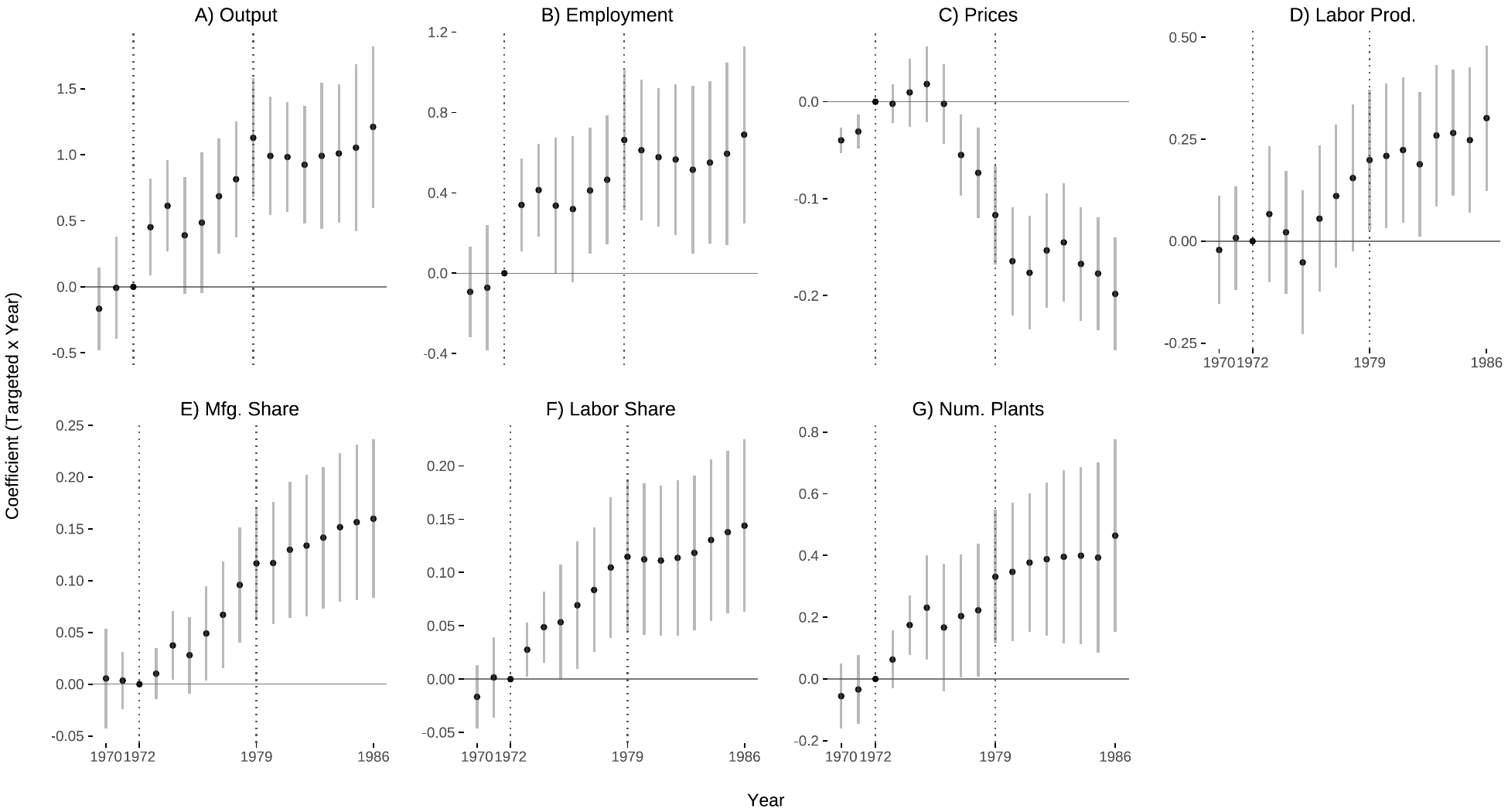}
}

\caption{Double Robust Estimates: Industrial Policy on Industrial Development, Five-Digit Panel}\label{fig:semiplot2}

\scriptsize
{
\setlength{\parindent}{2em}
    \input{semidid_5digit_plot_note.tex}
}

\end{figure}
\elandscape

\FloatBarrier

\blandscape
\begin{figure}[p]

{
\centering 
\includegraphics[height=.85\textwidth]{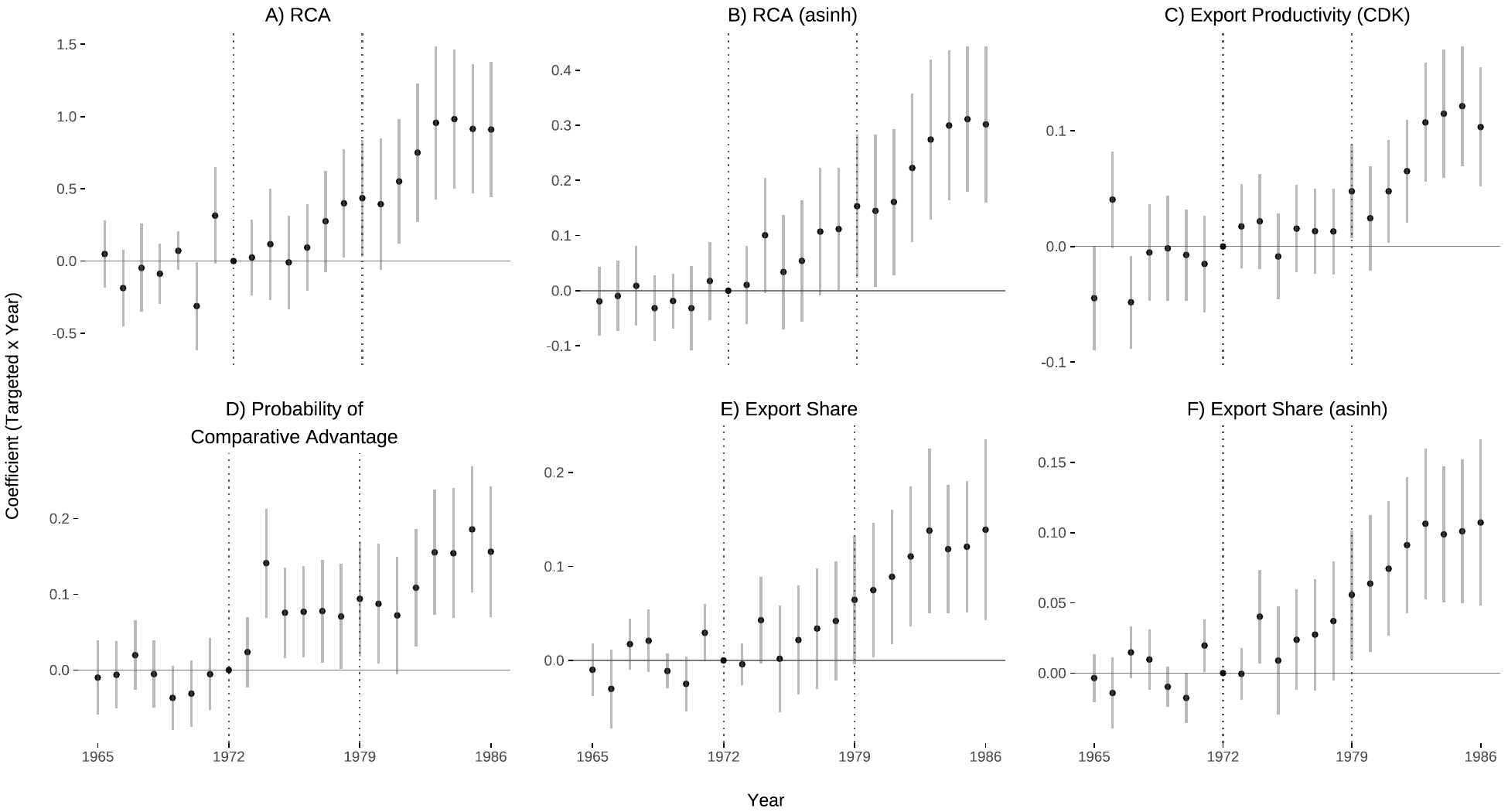}
}

\caption{Double Robust Estimates: Industrial Policy and Export Development}\label{fig:semitrade}

\scriptsize
{
\setlength{\parindent}{2em}
\input{semidid_sitc4_plot_note.tex}
}

\end{figure}
\elandscape

\FloatBarrier

\beginc

\begind

\begin{figure}

{
\centering 
\includegraphics{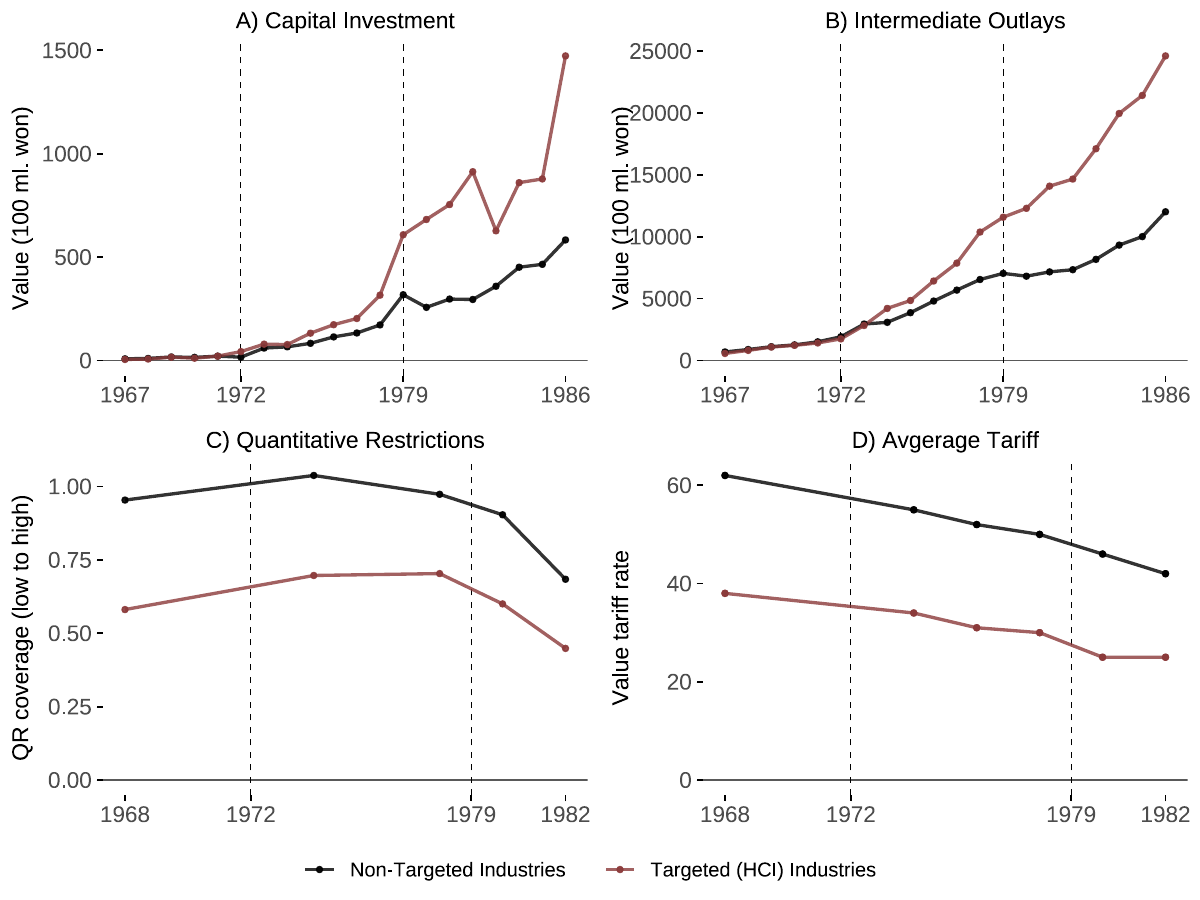} 
}

\caption{Average Aggregate Investment and Trade Policy}\label{fig:appendixaggregatepolicy}

\scriptsize
{
\setlength{\parindent}{2em}
    \input{gg_gridinvest_note.tex}
}

\end{figure}

\FloatBarrier

\blandscape
\begin{figure}

{
\centering 
\includegraphics[height=.8\textwidth]{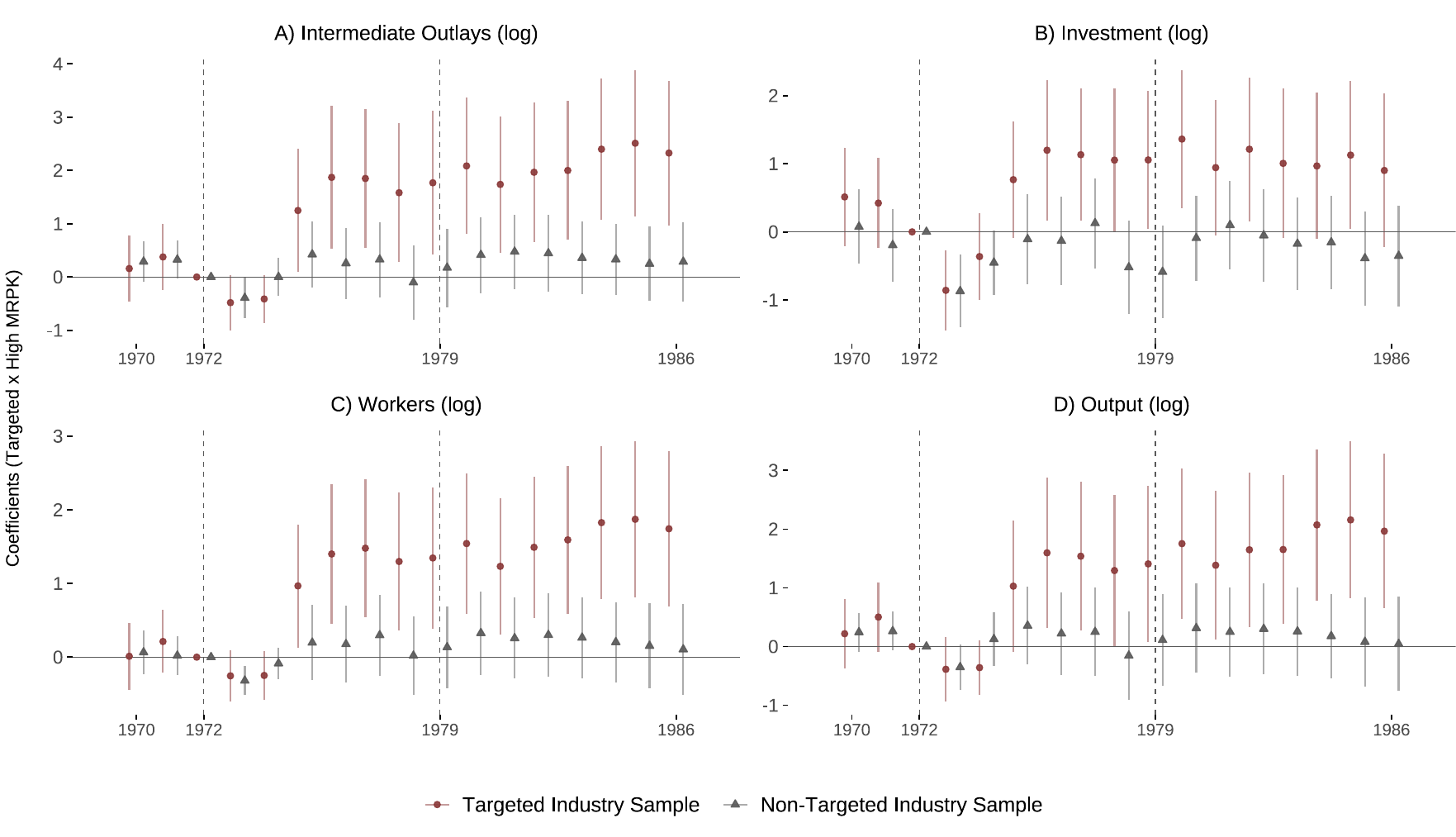}
}

\caption{Input Use and Marginal Revenue Product of Capital}\label{fig:appendixmrpkpolicy}

\scriptsize
{
\setlength{\parindent}{2em}
    \input{gg_mprk_plot_note.tex}
}
\end{figure}

\elandscape

\FloatBarrier

\begin{figure}[h]

{
\centering 
\includegraphics{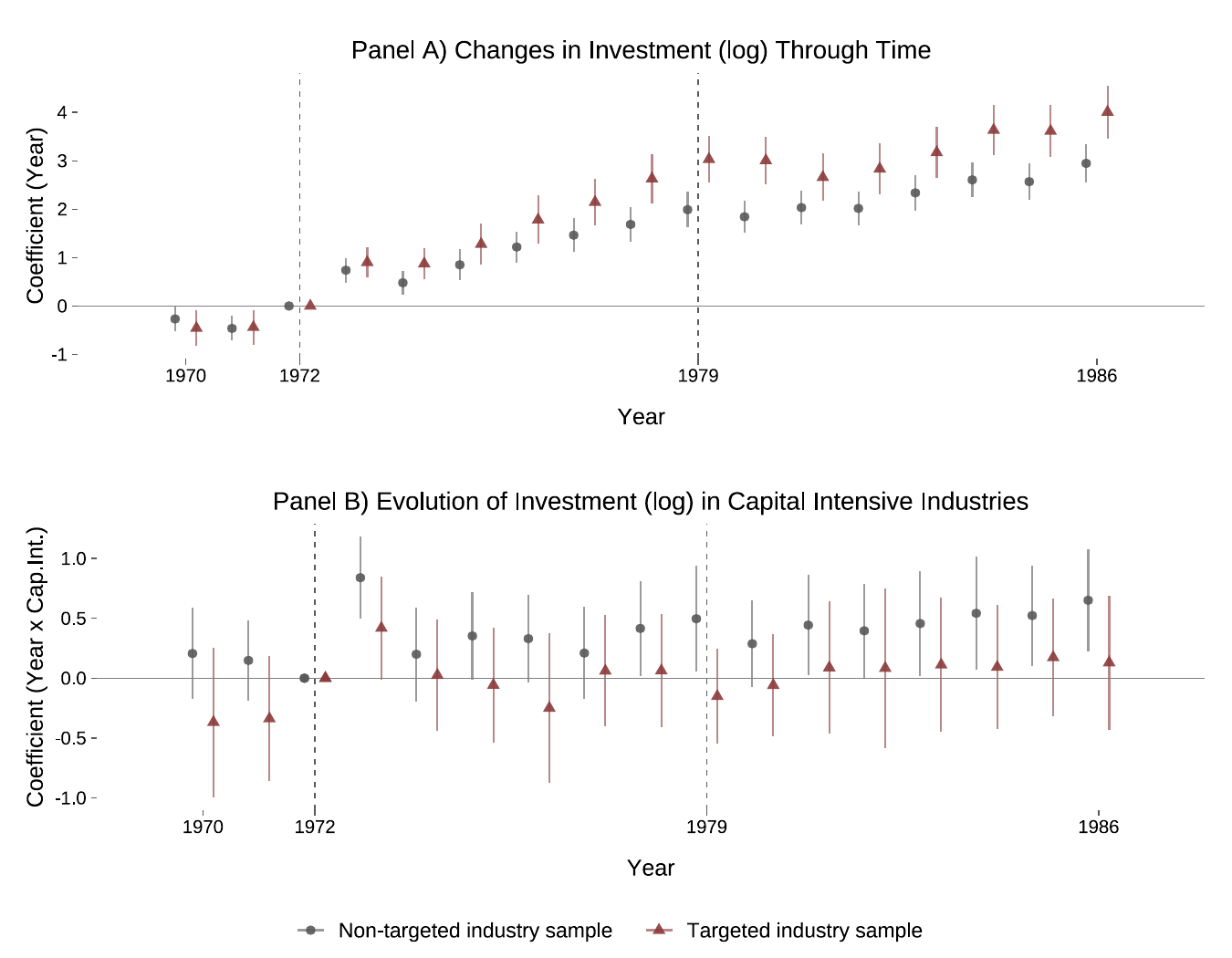}
}

\caption{Crowding Out and Investment by Treatment Status}\label{fig:appendixcrowdingout}

\scriptsize
{
\setlength{\parindent}{2em}
    \input{combined_crowdout_plot_note.tex}
}

\end{figure}

\FloatBarrier

\begin{figure}[h]

{
\centering 
\includegraphics[width=.9\textwidth]{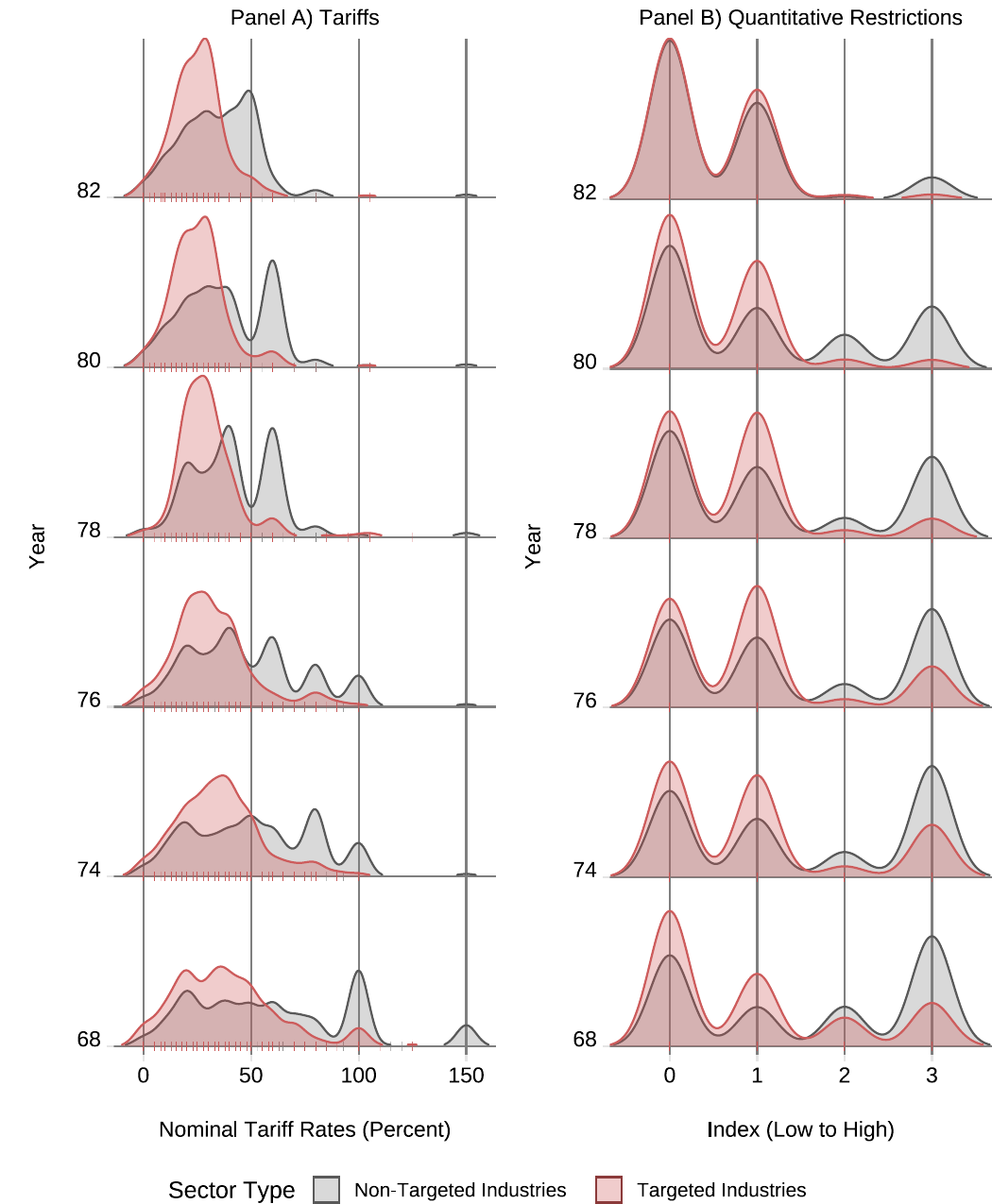}
}

\caption{Changes in Distribution of Trade Policies, 1968-1982}\label{fig:appendixtraderidgeplot}

\scriptsize
{
\setlength{\parindent}{2em}
    \input{trade_ridge_plots_note.tex}
}

\end{figure}

\FloatBarrier

\begine

\begin{figure}[p]

{
\centering 
\includegraphics{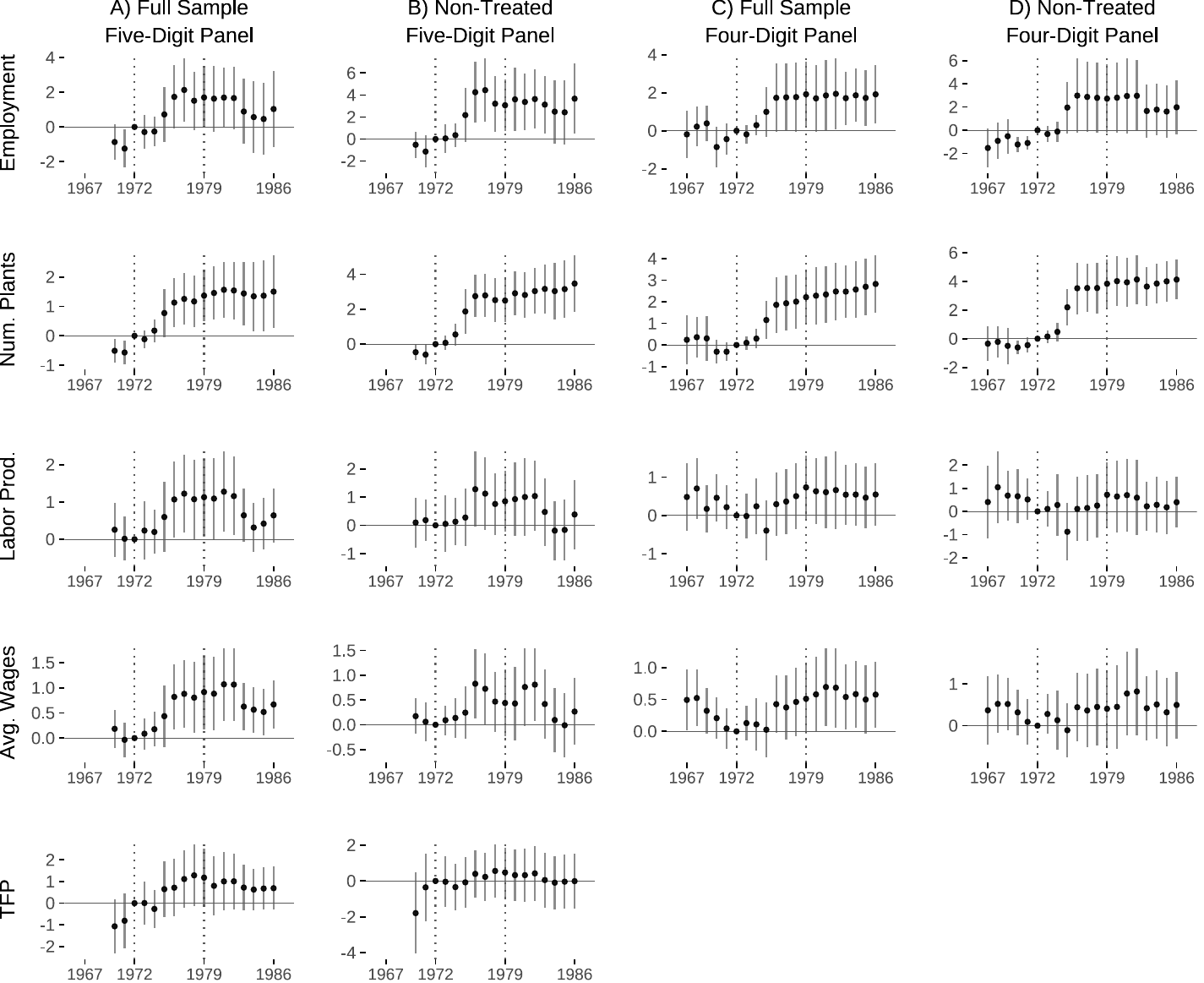} 
}

\caption{Direct Forward Linkages and Development Outcomes}\label{fig:appendixmoredevlinks}

\scriptsize
{
\setlength{\parindent}{2em}
    \input{gg_devlink_grid_note.tex}
}
\end{figure}

\FloatBarrier

\begin{figure}[p]

{
\centering 
\includegraphics{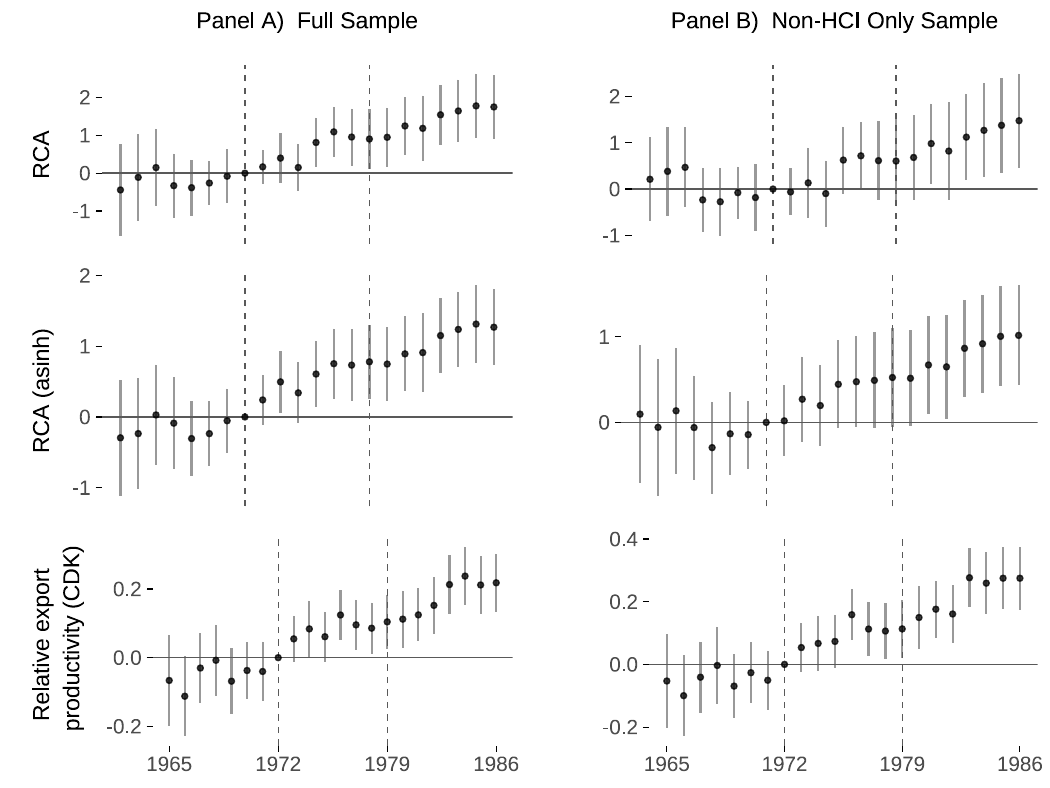} 
}

\caption{Total Forward Linkages and Export Development}\label{fig:appendixtradetotallinkage}

\scriptsize
{
\setlength{\parindent}{2em}
    \input{gg_rcatotallink_grid_note.tex}
}

\end{figure}

\FloatBarrier

\blandscape
\centering 
\begin{figure}[h]

{
  
    \includegraphics[height=.9\textwidth]{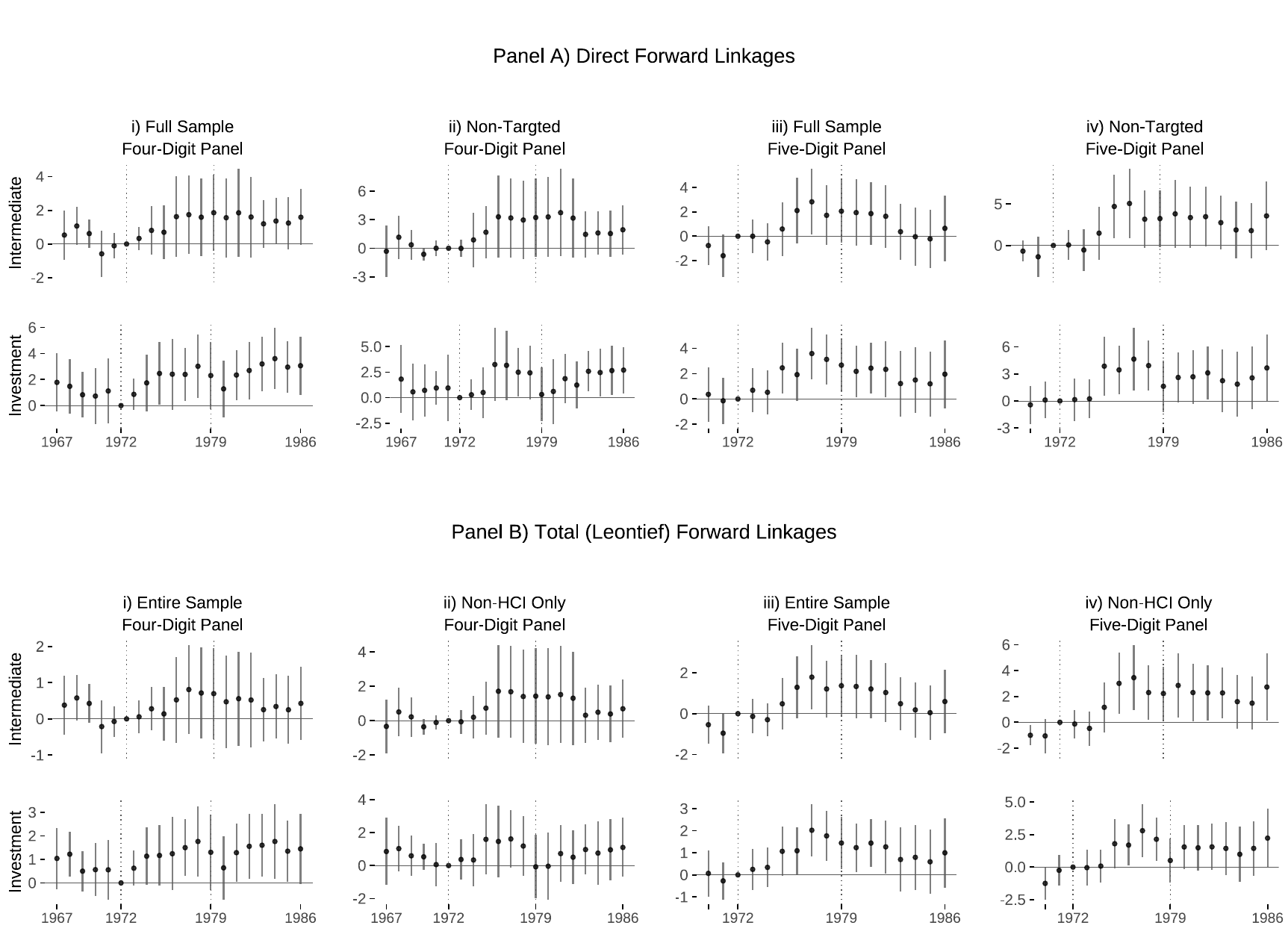} 
}
\caption{Linkage Mechanisms - Direct Forward Linkages, Intermediate Outlays, and Investment}\label{fig:appendixmechanismlinkage}

\scriptsize
{
\raggedright
\setlength{\parindent}{2em}
    \input{gg_mechanismlink_grid_note.tex}
}

\end{figure}

\elandscape

\FloatBarrier

\beginf

\begin{figure}[h]

{
\centering 
\includegraphics{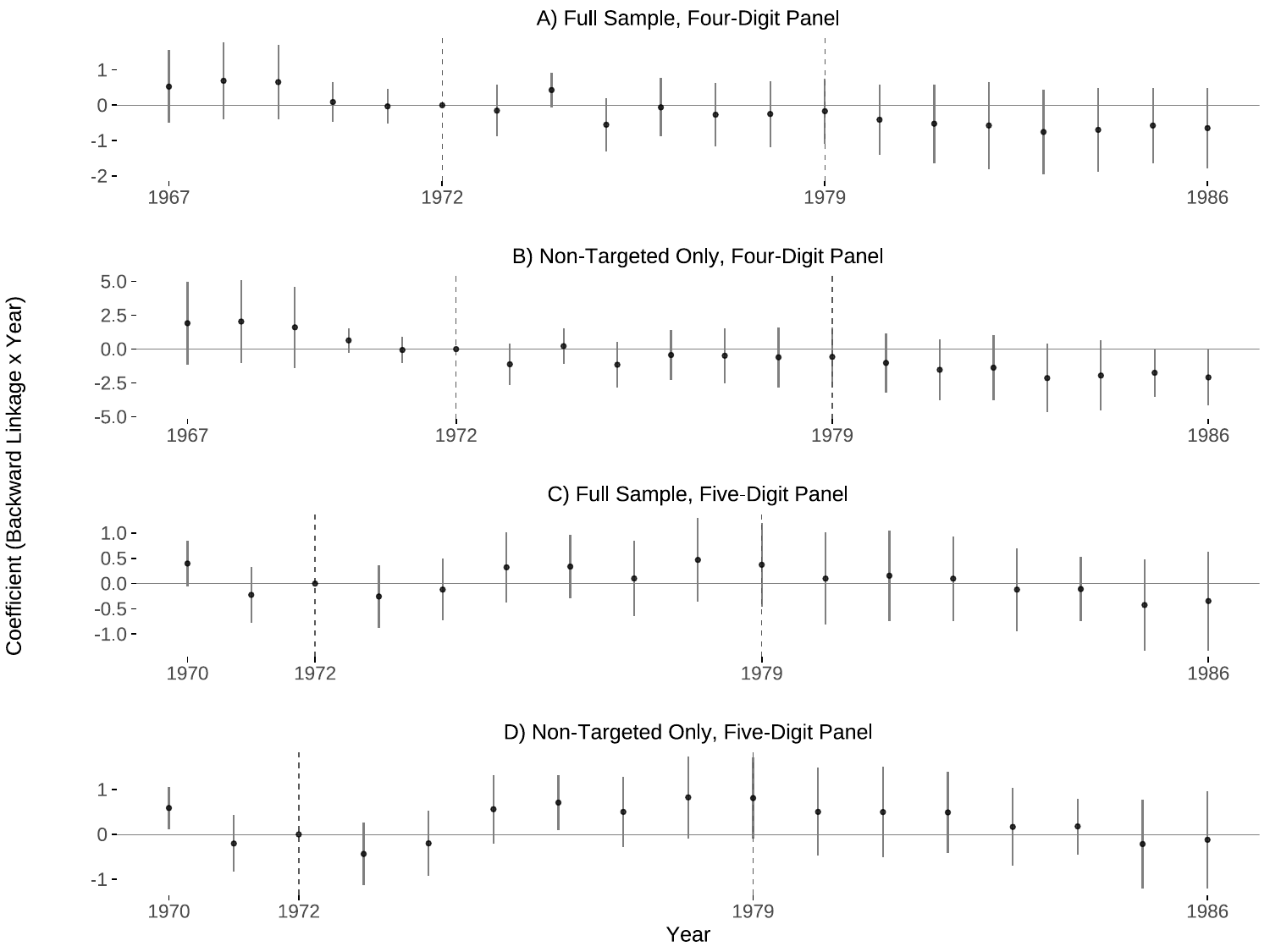}
}

\caption{Relationship Between Direct Backward Linkages on Upstream Output}\label{fig:appendixbacklinks}

\scriptsize
{
\setlength{\parindent}{2em}
    \input{gg_backwardlink_grid_note.tex}
}

\end{figure}

\FloatBarrier

\begin{figure}[h]

{
\centering 
\includegraphics{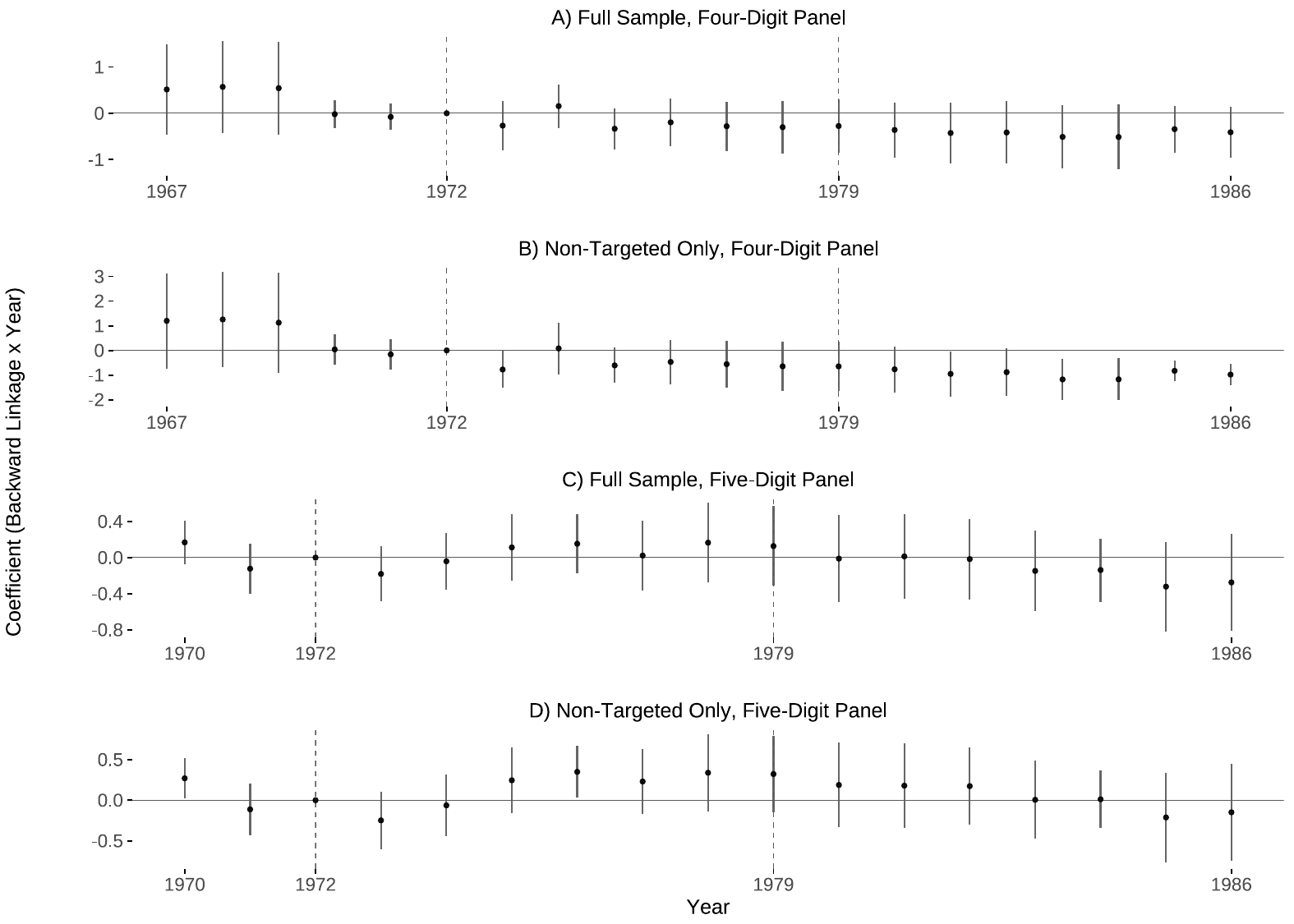}
}

\caption{Relationship Between Total Backward Linkages and Upstream Output}\label{fig:appendixbacklinkstotal}

\scriptsize
{
\setlength{\parindent}{2em}
    \input{gg_backwardlink_lf_grid_note.tex}
}

\end{figure}

\FloatBarrier

\beging

\begin{figure}[h]

{
\centering 
\includegraphics[width=1.1\textwidth]{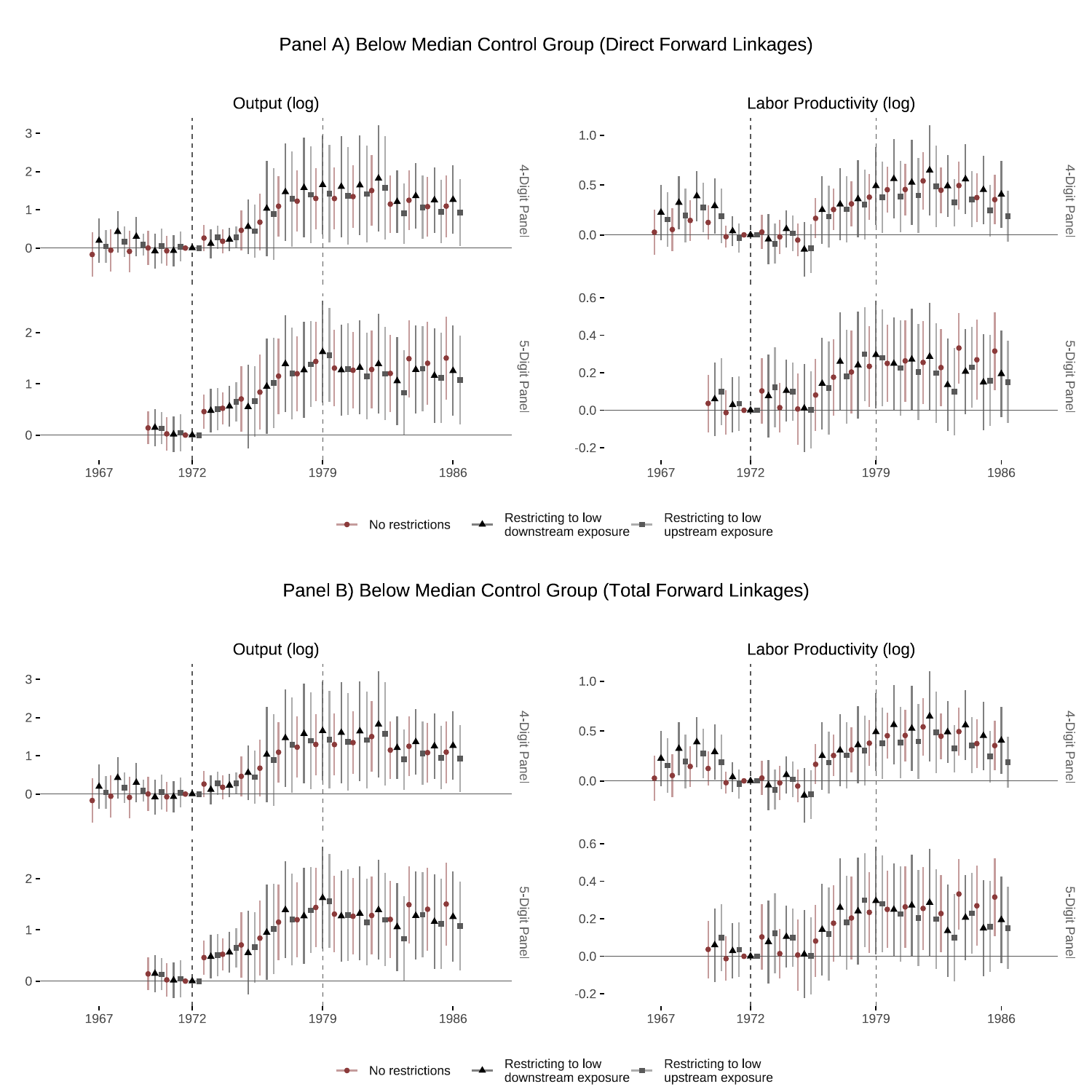}
}

\caption{Robustness: Impact of Industrial Policy on Development, Restricting to Control Industries with Low Linkages}\label{fig:appendixsutvalinkexposure}

\scriptsize
{
\setlength{\parindent}{2em}
    \input{gg_io_exposure_figure_note.tex}
}
\end{figure}

\FloatBarrier

\begin{figure}[h]

{
\centering 
\includegraphics{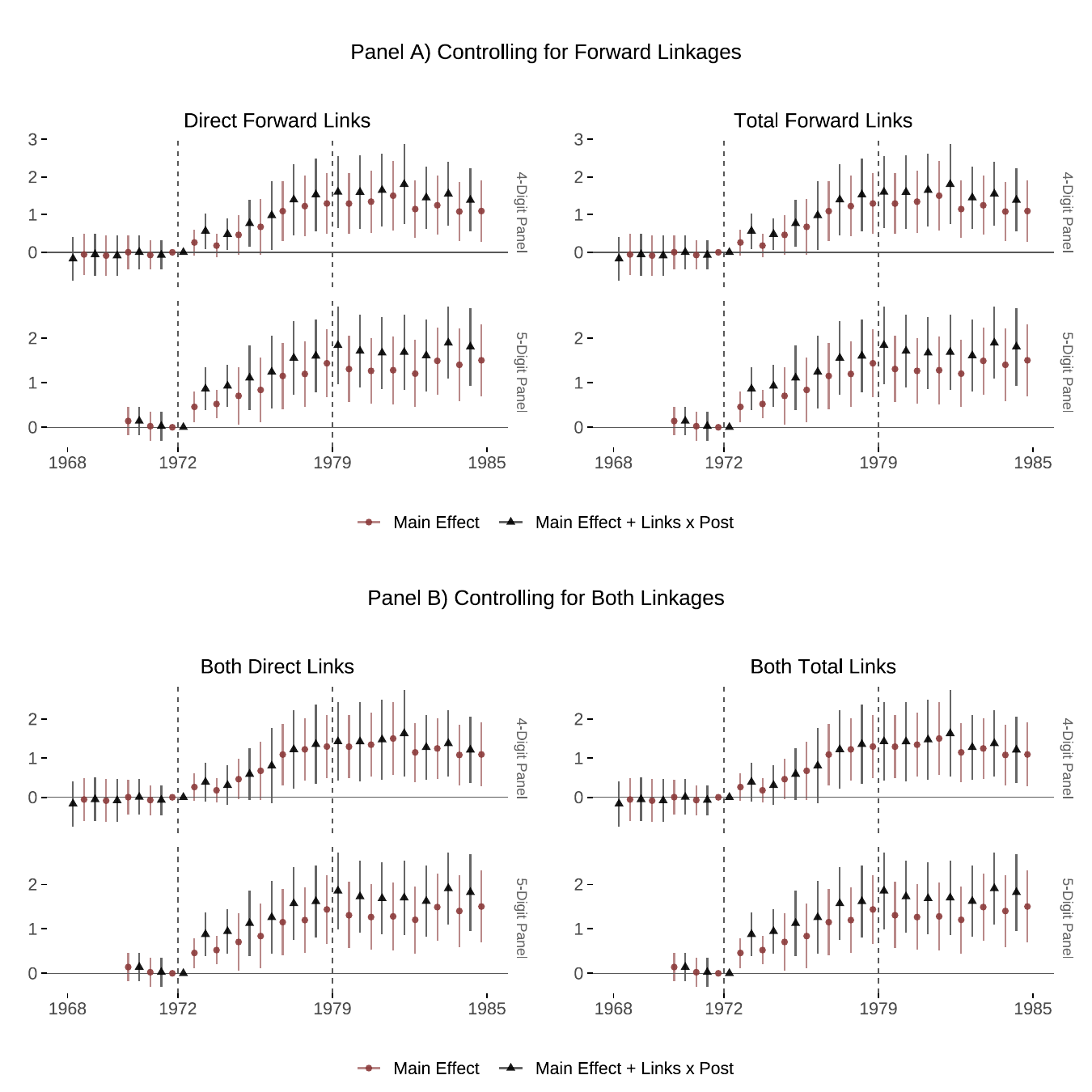}
}

\caption{Robustness: Impact of Industrial Policy on Development, Controlling for Non-Treated Linkages}\label{fig:appendixsutvapluslinks}

\scriptsize
{
\setlength{\parindent}{2em}
    \input{gg_control_io_figure_note.tex}
}

\end{figure}

\FloatBarrier

\begin{figure}[h]

{
\centering 
\includegraphics{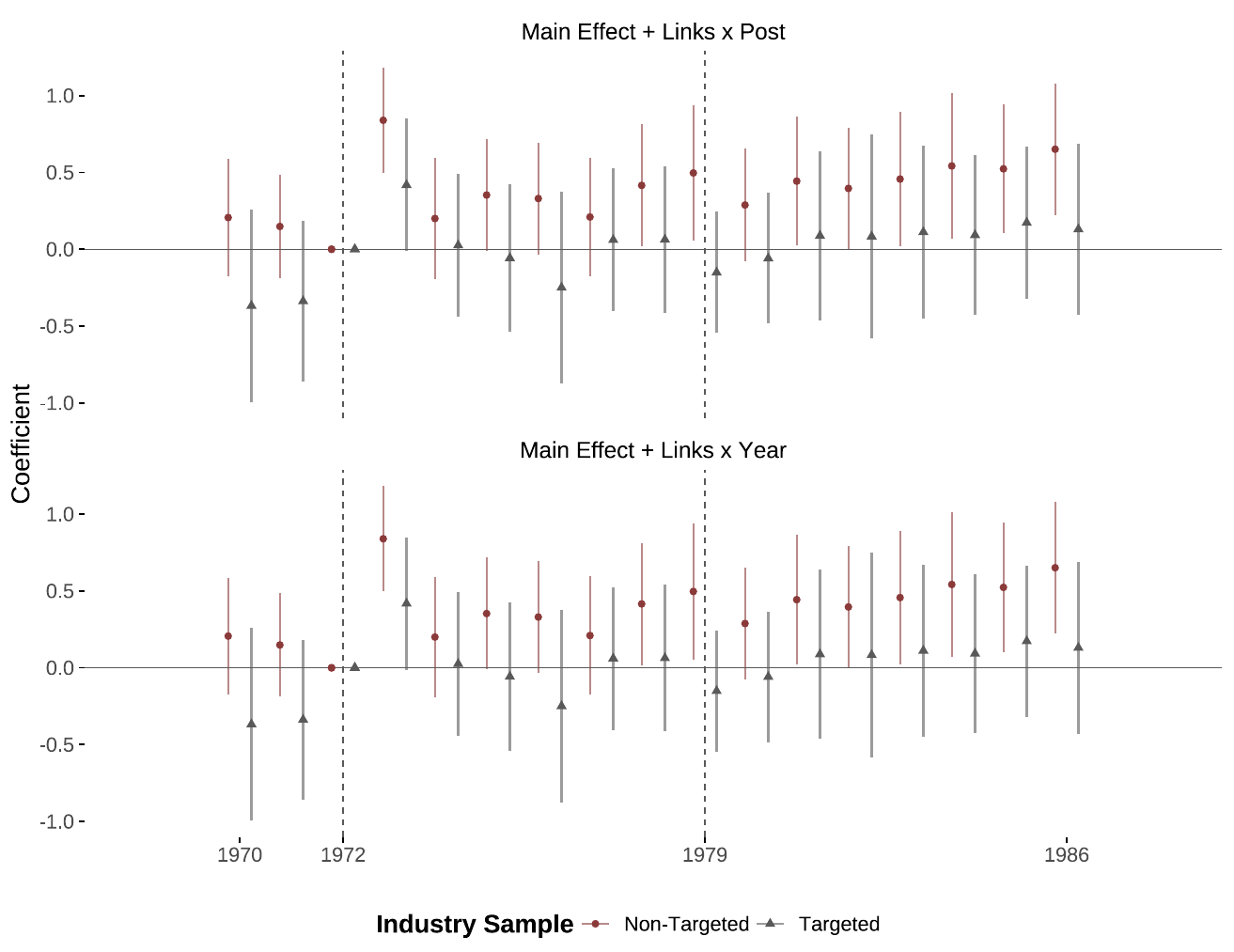}
}

\caption{Robustness: Relationship Between Investment and Capital Intensity, HCI Versus Non-HCI Industry}\label{fig:appendixcrowdingoutio}

\scriptsize
{
\setlength{\parindent}{2em}
    \input{gg_crowdingout_io_figure_note.tex}
}

\end{figure}

\FloatBarrier

\clearpage

\setcounter{table}{0}

\begina

\renewcommand{\arraystretch}{1}
\begin{table}
\centering
\caption{\label{tab:appendixtabledescriptive}Pre-HCI Drive Statistics By Treatment Status}
\centering
\resizebox{\ifdim\width>\linewidth\linewidth\else\width\fi}{!}{
\fontsize{9}{11}\selectfont
\begin{threeparttable}
\begin{tabular}[t]{>{\raggedright\arraybackslash}p{15em}lcccc}
\toprule
\multicolumn{2}{c}{ } & \multicolumn{2}{c}{A) Four-Digit Panel (1967-1972)} & \multicolumn{2}{c}{B) Five-Digit Panel (1970-1972)} \\
\cmidrule(l{3pt}r{3pt}){3-4} \cmidrule(l{3pt}r{3pt}){5-6}
Variable & Industry & Mean & N & Mean & N\\
\midrule
\addlinespace[0.2em]
\multicolumn{6}{l}{\textbf{i) Industrial Statistics}}\\
\hspace{1em} & Non-HCI & 74.17 & 330 & 69.33 & 528\\

\hspace{1em}\multirow[t]{-2}{15em}{\raggedright\arraybackslash Average Size} & HCI & 64.68 & 198 & 70.20 & 306\\

\hspace{1em} & Non-HCI & 406.48 & 330 & 106.35 & 528\\

\hspace{1em}\multirow[t]{-2}{15em}{\raggedright\arraybackslash Establishments} & HCI & 162.23 & 198 & 50.22 & 306\\

\hspace{1em} & Non-HCI & 184316.31 & 330 & 75962.19 & 528\\

\hspace{1em}\multirow[t]{-2}{15em}{\raggedright\arraybackslash Gross Output} & HCI & 154481.13 & 198 & 61968.10 & 306\\

\hspace{1em} & Non-HCI & 1434.22 & 330 & 1759.05 & 528\\

\hspace{1em}\multirow[t]{-2}{15em}{\raggedright\arraybackslash Investment} & HCI & 1741.88 & 198 & 2673.48 & 306\\

\hspace{1em} & Non-HCI & 7.50 & 330 & 7.20 & 528\\

\hspace{1em}\multirow[t]{-2}{15em}{\raggedright\arraybackslash Labor Productivity} & HCI & 8.97 & 198 & 7.47 & 306\\

\hspace{1em} & Non-HCI & 1.39 & 330 & 0.42 & 528\\

\hspace{1em}\multirow[t]{-2}{15em}{\raggedright\arraybackslash Labor Share} & HCI & 0.72 & 198 & 0.24 & 306\\

\hspace{1em} & Non-HCI & 9.70 & 330 & 10.97 & 528\\

\hspace{1em}\multirow[t]{-2}{15em}{\raggedright\arraybackslash Prices} & HCI & 29.20 & 198 & 29.13 & 306\\

\hspace{1em} & Non-HCI & 86472.29 & 330 & 31513.33 & 528\\

\hspace{1em}\multirow[t]{-2}{15em}{\raggedright\arraybackslash Value Added} & HCI & 52987.78 & 198 & 22038.69 & 306\\

\hspace{1em} & Non-HCI & 1.36 & 330 & 0.42 & 528\\

\hspace{1em}\multirow[t]{-2}{15em}{\raggedright\arraybackslash Value Added Share} & HCI & 0.81 & 198 & 0.26 & 306\\

\hspace{1em} & Non-HCI & 12983.87 & 330 & 4117.89 & 528\\

\hspace{1em}\multirow[t]{-2}{15em}{\raggedright\arraybackslash Workers} & HCI & 6775.03 & 198 & 2351.84 & 306\\

\addlinespace[0.2em]
\multicolumn{6}{l}{\textbf{ii) Linkage Exposure to HCI Sectors}}\\
\hspace{1em} & Non-HCI & 0.09 & 330 & 0.15 & 528\\

\hspace{1em}\multirow[t]{-2}{15em}{\raggedright\arraybackslash Backward Linkage} & HCI & 0.16 & 198 & 0.20 & 306\\

\hspace{1em} & Non-HCI & 0.11 & 330 & 0.10 & 528\\

\hspace{1em}\multirow[t]{-2}{15em}{\raggedright\arraybackslash Forward Linkage} & HCI & 0.31 & 198 & 0.34 & 306\\

\addlinespace[0.2em]
\multicolumn{6}{l}{\textbf{iii) Trade Statistics (SITC trade data, 1965-1972)}}\\
\hspace{1em} & Non-HCI & 1.27 & 3464 &  & \\

\hspace{1em}\multirow[t]{-2}{15em}{\raggedright\arraybackslash RCA (Balassa)} & HCI & 0.40 & 1448 &  & \\

\hspace{1em} & Non-HCI & 0.14 & 3464 &  & \\

\hspace{1em}\multirow[t]{-2}{15em}{\raggedright\arraybackslash Export Share} & HCI & 0.09 & 1448 &  & \\

\hspace{1em} & Non-HCI & 0.13 & 3464 &  & \\

\hspace{1em}\multirow[t]{-2}{15em}{\raggedright\arraybackslash Import Share} & HCI & 0.24 & 1448 &  & \\
\bottomrule
\end{tabular}
\begin{tablenotes}[para]
\item \textit{\hspace{1em}\textit{Notes.}} 
\item Table reports pre-1973 statistics for a selection of core industrial  variables. Panel A shows statistics for aggregated ('long') 4-digit industrial panel, 1967 to 1972. Panel B shows statistics for disaggregated ('short') 5-digit  industrial panel, 1970 to 1972. Part i) of table reports Mining and Manufacturing Survey/Census (MMS) outcomes, with the exception of prices, which come from the  Bank of Korea publications. Part ii) shows data from the 1970 input-output tables  published by the Bank of Korea (1970), harmonized and matched to industry-level  data. Part iii) shows trade variables (from UN-Comtrade). All values are deflated using 2010 baseline won, except for real USD trade values.
\end{tablenotes}
\end{threeparttable}}
\end{table}
\FloatBarrier

\beginc

\blandscape
\begin{table}
\centering
\caption{\label{tab:appendixprobrca}Probability of Attaining Comparative Advantage in Targeted Industry, South Korea v. Other Countries, Post-1972}
\centering
\fontsize{9}{11}\selectfont
\begin{threeparttable}
\begin{tabular}[t]{lcccccccc}
\toprule
\multicolumn{1}{c}{\bgroup\fontsize{10}{12}\selectfont \egroup{}} & \multicolumn{8}{c}{\bgroup\fontsize{10}{12}\selectfont Outcomes: Probability of Comparative Advantage\egroup{}} \\
\cmidrule(l{3pt}r{3pt}){2-9}
\multicolumn{1}{c}{} & \multicolumn{4}{c}{Estimates with OLS} & \multicolumn{4}{c}{Estimates with PPML} \\
\cmidrule(l{3pt}r{3pt}){2-5} \cmidrule(l{3pt}r{3pt}){6-9}
\multicolumn{1}{c}{} & \multicolumn{2}{c}{Full Sample} & \multicolumn{1}{c}{Similar GDP} & \multicolumn{1}{c}{Same GDP} & \multicolumn{2}{c}{Full Sample} & \multicolumn{1}{c}{Similar GDP} & \multicolumn{1}{c}{Same GDP} \\
\cmidrule(l{3pt}r{3pt}){2-3} \cmidrule(l{3pt}r{3pt}){4-4} \cmidrule(l{3pt}r{3pt}){5-5} \cmidrule(l{3pt}r{3pt}){6-7} \cmidrule(l{3pt}r{3pt}){8-8} \cmidrule(l{3pt}r{3pt}){9-9}
 & (1) & (2) & (3) & (4) & (5) & (6) & (7) & (8)\\
\midrule
\hspace{1em}Korea & 0.131 & 0.110*** & 0.137 & 0.119 & 1.002*** & 1.145*** & 1.085*** & 0.853***\\
\hspace{1em} & (.) & (0.011) & (.) & (.) & (0.141) & (0.094) & (0.143) & (0.212)\\
\hspace{1em}GDP per capita &  & 0.047*** &  &  &  & 0.664*** &  & \\
\hspace{1em} &  & (0.008) &  &  &  & (0.074) &  & \\
\addlinespace[0.3em]
\multicolumn{9}{l}{\textbf{}}\\
\hspace{1em}\hspace{1em}Industry X Year Effect & Yes & Yes & Yes & Yes & Yes & Yes & Yes & Yes\\
\hspace{1em}\hspace{1em}\(R^2\) & 0.025 & 0.089 & 0.065 & 0.053 & 0.052 & 0.162 & 0.082 & 0.064\\
\hspace{1em}\hspace{1em}Observations & 251160 & 251160 & 76440 & 24570 & 246652 & 246652 & 55720 & 13824\\
\hspace{1em}\hspace{1em}Mean of Dependent Variable & 0.078 & 0.078 & 0.075 & 0.102 & 0.079 & 0.079 & 0.103 & 0.180\\
\hspace{1em}\hspace{1em}Clusters (Country-Industry) & 92 x 182 & 92 x 182 & 28 x 182 & 9 x 182 & 92 x 182 & 92 x 182 & 28 x 178 & 9 x 160\\
\bottomrule
\end{tabular}
\begin{tablenotes}[para]
\item \textit{\hspace{1em}\textit{Notes.}} 
\item The probability of attaining RCA (RCA>1) in HCI products for  Korea versus other countries in the post-1972 period. Regressions include industry-by-year  effects. Data is restricted to treated industries. Two-way standard errors are  clustered at the industry and country levels.
\end{tablenotes}
\end{threeparttable}
\end{table}
\elandscape

\FloatBarrier

\begind

\blandscape
\begin{table}
\centering
\caption{\label{tab:appendixindustrylbd}Robustness: Learning in Industrial-Level Data, by Treatment Status}
\centering
\resizebox{\ifdim\width>\linewidth\linewidth\else\width\fi}{!}{
\fontsize{9}{11}\selectfont
\begin{threeparttable}
\begin{tabular}[t]{lcccccccccccccc}
\toprule
\multicolumn{1}{c}{} & \multicolumn{14}{c}{Outcomes} \\
\cmidrule(l{2pt}r{2pt}){2-15}
\multicolumn{1}{c}{ } & \multicolumn{2}{c}{Price (log)} & \multicolumn{2}{c}{Unit Cost (log)} & \multicolumn{2}{c}{Unit Cost (revenue, log)} & \multicolumn{2}{c}{TFP (OP)} & \multicolumn{2}{c}{TFP (ACF)} & \multicolumn{2}{c}{TFP (LP)} & \multicolumn{2}{c}{TFP (W)} \\
\cmidrule(l{3pt}r{3pt}){2-3} \cmidrule(l{3pt}r{3pt}){4-5} \cmidrule(l{3pt}r{3pt}){6-7} \cmidrule(l{3pt}r{3pt}){8-9} \cmidrule(l{3pt}r{3pt}){10-11} \cmidrule(l{3pt}r{3pt}){12-13} \cmidrule(l{3pt}r{3pt}){14-15}
\multicolumn{1}{c}{\bgroup\fontsize{8}{10}\selectfont \textbackslash{}textit\{Alternative measures:\}\egroup{}} & \multicolumn{1}{c}{\bgroup\fontsize{8}{10}\selectfont \makecell[c]{Experience \\per worker}\egroup{}} & \multicolumn{1}{c}{\bgroup\fontsize{8}{10}\selectfont \makecell[c]{Experience \\(alternative)}\egroup{}} & \multicolumn{1}{c}{\bgroup\fontsize{8}{10}\selectfont \makecell[c]{Experience \\per worker}\egroup{}} & \multicolumn{1}{c}{\bgroup\fontsize{8}{10}\selectfont \makecell[c]{Experience \\(alternative)}\egroup{}} & \multicolumn{1}{c}{\bgroup\fontsize{8}{10}\selectfont \makecell[c]{Experience \\per worker}\egroup{}} & \multicolumn{1}{c}{\bgroup\fontsize{8}{10}\selectfont \makecell[c]{Experience \\(alternative)}\egroup{}} & \multicolumn{1}{c}{\bgroup\fontsize{8}{10}\selectfont \makecell[c]{Experience \\per worker}\egroup{}} & \multicolumn{1}{c}{\bgroup\fontsize{8}{10}\selectfont \makecell[c]{Experience \\(alternative)}\egroup{}} & \multicolumn{1}{c}{\bgroup\fontsize{8}{10}\selectfont \makecell[c]{Experience \\per worker}\egroup{}} & \multicolumn{1}{c}{\bgroup\fontsize{8}{10}\selectfont \makecell[c]{Experience \\(alternative)}\egroup{}} & \multicolumn{1}{c}{\bgroup\fontsize{8}{10}\selectfont \makecell[c]{Experience \\per worker}\egroup{}} & \multicolumn{1}{c}{\bgroup\fontsize{8}{10}\selectfont \makecell[c]{Experience \\(alternative)}\egroup{}} & \multicolumn{1}{c}{\bgroup\fontsize{8}{10}\selectfont \makecell[c]{Experience \\per worker}\egroup{}} & \multicolumn{1}{c}{\bgroup\fontsize{8}{10}\selectfont \makecell[c]{Experience \\(alternative)}\egroup{}} \\
 & (1) & (2) & (3) & (4) & (5) & (6) & (7) & (8) & (9) & (10) & (11) & (12) & (13) & (14)\\
\midrule
Experience & -0.197*** & -0.155*** & -0.101*** & -0.110*** & -0.101*** & -0.115*** & 0.355*** & 0.409*** & 0.403*** & 0.456*** & 0.439*** & 0.459*** & 0.428*** & 0.444***\\
 & (0.029) & (0.027) & (0.015) & (0.014) & (0.015) & (0.015) & (0.060) & (0.055) & (0.060) & (0.061) & (0.059) & (0.058) & (0.065) & (0.062)\\
Targeted \(\times\) Experience & -0.058*** & -0.061*** & -0.045*** & -0.042*** & -0.043*** & -0.039*** & 0.039 & 0.050** & 0.036 & 0.039 & 0.087*** & 0.138*** & 0.092*** & 0.144***\\
 & (0.014) & (0.014) & (0.010) & (0.010) & (0.011) & (0.010) & (0.031) & (0.025) & (0.031) & (0.024) & (0.032) & (0.026) & (0.033) & (0.026)\\
\addlinespace[0.75em]
\multicolumn{15}{p{\linewidth}}{\textbf{Controls}}\\
\hspace{1em}Controls for Size & Yes & Yes & Yes & Yes & Yes & Yes & Yes & Yes & Yes & Yes & Yes & Yes & Yes & Yes\\
\hspace{1em}Controls for Capital Intensity & Yes & Yes & Yes & Yes & Yes & Yes & Yes & Yes & Yes & Yes & Yes & Yes & Yes & Yes\\
\hspace{1em}Controls for Intermediates & Yes & Yes & Yes & Yes & Yes & Yes & Yes & Yes & Yes & Yes & Yes & Yes & Yes & Yes\\
\hspace{1em}Controls for Investment & Yes & Yes & Yes & Yes & Yes & Yes & Yes & Yes & Yes & Yes & Yes & Yes & Yes & Yes\\
\addlinespace[0.75em]
\multicolumn{15}{p{\linewidth}}{\textbf{ }}\\
\hspace{1em}Industry Effects & Yes & Yes & Yes & Yes & Yes & Yes & Yes & Yes & Yes & Yes & Yes & Yes & Yes & Yes\\
\hspace{1em}Year Effects & Yes & Yes & Yes & Yes & Yes & Yes & Yes & Yes & Yes & Yes & Yes & Yes & Yes & Yes\\
\hspace{1em}R2 & 0.961 & 0.960 & 0.899 & 0.901 & 0.903 & 0.905 & 0.981 & 0.981 & 0.878 & 0.882 & 0.986 & 0.986 & 0.990 & 0.990\\
\hspace{1em}Observations & 3427 & 3429 & 3427 & 3429 & 3426 & 3428 & 3426 & 3428 & 3426 & 3428 & 3426 & 3428 & 3426 & 3428\\
\hspace{1em}Clusters & 263 & 263 & 263 & 263 & 263 & 263 & 263 & 263 & 263 & 263 & 263 & 263 & 263 & 263\\
\addlinespace[0.75em]
\multicolumn{15}{p{\linewidth}}{\textbf{Combined Effects}}\\
\hline
\hspace{1em}Linear Combination & -0.255 & -0.216 & -0.146 & -0.152 & -0.145 & -0.154 & 0.394 & 0.459 & 0.439 & 0.495 & 0.526 & 0.597 & 0.519 & 0.588\\
\hspace{1em}(St.Err.) & (0.029) & (0.026) & (0.016) & (0.016) & (0.016) & (0.016) & (0.051) & (0.056) & (0.051) & (0.059) & (0.054) & (0.061) & (0.057) & (0.064)\\
\bottomrule
\end{tabular}
\begin{tablenotes}[para]
\item \textit{\hspace{1em}\textit{Notes.}} 
\item This table shows the robustness of industry-level estimates from equation (5) for alternative outcomes.  Unit Cost is the baseline unit cost measure: (log) total real intermediate cost per real gross output; Unit Cost (revenue) is measured using total real intermediate costs per  unit of real revenue. TFP outcomes are estimated using Ackerberg-Caves-Frazer  (ACF), Levinsohn-Petrin (LP), Olley-Pakes (OP), and Wooldridge (W) methods.  Table shows estimates for each outcome using two alternative Experience measures: (log) Experience per worker, and Experience (alternative), which is experience  calculated using cumulative value added units. All equations control for size and scale:  (log) average plant size and total industry employment. Additional controls include  (log): capital intensity, skill ratio, investment per worker, and intermediate  input intensity per worker. Linear Combination, at the bottom, gives the combined effects for Experience for targeted industries. All specifications are estimated  using industry and year fixed effects. Standard errors are clustered at the industry level.
\end{tablenotes}
\end{threeparttable}}
\end{table}
\elandscape

\FloatBarrier

\blandscape
\begin{table}
\centering
\caption{\label{tab:appendixmicrolbd}Robustness: Plant and Industry-Level Learning, by Treatment Status}
\centering
\resizebox{\ifdim\width>\linewidth\linewidth\else\width\fi}{!}{
\fontsize{9}{11}\selectfont
\begin{threeparttable}
\begin{tabular}[t]{lcccccccccc}
\toprule
\multicolumn{1}{c}{ } & \multicolumn{5}{c}{Panel A) Experience} & \multicolumn{5}{c}{Panel B) Experience (per worker)} \\
\cmidrule(l{3pt}r{3pt}){2-6} \cmidrule(l{3pt}r{3pt}){7-11}
\multicolumn{1}{c}{ } & \multicolumn{1}{c}{ } & \multicolumn{4}{c}{TFP} & \multicolumn{1}{c}{ } & \multicolumn{4}{c}{TFP} \\
\cmidrule(l{3pt}r{3pt}){3-6} \cmidrule(l{3pt}r{3pt}){8-11}
\multicolumn{1}{c}{} & \multicolumn{1}{c}{\makecell[c]{Unit cost\\(revenue)}} & \multicolumn{1}{c}{(ACF)} & \multicolumn{1}{c}{(LP)} & \multicolumn{1}{c}{(W)} & \multicolumn{1}{c}{(OP)} & \multicolumn{1}{c}{\makecell[c]{Unit cost\\(revenue)}} & \multicolumn{1}{c}{(ACF)} & \multicolumn{1}{c}{(LP)} & \multicolumn{1}{c}{(W)} & \multicolumn{1}{c}{(OP)} \\
\cmidrule(l{3pt}r{3pt}){2-2} \cmidrule(l{3pt}r{3pt}){3-3} \cmidrule(l{3pt}r{3pt}){4-4} \cmidrule(l{3pt}r{3pt}){5-5} \cmidrule(l{3pt}r{3pt}){6-6} \cmidrule(l{3pt}r{3pt}){7-7} \cmidrule(l{3pt}r{3pt}){8-8} \cmidrule(l{3pt}r{3pt}){9-9} \cmidrule(l{3pt}r{3pt}){10-10} \cmidrule(l{3pt}r{3pt}){11-11}
  & (1) & (2) & (3) & (4) & (5) & (6) & (7) & (8) & (9) & (10)\\
\midrule
Plant Experience & -0.072*** & 0.485*** & 0.484*** & 0.484*** & 0.486*** & -0.076*** & 0.493*** & 0.491*** & 0.497*** & 0.492***\\
 & (0.002) & (0.011) & (0.011) & (0.011) & (0.011) & (0.003) & (0.012) & (0.013) & (0.012) & (0.012)\\
Targeted x Plant Experience & -0.007*** & 0.014* & 0.024*** & 0.030*** & 0.020*** & -0.005** & 0.004 & 0.013 & 0.005 & 0.012\\
 & (0.002) & (0.008) & (0.008) & (0.008) & (0.007) & (0.002) & (0.010) & (0.011) & (0.011) & (0.011)\\
Industry Experience & -0.011*** & 0.022 & 0.037** & 0.037** & 0.030* & 0.000 & -0.006 & 0.001 & -0.001 & -0.004\\
 & (0.003) & (0.014) & (0.016) & (0.017) & (0.016) & (0.003) & (0.012) & (0.015) & (0.016) & (0.015)\\
Targeted x Industry Experience & -0.004** & 0.030*** & 0.018** & 0.016** & 0.021*** & -0.010*** & 0.070*** & 0.062*** & 0.070*** & 0.061***\\
 & (0.001) & (0.009) & (0.007) & (0.007) & (0.007) & (0.003) & (0.012) & (0.011) & (0.011) & (0.011)\\
\addlinespace[0.75em]
\multicolumn{11}{p{\linewidth}}{\textbf{Controls}}\\
\hspace{1em}Control for Plant Size & Yes & Yes & Yes & Yes & Yes & Yes & Yes & Yes & Yes & Yes\\
\hspace{1em}Control for Capital & Yes & Yes & Yes & Yes & Yes & Yes & Yes & Yes & Yes & Yes\\
\hspace{1em}Control for Skill Ratio & Yes & Yes & Yes & Yes & Yes & Yes & Yes & Yes & Yes & Yes\\
\hspace{1em}Control for Investment & Yes & Yes & Yes & Yes & Yes & Yes & Yes & Yes & Yes & Yes\\
\hspace{1em}Control for Intermediates & Yes & Yes & Yes & Yes & Yes & Yes & Yes & Yes & Yes & Yes\\
\hspace{1em}Polynomial Controls & No & No & No & No & No & No & No & No & No & No\\
\addlinespace[0.75em]
\multicolumn{11}{p{\linewidth}}{\textbf{ }}\\
\hspace{1em}Plant Effect & Yes & Yes & Yes & Yes & Yes & Yes & Yes & Yes & Yes & Yes\\
\hspace{1em}Year Effect & Yes & Yes & Yes & Yes & Yes & Yes & Yes & Yes & Yes & Yes\\
\hspace{1em}Industry Effect & Yes & Yes & Yes & Yes & Yes & Yes & Yes & Yes & Yes & Yes\\
\hspace{1em}\(R^2\) & 0.890 & 0.691 & 0.682 & 0.678 & 0.600 & 0.890 & 0.689 & 0.680 & 0.675 & 0.597\\
\hspace{1em}Observations & 250853 & 235940 & 235940 & 235940 & 235940 & 250853 & 235940 & 235940 & 235940 & 235940\\
\hspace{1em}Clusters (Industry and Plant) & 489 x 59978 & 489 x 57942 & 489 x 57942 & 489 x 57942 & 489 x 57942 & 489 x 59978 & 489 x 57942 & 489 x 57942 & 489 x 57942 & 489 x 57942\\
\addlinespace[0.75em]
\multicolumn{11}{p{\linewidth}}{\textbf{Combined Effects}}\\
\hline
\hspace{1em}Linear Combination (Plant-Level) & -0.078 & 0.499 & 0.508 & 0.514 & 0.506 & -0.081 & 0.497 & 0.504 & 0.502 & 0.504\\
\hspace{1em}(St.Err.) & (0.003) & (0.013) & (0.013) & (0.013) & (0.012) & (0.003) & (0.014) & (0.014) & (0.015) & (0.014)\\
\hspace{1em}Linear Combination (Industry-Level) & -0.014 & 0.052 & 0.056 & 0.053 & 0.051 & -0.010 & 0.064 & 0.063 & 0.069 & 0.058\\
\hspace{1em}(St.Err.) & (0.003) & (0.014) & (0.017) & (0.018) & (0.017) & (0.003) & (0.016) & (0.018) & (0.018) & (0.018)\\
\bottomrule
\end{tabular}
\begin{tablenotes}[para]
\item \textit{\hspace{1em}\textit{Notes.}} 
\item This table shows the robustness of plant-level estimates from equation (5) for alternative outcomes.  Unit Cost is measured using total real intermediate costs per unit of (real) revenue. TFP outcomes are  estimated using Ackerberg-Caves-Frazer (ACF), Levinsohn-Petrin (LP), Olley-Pakes  (OP), and Wooldridge (W) methods. Panel A shows estimates for log Experience, and  Panel B shows log Experience per worker. 'Plant Experience' refers to plant-level  cumulative learning, and 'Industry Experience' refers to industry-level learning,  calculated at the 4-digit industry level. All equations control for log plant size  (workers). Additional controls include (log): capital intensity, skill ratio, investment  per worker, and intermediate input intensity per worker. Linear Combination, at  the bottom, gives the combined effects for Plant and Industry Experience for HCI  establishments. All specifications are estimated using plant, industry, and year  fixed effects. Two-way standard errors are clustered at the industry and plant levels.
\end{tablenotes}
\end{threeparttable}}
\end{table}
\elandscape

\blandscape
\begin{table}
\centering
\caption{\label{tab:appendixtradepolicy}Differences in Trade Policy by Treatment Status, 1968-1982}
\centering
\resizebox{\ifdim\width>\linewidth\linewidth\else\width\fi}{!}{
\fontsize{9}{11}\selectfont
\begin{threeparttable}
\begin{tabular}[t]{lcccccccccccc}
\toprule
\multicolumn{1}{c}{\bgroup\fontsize{9}{11}\selectfont  \egroup{}} & \multicolumn{8}{c}{\bgroup\fontsize{9}{11}\selectfont Outcomes: Levels\egroup{}} & \multicolumn{4}{c}{\bgroup\fontsize{9}{11}\selectfont Outcomes: Changes\egroup{}} \\
\cmidrule(l{3pt}r{3pt}){2-9} \cmidrule(l{3pt}r{3pt}){10-13}
\multicolumn{1}{c}{\bgroup\fontsize{9}{11}\selectfont  \egroup{}} & \multicolumn{4}{c}{\bgroup\fontsize{9}{11}\selectfont Tariff Rate (log)\egroup{}} & \multicolumn{4}{c}{\bgroup\fontsize{9}{11}\selectfont \makecell[c]{Quantitative\\Restrictions (log)}\egroup{}} & \multicolumn{2}{c}{\bgroup\fontsize{9}{11}\selectfont Tariff Rate (log)\egroup{}} & \multicolumn{2}{c}{\bgroup\fontsize{9}{11}\selectfont \makecell[c]{Quantitative\\Restrictions (log)}\egroup{}} \\
\cmidrule(l{3pt}r{3pt}){2-5} \cmidrule(l{3pt}r{3pt}){6-9} \cmidrule(l{3pt}r{3pt}){10-11} \cmidrule(l{3pt}r{3pt}){12-13}
  & (1) & (2) & (3) & (4) & (5) & (6) & (7) & (8) & (9) & (10) & (11) & (12)\\
\midrule
\addlinespace[1em]
\multicolumn{13}{l}{\textbf{Panel A) Output Protection}}\\
\hline
\hspace{1em}Targeted & -0.438*** & -0.492*** & -0.430*** & -0.483*** & -0.146** & -0.190*** & -0.138** & -0.182*** & 0.012 & 0.012 & 0.028 & 0.022\\
\hspace{1em} & (0.123) & (0.104) & (0.123) & (0.104) & (0.057) & (0.052) & (0.058) & (0.053) & (0.022) & (0.024) & (0.017) & (0.014)\\
\addlinespace[0.3em]
\multicolumn{13}{l}{\textbf{}}\\
\hspace{1em}\hspace{1em}Year Effect & Yes & Yes & Yes & Yes & Yes & Yes & Yes & Yes & Yes & Yes & Yes & \vphantom{1} Yes\\
\hspace{1em}\hspace{1em}Controls &  & Yes &  & Yes &  & Yes &  & Yes &  & Yes &  & \vphantom{1} Yes\\
\hspace{1em}\hspace{1em}Sample & Full & Full & Post-1973 & Post-1973 & Full & Full & Post-1973 & Post-1973 & Full & Full & Full & \vphantom{1} Full\\
\hspace{1em}\hspace{1em}\(R^2\) & 0.149 & 0.527 & 0.131 & 0.533 & 0.088 & 0.265 & 0.097 & 0.291 & 0.160 & 0.203 & 0.062 & 0.250\\
\hspace{1em}\hspace{1em}Observations & 522 & 516 & 435 & 430 & 435 & 430 & 348 & 344 & 261 & 258 & 435 & 430\\
\hspace{1em}\hspace{1em}Clusters & 87 & 86 & 87 & 86 & 87 & 86 & 87 & 86 & 87 & 86 & 87 & \vphantom{1} 86\\
\addlinespace[1em]
\multicolumn{13}{l}{\textbf{Panel B) Exposure to Input Protection}}\\
\hspace{1em}Targeted & -0.199** & -0.314*** & -0.234** & -0.356*** & -0.044* & -0.070*** & -0.051* & -0.076*** & -0.042*** & -0.052*** & -0.021** & -0.015\\
\hspace{1em} & (0.098) & (0.076) & (0.102) & (0.079) & (0.026) & (0.024) & (0.026) & (0.023) & (0.012) & (0.014) & (0.009) & (0.011)\\
\addlinespace[0.3em]
\multicolumn{13}{l}{\textbf{}}\\
\hspace{1em}\hspace{1em}Year Effect & Yes & Yes & Yes & Yes & Yes & Yes & Yes & Yes & Yes & Yes & Yes & Yes\\
\hspace{1em}\hspace{1em}Controls &  & Yes &  & Yes &  & Yes &  & Yes &  & Yes &  & Yes\\
\hspace{1em}\hspace{1em}Sample & Full & Full & Post-1973 & Post-1973 & Full & Full & Post-1973 & Post-1973 & Full & Full & Full & Full\\
\hspace{1em}\hspace{1em}\(R^2\) & 0.123 & 0.256 & 0.091 & 0.243 & 0.158 & 0.252 & 0.169 & 0.265 & 0.184 & 0.297 & 0.198 & 0.263\\
\hspace{1em}\hspace{1em}Observations & 522 & 516 & 435 & 430 & 435 & 430 & 348 & 344 & 435 & 430 & 261 & 258\\
\hspace{1em}\hspace{1em}Clusters & 87 & 86 & 87 & 86 & 87 & 86 & 87 & 86 & 87 & 86 & 87 & 86\\
\bottomrule
\end{tabular}
\begin{tablenotes}[para]
\item \textit{\hspace{1em}\textit{Notes.}} 
\item Table shows trade policy by treatment status (targeted vs.  non-targeted), using nominal trade policy data for 1968-1982 (intermittent). Columns  (1-8) show estimates in levels and columns (9-12) show changes. All regressions  are at the 4-digit industry level. Columns (1-4) report estimates for log tariffs.  Columns (5-8) report estimates for log quantitative restriction coverage. Columns  (9-10) show estimates for changes in log tariff rates. Columns (11-12) show  estimates for changes in log quantitative restrictions. Panel A presents tariff and quantitative restriction outcomes  for output market protection (industry-level): the average level or change in tariff  or quantitative restriction coverage. Panel B shows outcomes for input protection.  Exposure to input protection is calculated using the weighted sum of tariffs or  QRs for an industry's input basket, with weights taken from the 1970 input-output  accounts. See text for calculation. Sample refers to whether all five periods are used, or whether only post-HCI (1973) observations are used.
\end{tablenotes}
\end{threeparttable}}
\end{table}
\elandscape

\FloatBarrier

\begine

\renewcommand{\arraystretch}{.9}
\begin{table}
\centering
\caption{\label{tab:appendixprepostlfoutput}Total Linkage Exposure and Output}
\centering
\resizebox{\ifdim\width>\linewidth\linewidth\else\width\fi}{!}{
\fontsize{9}{11}\selectfont
\begin{threeparttable}
\begin{tabular}[t]{lcccc}
\toprule
\multicolumn{1}{c}{\bgroup\fontsize{10}{12}\selectfont  \egroup{}} & \multicolumn{4}{c}{\bgroup\fontsize{10}{12}\selectfont Outcome: Value Added (log)\egroup{}} \\
\cmidrule(l{2pt}r{2pt}){2-5}
\multicolumn{1}{c}{\bgroup\fontsize{10}{12}\selectfont  \egroup{}} & \multicolumn{2}{c}{\bgroup\fontsize{10}{12}\selectfont A) Five-Digit Panel (1970-1986)\egroup{}} & \multicolumn{2}{c}{\bgroup\fontsize{10}{12}\selectfont B) Four-Digit Panel (1967-1986)\egroup{}} \\
\cmidrule(l{3pt}r{3pt}){2-3} \cmidrule(l{3pt}r{3pt}){4-5}
\multicolumn{1}{c}{\em{ }} & \multicolumn{1}{c}{\em{Full Sample}} & \multicolumn{1}{c}{\em{Non-HCI Sample}} & \multicolumn{1}{c}{\em{Full Sample}} & \multicolumn{1}{c}{\em{Non-HCI Sample}} \\
\cmidrule(l{3pt}r{3pt}){2-2} \cmidrule(l{3pt}r{3pt}){3-3} \cmidrule(l{3pt}r{3pt}){4-4} \cmidrule(l{3pt}r{3pt}){5-5}
 & (1) & (2) & (3) & (4)\\
\midrule
\hspace{1em}Post \(\times\) Forward Linkage & 1.909*** & 3.388*** & 0.988** & 1.512*\\
\hspace{1em} & (0.516) & (0.857) & (0.485) & (0.853)\\
\hspace{1em}Post \(\times\) Backward Linkage & -0.0536 & 0.0452 & -0.574 & -1.316**\\
\hspace{1em} & (0.175) & (0.197) & (0.393) & (0.511)\\
\addlinespace[0.25em]
\multicolumn{5}{l}{\textbf{ }}\\
\hspace{1em}\hspace{1em}Industry Effects & Yes & Yes & Yes & Yes\\
\hspace{1em}\hspace{1em}Year Effects & Yes & Yes & Yes & Yes\\
\hspace{1em}\hspace{1em}Targeted X Year & Yes & No & Yes & No\\
\hspace{1em}\hspace{1em}\(R^2\) & 0.777 & 0.765 & 0.849 & 0.828\\
\hspace{1em}\hspace{1em}Observations & 4720 & 2986 & 1750 & 1096\\
\hspace{1em}\hspace{1em}Clusters & 278 & 176 & 88 & 55\\
\bottomrule
\end{tabular}
\begin{tablenotes}[para]
\item \textit{\hspace{1em}\textit{Notes.}} 
\item This table shows average differences-in-differences estimates , before and after 1973. Estimates come from the main DD linkage specification. Both linkage  interactions (forward and backward) are shown. Note that dynamic figures present  only estimates for the linkage of interest.
\end{tablenotes}
\end{threeparttable}}
\end{table}

\FloatBarrier

\renewcommand{\arraystretch}{.9}
\begin{table}
\centering
\caption{\label{tab:appendixprepostlfprices}Total Linkage Exposure and Output Prices}
\centering
\fontsize{9}{11}\selectfont
\begin{threeparttable}
\begin{tabular}[t]{lcccc}
\toprule
\multicolumn{1}{c}{\bgroup\fontsize{10}{12}\selectfont  \egroup{}} & \multicolumn{4}{c}{\bgroup\fontsize{10}{12}\selectfont Outcome: Prices (log)\egroup{}} \\
\cmidrule(l{2pt}r{2pt}){2-5}
\multicolumn{1}{c}{\bgroup\fontsize{10}{12}\selectfont  \egroup{}} & \multicolumn{2}{c}{\bgroup\fontsize{10}{12}\selectfont A) Five-Digit Panel (1970-1986)\egroup{}} & \multicolumn{2}{c}{\bgroup\fontsize{10}{12}\selectfont B) Four-Digit Panel (1967-1986)\egroup{}} \\
\cmidrule(l{1pt}r{1pt}){2-3} \cmidrule(l{1pt}r{1pt}){4-5}
\multicolumn{1}{c}{\em{}} & \multicolumn{1}{c}{\em{Full Sample}} & \multicolumn{1}{c}{\em{Non-HCI Sample}} & \multicolumn{1}{c}{\em{Full Sample}} & \multicolumn{1}{c}{\em{Non-HCI Sample}} \\
\cmidrule(l{3pt}r{3pt}){2-2} \cmidrule(l{3pt}r{3pt}){3-3} \cmidrule(l{3pt}r{3pt}){4-4} \cmidrule(l{3pt}r{3pt}){5-5}
 & (1) & (2) & (3) & (4)\\
\midrule
Post \(\times\) Forward Linkage & -0.289*** & -0.406*** & -0.344*** & -0.421***\\
 & (0.0726) & (0.0883) & (0.111) & (0.158)\\
Post \(\times\) Backward Linkage & 0.0500*** & 0.0463*** & 0.103 & 0.241***\\
 & (0.0127) & (0.0100) & (0.0651) & (0.0318)\\
\addlinespace[0.25em]
\multicolumn{5}{l}{\textbf{ }}\\
\hspace{1em}Industry Effects & Yes & Yes & Yes & Yes\\
\hspace{1em}Year Effects & Yes & Yes & Yes & Yes\\
\hspace{1em}Targeted X Year & Yes & No & Yes & No\\
\hspace{1em}\(R^2\) & 0.958 & 0.943 & 0.963 & 0.959\\
\hspace{1em}Observations & 4721 & 2987 & 1751 & 1097\\
\hspace{1em}Clusters & 278 & 176 & 88 & 55\\
\bottomrule
\end{tabular}
\begin{tablenotes}[para]
\item \textit{\hspace{1em}\textit{Notes.}} 
\item This table shows average differences-in-differences estimates,  before and after 1973. Estimates come from the main DD linkage specification. Both linkage  interactions (forward and backward) are shown. Note that dynamic figures present  only estimates for the linkage of interest.
\end{tablenotes}
\end{threeparttable}
\end{table}

\FloatBarrier

\blandscape
\renewcommand{\arraystretch}{.9}
\begin{table}
\centering
\caption{\label{tab:appprepostlinkmoredev}Direct Linkage Exposure and Industrial Development Outcomes}
\centering
\resizebox{\ifdim\width>\linewidth\linewidth\else\width\fi}{!}{
\fontsize{9}{11}\selectfont
\begin{threeparttable}
\begin{tabular}[t]{lcccccccccccccccccc}
\toprule
\multicolumn{1}{c}{\bgroup\fontsize{10}{12}\selectfont  \egroup{}} & \multicolumn{10}{c}{\bgroup\fontsize{10}{12}\selectfont Panel A) Five-Digit Panel (1970-1986)\egroup{}} & \multicolumn{8}{c}{\bgroup\fontsize{10}{12}\selectfont Panel B) Four-Digit Panel (1967-1986)\egroup{}} \\
\cmidrule(l{1pt}r{1pt}){2-11} \cmidrule(l{1pt}r{1pt}){12-19}
\multicolumn{1}{c}{\bgroup\fontsize{10}{12}\selectfont \egroup{}} & \multicolumn{10}{c}{\bgroup\fontsize{10}{12}\selectfont \makecell[c]{Outcomes (log)\\}\egroup{}} & \multicolumn{8}{c}{\bgroup\fontsize{10}{12}\selectfont \makecell[c]{Outcomes (log)\\}\egroup{}} \\
\multicolumn{1}{c}{\bgroup\fontsize{10}{12}\selectfont \egroup{}} & \multicolumn{2}{c}{\bgroup\fontsize{10}{12}\selectfont Employment\egroup{}} & \multicolumn{2}{c}{\bgroup\fontsize{10}{12}\selectfont Num. Plants\egroup{}} & \multicolumn{2}{c}{\bgroup\fontsize{10}{12}\selectfont Labor Prod.\egroup{}} & \multicolumn{2}{c}{\bgroup\fontsize{10}{12}\selectfont TFP (ACF)\egroup{}} & \multicolumn{2}{c}{\bgroup\fontsize{10}{12}\selectfont Avg. Wage.\egroup{}} & \multicolumn{2}{c}{\bgroup\fontsize{10}{12}\selectfont Employment\egroup{}} & \multicolumn{2}{c}{\bgroup\fontsize{10}{12}\selectfont Num. Plants\egroup{}} & \multicolumn{2}{c}{\bgroup\fontsize{10}{12}\selectfont Labor Prod.\egroup{}} & \multicolumn{2}{c}{\bgroup\fontsize{10}{12}\selectfont Avg. Wage.\egroup{}} \\
\cmidrule(l{3pt}r{3pt}){2-3} \cmidrule(l{3pt}r{3pt}){4-5} \cmidrule(l{3pt}r{3pt}){6-7} \cmidrule(l{3pt}r{3pt}){8-9} \cmidrule(l{3pt}r{3pt}){10-11} \cmidrule(l{3pt}r{3pt}){12-13} \cmidrule(l{3pt}r{3pt}){14-15} \cmidrule(l{3pt}r{3pt}){16-17} \cmidrule(l{3pt}r{3pt}){18-19}
 & (1) & (2) & (3) & (4) & (5) & (6) & (7) & (8) & (9) & (10) & (1) & (2) & (3) & (4) & (5) & (6) & (7) & (8)\\
\midrule
Post \(\times\) Forward Linkage & 1.788*** & 3.403*** & 1.499*** & 2.828*** & 0.699** & 0.473 & 1.353*** & 0.869 & 0.629*** & 0.330 & 1.645** & 2.925** & 1.897*** & 3.564*** & 0.0722 & -0.283 & 0.182 & 0.107\\
 & (0.685) & (1.108) & (0.393) & (0.562) & (0.325) & (0.390) & (0.499) & (0.588) & (0.220) & (0.233) & (0.777) & (1.143) & (0.639) & (0.677) & (0.297) & (0.424) & (0.173) & (0.219)\\
Post \(\times\) Backward Linkage & -0.0917 & 0.116 & 0.105 & 0.177 & 0.116 & 0.104 & 0.0542 & 0.00243 & 0.0424 & 0.0307 & -0.373 & -1.084 & 0.279 & 0.0609 & -0.137 & -0.890 & -0.140 & -0.702**\\
 & (0.258) & (0.279) & (0.104) & (0.115) & (0.105) & (0.123) & (0.0926) & (0.0680) & (0.0736) & (0.0866) & (0.323) & (0.727) & (0.210) & (0.397) & (0.287) & (0.539) & (0.193) & (0.319)\\
Industry Effects & Yes & Yes & Yes & Yes & Yes & Yes & Yes & Yes & Yes & Yes & Yes & Yes & Yes & Yes & Yes & Yes & Yes & Yes\\
Year Effects & Yes & Yes & Yes & Yes & Yes & Yes & Yes & Yes & Yes & Yes & Yes & Yes & Yes & Yes & Yes & Yes & Yes & Yes\\
Targeted \(\times\) Year & Yes & No & Yes & No & Yes & No & Yes & No & Yes & No & Yes & No & Yes & No & Yes & No & Yes & No\\
\(R^2\) & 0.797 & 0.807 & 0.867 & 0.873 & 0.750 & 0.687 & 0.706 & 0.674 & 0.774 & 0.714 & 0.853 & 0.848 & 0.892 & 0.895 & 0.847 & 0.777 & 0.853 & 0.791\\
Observations & 4726 & 2992 & 4726 & 2992 & 4714 & 2981 & 4214 & 2682 & 4721 & 2987 & 1760 & 1100 & 1760 & 1100 & 1750 & 1096 & 1751 & 1097\\
Clusters & 278 & 176 & 278 & 176 & 278 & 176 & 264 & 167 & 278 & 176 & 88 & 55 & 88 & 55 & 88 & 55 & 88 & 55\\
Sample & Full & Non-HCI & Full & Non-HCI & Full & Non-HCI & Full & Non-HCI & Full & Non-HCI & Full & Non-HCI & Full & Non-HCI & Full & Non-HCI & Full & Non-HCI\\
\bottomrule
\end{tabular}
\begin{tablenotes}[para]
\item \textit{\hspace{1em}\textit{Notes.}} 
\item This table shows average differences-in-differences estimates,  before and after 1973. Estimates come from the main DD linkage specification. Both linkage  interactions (forward and backward) are shown. Note that dynamic figures present  only estimates for the linkage of interest.
\end{tablenotes}
\end{threeparttable}}
\end{table}
\elandscape

\blandscape
\renewcommand{\arraystretch}{.9}
\begin{table}
\centering
\caption{\label{tab:appprepostlfmoredev}Total Linkage Exposure and Industrial Development}
\centering
\resizebox{\ifdim\width>\linewidth\linewidth\else\width\fi}{!}{
\fontsize{9}{11}\selectfont
\begin{threeparttable}
\begin{tabular}[t]{lcccccccccccccccccc}
\toprule
\multicolumn{1}{c}{\bgroup\fontsize{10}{12}\selectfont  \egroup{}} & \multicolumn{10}{c}{\bgroup\fontsize{10}{12}\selectfont Panel A) Five-Digit Panel (1970-1986)\egroup{}} & \multicolumn{8}{c}{\bgroup\fontsize{10}{12}\selectfont Panel B) Four-Digit Panel (1967-1986)\egroup{}} \\
\cmidrule(l{1pt}r{1pt}){2-11} \cmidrule(l{1pt}r{1pt}){12-19}
\multicolumn{1}{c}{\bgroup\fontsize{10}{12}\selectfont \egroup{}} & \multicolumn{10}{c}{\bgroup\fontsize{10}{12}\selectfont \makecell[c]{Outcomes (log)\\}\egroup{}} & \multicolumn{8}{c}{\bgroup\fontsize{10}{12}\selectfont \makecell[c]{Outcomes (log)\\}\egroup{}} \\
\multicolumn{1}{c}{\bgroup\fontsize{10}{12}\selectfont \egroup{}} & \multicolumn{2}{c}{\bgroup\fontsize{10}{12}\selectfont Employment\egroup{}} & \multicolumn{2}{c}{\bgroup\fontsize{10}{12}\selectfont Num. Plants\egroup{}} & \multicolumn{2}{c}{\bgroup\fontsize{10}{12}\selectfont Labor Prod.\egroup{}} & \multicolumn{2}{c}{\bgroup\fontsize{10}{12}\selectfont TFP (ACF)\egroup{}} & \multicolumn{2}{c}{\bgroup\fontsize{10}{12}\selectfont Avg. Wage.\egroup{}} & \multicolumn{2}{c}{\bgroup\fontsize{10}{12}\selectfont Employment\egroup{}} & \multicolumn{2}{c}{\bgroup\fontsize{10}{12}\selectfont Num. Plants\egroup{}} & \multicolumn{2}{c}{\bgroup\fontsize{10}{12}\selectfont Labor Prod.\egroup{}} & \multicolumn{2}{c}{\bgroup\fontsize{10}{12}\selectfont Avg. Wage.\egroup{}} \\
\cmidrule(l{3pt}r{3pt}){2-3} \cmidrule(l{3pt}r{3pt}){4-5} \cmidrule(l{3pt}r{3pt}){6-7} \cmidrule(l{3pt}r{3pt}){8-9} \cmidrule(l{3pt}r{3pt}){10-11} \cmidrule(l{3pt}r{3pt}){12-13} \cmidrule(l{3pt}r{3pt}){14-15} \cmidrule(l{3pt}r{3pt}){16-17} \cmidrule(l{3pt}r{3pt}){18-19}
 & (1) & (2) & (3) & (4) & (5) & (6) & (7) & (8) & (9) & (10) & (1) & (2) & (3) & (4) & (5) & (6) & (7) & (8)\\
\midrule
Post \(\times\) Forward Linkage & 1.139*** & 2.449*** & 0.882*** & 1.868*** & 0.534*** & 0.550** & 0.821*** & 0.737** & 0.432*** & 0.382*** & 0.632 & 1.404* & 0.747** & 1.919*** & 0.0721 & -0.172 & 0.107 & 0.0401\\
 & (0.386) & (0.637) & (0.227) & (0.338) & (0.182) & (0.227) & (0.252) & (0.300) & (0.124) & (0.138) & (0.421) & (0.779) & (0.371) & (0.571) & (0.175) & (0.298) & (0.104) & (0.162)\\
Post \(\times\) Backward Linkage & -0.0856 & 0.0311 & 0.0287 & 0.0803 & 0.0482 & 0.0289 & -0.0119 & -0.0194 & 0.0148 & 0.000387 & -0.352 & -0.731** & -0.0472 & -0.188 & -0.151 & -0.508*** & -0.120 & -0.366***\\
 & (0.136) & (0.143) & (0.0552) & (0.0573) & (0.0567) & (0.0682) & (0.0460) & (0.0453) & (0.0390) & (0.0468) & (0.229) & (0.347) & (0.117) & (0.212) & (0.168) & (0.176) & (0.112) & (0.117)\\
Industry Effects & Yes & Yes & Yes & Yes & Yes & Yes & Yes & Yes & Yes & Yes & Yes & Yes & Yes & Yes & Yes & Yes & Yes & Yes\\
Year Effects & Yes & Yes & Yes & Yes & Yes & Yes & Yes & Yes & Yes & Yes & Yes & Yes & Yes & Yes & Yes & Yes & Yes & Yes\\
Targeted \(\times\) Year & Yes & No & Yes & No & Yes & No & Yes & No & Yes & No & Yes & No & Yes & No & Yes & No & Yes & No\\
\(R^2\) & 0.798 & 0.808 & 0.867 & 0.874 & 0.750 & 0.687 & 0.706 & 0.675 & 0.775 & 0.715 & 0.855 & 0.852 & 0.890 & 0.894 & 0.849 & 0.786 & 0.855 & 0.803\\
Observations & 4726 & 2992 & 4726 & 2992 & 4714 & 2981 & 4214 & 2682 & 4721 & 2987 & 1760 & 1100 & 1760 & 1100 & 1750 & 1096 & 1751 & 1097\\
Clusters & 278 & 176 & 278 & 176 & 278 & 176 & 264 & 167 & 278 & 176 & 88 & 55 & 88 & 55 & 88 & 55 & 88 & 55\\
Sample & Full & Non-HCI & Full & Non-HCI & Full & Non-HCI & Full & Non-HCI & Full & Non-HCI & Full & Non-HCI & Full & Non-HCI & Full & Non-HCI & Full & Non-HCI\\
\bottomrule
\end{tabular}
\begin{tablenotes}[para]
\item \textit{\hspace{1em}\textit{Notes.}} 
\item This table shows average differences-in-differences estimates, before and after 1973. Estimates come from the main DD linkage specification. Both linkage interactions (forward and backward) are shown. Note that dynamic figures present  only estimates for the linkage of interest.
\end{tablenotes}
\end{threeparttable}}
\end{table}
\elandscape

\clearpage
\part*{Online Supplemental Appendix}
\section*{Note on Supplemental Appendix}

This paper includes references to an Online Supplemental Appendix that contains additional robustness checks, historical details, and extended analysis. The Supplemental Appendix is available as a separate document and includes sections on:

\begin{itemize}
\item Additional historical background and policy details
\item Extended robustness checks and alternative specifications  
\item Additional data sources and construction details
\item Supplemental figures and tables
\end{itemize}

References to supplemental appendix sections refer to this separate document.

\end{document}